\documentclass[aip,graphicx]{revtex4-1}
\usepackage{booktabs,array,multirow,tabularx,siunitx}
\usepackage{graphicx}
\usepackage{xcolor}
\usepackage{amssymb}
\usepackage{braket}
\usepackage{amsmath}
\usepackage{braket}
\usepackage{bm}
\usepackage{url}
\usepackage{hyperref}
\usepackage{breakurl}
\usepackage{auto-pst-pdf}
\usepackage[T1]{fontenc}

\newcommand{\NECI}{{\tt NECI} }
\newcommand{\OMolcas}{{\tt OpenMolcas} }
\newcommand{\Molpro}{{\tt Molpro} }
\newcommand{\VASP}{{\tt VASP} }
\newcommand{\PySCF}{{\tt PySCF} }
\newcommand{\abs}[1]{\lvert #1 \rvert}
\providecommand{\calC}[0]{\ensuremath{{\cal{C}}}}

\begin{document}

\title{NECI: \emph{N}-Electron Configuration Interaction with an emphasis on state-of-the-art stochastic methods}

\author{Kai Guther}
\email{k.guther@fkf.mpg.de}
\affiliation{Max Planck Institute for Solid State Research, Heisenbergstr. 1,
  70569 Stuttgart, Germany}
\author{Robert J. Anderson}
\affiliation{Department of Physics, King's College London, Strand, London
  WC2R 2LS, United Kingdom}
\author{Nick S. Blunt}
\affiliation{Department of Chemistry, University of Cambridge, Lensfield
  Road, Cambridge CB2 1EW, United Kingdom}
\author{Nikolay A. Bogdanov}
\affiliation{Max Planck Institute for Solid State Research, Heisenbergstr. 1,
  70569 Stuttgart, Germany}
\author{Deidre Cleland}
\affiliation{CSIRO Data61, Docklands VIC 3008, Australia}
\author{Nike Dattani}
\affiliation{Department of Electrical and Computer
Engineering, University of Waterloo, 200 University Avenue, Waterloo,
Canada}
\author{Werner Dobrautz}
\affiliation{Max Planck Institute for Solid State Research, Heisenbergstr. 1,
  70569 Stuttgart, Germany}
\author{Khaldoon Ghanem}
\affiliation{Max Planck Institute for Solid State Research, Heisenbergstr. 1,
  70569 Stuttgart, Germany}
\author{Peter Jeszenski}
\affiliation{Centre for Theoretical Chemistry and Physics, NZ Institute for
  Advanced Study, Massey University, New Zealand}
\author{Niklas Liebermann}
\affiliation{Max Planck Institute for Solid State Research, Heisenbergstr. 1,
  70569 Stuttgart, Germany}
\author{Giovanni Li Manni}
\affiliation{Max Planck Institute for Solid State Research, Heisenbergstr. 1,
  70569 Stuttgart, Germany}
\author{Alexander Y. Lozovoi}
\affiliation{Max Planck Institute for Solid State Research, Heisenbergstr. 1,
  70569 Stuttgart, Germany}
\author{Hongjun Luo}
\affiliation{Max Planck Institute for Solid State Research, Heisenbergstr. 1,
  70569 Stuttgart, Germany}
\author{Dongxia Ma}
\affiliation{Max Planck Institute for Solid State Research, Heisenbergstr. 1,
  70569 Stuttgart, Germany}
\author{Florian Merz}
\affiliation{Lenovo HPC\&AI Innovation Center, Meitnerstr. 9, 70563
  Stuttgart}
\author{Catherine Overy}
\affiliation{Department of Chemistry, University of Cambridge, Lensfield
  Road, Cambridge CB2 1EW, United Kingdom}
\author{Markus Rampp}
\affiliation{Max Planck Computing and Data Facility (MPCDF), Gie{\ss}enbachstr.
  2, 85748 Garching, Germany}
\author{Pradipta Kumar Samanta}
\affiliation{Max Planck Institute for Solid State Research, Heisenbergstr. 1,
  70569 Stuttgart, Germany}
\author{Lauretta R. Schwarz}
\affiliation{Max Planck Institute for Solid State Research, Heisenbergstr. 1,
  70569 Stuttgart, Germany}
\affiliation{Department of Chemistry, University of Cambridge, Lensfield
  Road, Cambridge CB2 1EW, United Kingdom}
\author{James J. Shepherd}
\affiliation{Department of Chemistry \& Informatics Institute, University of
  Iowa}
\author{Simon D. Smart}
\affiliation{Department of Chemistry, University of Cambridge, Lensfield
  Road, Cambridge CB2 1EW, United Kingdom}
\author{Eugenio Vitale}
\affiliation{Max Planck Institute for Solid State Research, Heisenbergstr. 1,
  70569 Stuttgart, Germany}
\author{Oskar Weser}
\affiliation{Max Planck Institute for Solid State Research, Heisenbergstr. 1,
  70569 Stuttgart, Germany}
\affiliation{Department of Physics, King's College London, Strand, London
  WC2R 2LS, United Kingdom}
\author{George H. Booth}
\affiliation{Department of Physics, King's College London, Strand, London
  WC2R 2LS, United Kingdom}
\author{Ali Alavi}
\email{a.alavi@fkf.mpg.de}
\affiliation{Max Planck Institute for Solid State Research, Heisenbergstr. 1,
  70569 Stuttgart, Germany}
\affiliation{Department of Chemistry, University of Cambridge, Lensfield
  Road, Cambridge CB2 1EW, United Kingdom}

\date{\today}

\begin{abstract}
  We present \texttt{NECI}, a state-of-the-art implementation of the
  Full Configuration Interaction Quantum Monte Carlo algorithm, a method based on a stochastic
  application of the Hamiltonian matrix on a sparse sampling of the wave
  function. The program utilizes a very powerful parallelization and scales efficiently to more than
  24000 CPU cores. In this paper, we describe the core functionalities of \NECI and
  recent developments. This includes the capabilities to calculate ground and excited state
  energies, properties via the one- and two-body reduced density matrices, as well as spectral and Green's
  functions for \textit{ab initio} and model systems. A number of enhancements of the bare FCIQMC algorithm are available within
  \texttt{NECI}, allowing to use a partially deterministic formulation of the
  algorithm, working in a spin-adapted basis or supporting transcorrelated
  Hamiltonians. \NECI supports the FCIDUMP file format for integrals, supplying a convenient
  interface to numerous quantum chemistry programs and it is licensed under
  GPL-3.0.
  \\
  This article has been accepted by the Jouranl of Chemical Physics, after it
  is published, it will be found at \url{https://aip.scitation.org/journal/jcp}.
\end{abstract}

\maketitle

\section{\label{sec:intro}Introduction}
\NewDocumentCommand{\overarrow}{O{=} O{\uparrow} m}{  \overset{\makebox[0pt]{\begin{tabular}{@{}c@{}}#3\\[0pt]\ensuremath{#2}\end{tabular}}}{#1}
}
~\\
\NECI started off in the late 1990s as an exact diagonalisation code for model
quantum dots \cite{Alavi2000,ThompsonAlavi2002}, and has evolved into a code
to perform stochastic diagonalisation of large fermionic systems in finite but
large quantum chemical basis sets, using the Full Configuration Interaction
Quantum Monte Carlo (FCIQMC) algorithm \cite{BoothThomAlavi2009}. This
algorithm samples Slater determinant (i.e. antisymmetrized) Hilbert spaces
using signed {\em walkers}, by propagation of the walkers through stochastic
application of the second-quantized Hamiltonian onto the walker population. In
philosophy, it is similar to the continuum real-space Diffusion Monte Carlo
(DMC) algorithm.
However, unlike DMC, no fixed node approximation needs to be applied. Instead,
the nodal structure of the wavefunction, as encoded by the signed coefficients
of the sampled Slater determinants, emerges from the dynamics of the
simulation itself. However, being based on an FCI parametrization of the wave
function, the FCIQMC method exhibits a steep scaling with the number of electrons and is thus only suited for relatively 
small chemical systems compared to those accessible to DMC. While the common energy measures in FCIQMC methods, namely the projected, trial energies (cf section \ref{sec:trial_wf}) and the energy "shift", are not variational, a variational energy
can be computed from two parallel FCIQMC
calculations either directly (cf section \ref{sec:pt2}), or via the reduced density matrix (RDM) based energy estimator (cf
section \ref{sec:rdms}).

There are also similarities between the FCIQMC approach and the
Auxiliary-Field Quantum Monte Carlo (AFQMC) method\cite{sugar1981,koonin1986,Zhang2003}, both being stochastic projector
techniques formulated in second quantized spaces. The latter however works in an over-complete space of non-orthogonal Slater
determinants and relies on the phase-less approximation \cite{zhang1997} to eliminate the phase problem associated with the Hubbard-Stratonovich transformation
of the Coulomb interaction kernel, the quality of this approximation being reliant on the trial wavefunction used to constrain
the path. The objective of AFQMC is the measurement of observables such as the energy by sampling over the Hubbard-Stratonovich fields. 
FCIQMC on the other hand works in a fixed Slater determinant space
and relies on walker annihilation to overcome the fermion sign problem. The phase-less approximation renders the AFQMC method polynomial scaling, with an uncontrolled approximation, while i-FCIQMC, which is an in principle 
exact method, remains exponential scaling. Finally FCIQMC provides a direct measure of the CI amplitudes of the many-body wavefunction expressed in the given orbital basis, 
from which observables can be computed including elements of reduced density matrices (which do not commute with the Hamiltonian)  via pure estimators. Exact symmetry constraints, including total spin, can be incorporated into the formalism\cite{Dobrautz2019}. In this sense,
the FCIQMC method is closer in spirit to multi-reference CI methods used in quantum chemistry to study multi-reference problems rather
than the AFQMC method.

In its original formulation,
the algorithm is guaranteed to converge onto the ground-state wavefunction in the long imaginary-time propagation limit, provided a sufficient number of walkers is used. This number is generally
found to scale with the Hilbert space size, and is a manifestation of the
sign-problem in this method, essentially implying an exponential memory cost
in order to guarantee stable convergence onto the exact solution. In the
subsequent development of the {\em initiator} method
(i-FCIQMC)\cite{initiator-fciqmc}, this condition was relaxed to allow for
stable simulations at relatively low walker populations, much smaller than the
full Hilbert space size, albeit at the cost of a systematically improvable
bias. While the initiator adaptation removes the strict need for a minimum
walker number, it does not eliminate the exponential scaling of the method, such
that calculations become more and more challenging with increasing system
size. To give an idea of the capabilities of the \NECI implementation,
estimates for the accessible system sizes are given below. The rate of convergence of the initiator error with walker number has been found to be slow for large systems. This is a manifestation of size-inconsistency error which generally plagues linear Configuration Interaction methods. A very recent development of the {\em adaptive shift} method \cite{adaptive_shift}, mitigates this error substantially, enabling near-FCI quality results to be obtained for systems as large as benzene.

The development of the semi-stochastic method by Umrigar \textit{et al.} \cite{Petruzielo2012} and its further refinements \cite{Blunt2015} dramatically reduced the stochastic noise and hence improved the efficiency of the method.

\begin{figure}[t!]
\includegraphics[width=0.8\textwidth]{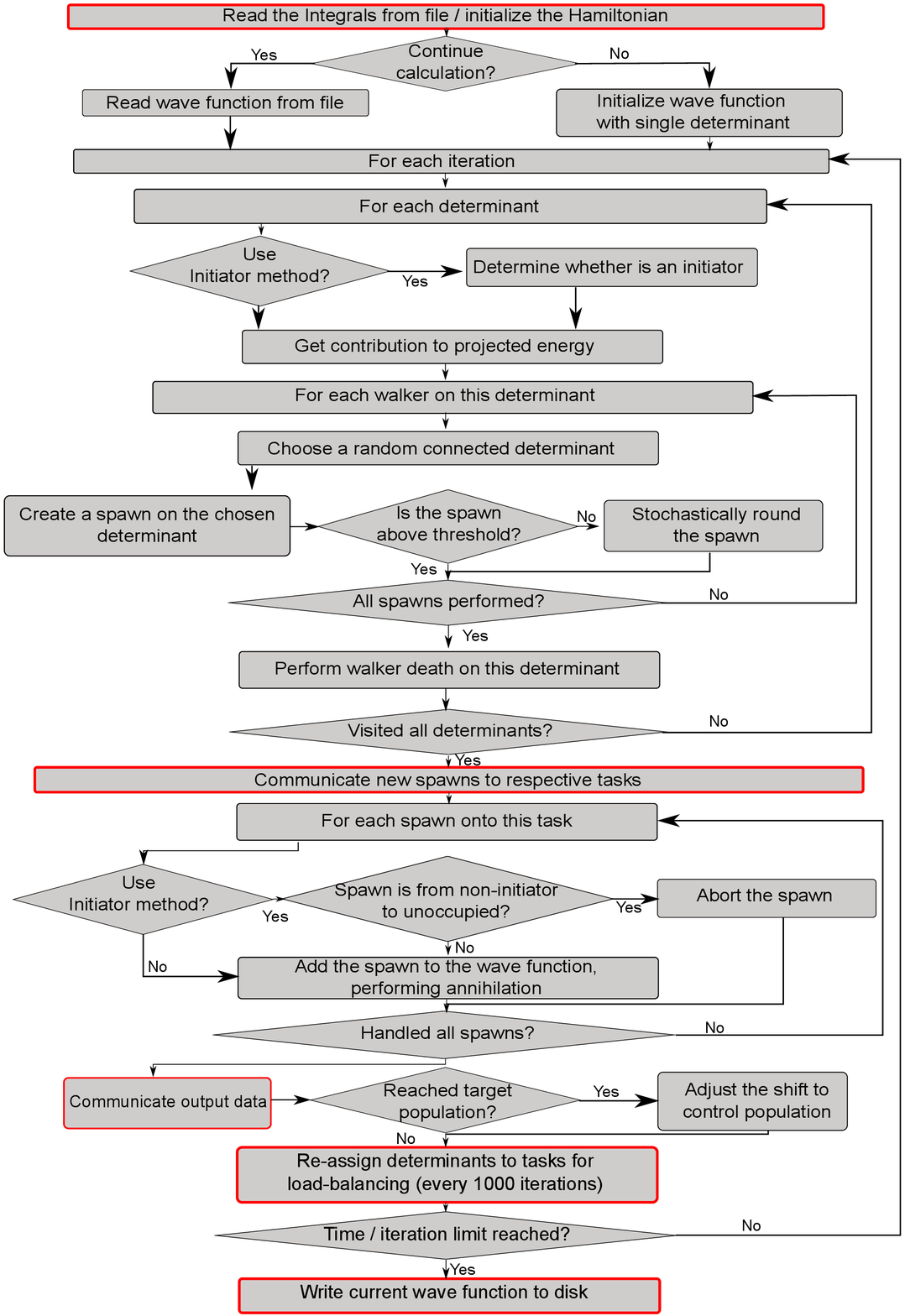}
\caption{Flow chart showing the basic initiator-FCIQMC implementation in \NECI. Marked in red are steps that require synchronisation between the MPI tasks and thus are not trivially parallelizable.}
\label{fig:fc}
\end{figure}

The FCIQMC algorithm, as well as its semi-stochastic and initiator versions,
are scalable on large parallel machines, thanks to the fact that walker
distribution can be distributed over many processors with relatively small
communication overhead. The methods, however, are not embarrassingly parallel,
owing to the annihilation step of the algorithm (see also figure ~\ref{fig:fc}).  For this reason,
parallelisation over very large numbers of processors is a highly non-trivial
task, but substantial progress has been made, and here we show that efficient
parallelisation up to more than 24000 CPU cores can be achieved with the current \NECI code.

The FCIQMC method has been generalised to excited states \cite{Blunt2015b} of
the same symmetry as the ground state and to the calculation of pure one-and two-particle reduced density matrices via the "replica-trick" \cite{Zhang1993,Overy2014,Blunt2014,Blunt2017} (and more recently three and four-particle RDMs \cite{Anderson2020}). The availability of RDMs enabled the development of the Stochastic CASSCF method \cite{LiManni2016, Thomas2015_2} for treating extremely large active spaces. More recently, a fully spin-adapted formulation of FCIQMC has been implemented based on the Graphical Unitary Group Approach \cite{Dobrautz2019}, which overcomes the previous limitations of spin-adaptation, which severely limited the  number of open-shell orbitals which could be handled. Other advanced  developments of FCIQMC in the NECI code include real-time propagation and application to spectroscopy \cite{guther2018}, Krylov-space FCIQMC \cite{Blunt2015}, and the
similarity transformed FCIQMC \cite{Luo2018, Dobrautz2019_b,Cohen2019,
Jezenski2019} which allows the direct incorporation of Jastrow and similar factors depending on explicit electron-electron variables into the wavefunction.

A number of stochastic methods have been developed as an extension or variation of the FCIQMC approach. These include density matrix quantum Monte Carlo (DMQMC), which allows the exact thermal density matrix to be sampled at any given temperature, and also allows straightforward estimation of general observables, including those which do not commute with the Hamiltonian\cite{Blunt2014, Malone2015}. Applications of DMQMC include providing accurate data for the warm dense electron gas\cite{Malone2016}. Although not implemented in NECI, DMQMC is available in the HANDE-QMC code\cite{hande}.

The FCIQMC method has lead to the development of a number of highly efficient deterministic selected CI methods, 
including the adaptive sampling CI method of Head-Gordon and co-workers \cite{tubman2016}, who also establish the connection with the much older perturbatively selected CIPSI method of Malrieu et al \cite{Huron1973} but with a modified search procedure, while the Heat-Bath CI method of Umrigar and coworkers \cite{holmes2016_b} was developed from the Heat-Bath excitation generation for FCIQMC \cite{holmes2016} together with an initiator-like criterion to select the connected determinants with extreme efficiency.  Later a sign-problem-free semi-stochastic evaluation of the Epstein-Nesbet perturbation energy was developed by Sharma et al \cite{sharma2017} to compute the missing dynamical correlation energy at second-order in a memory and CPU efficient manner. 
Other highly related developments to FCIQMC which originate in the numerical analysis literature include the Fast Randomised Iteration \cite{lim2017} and further developments by Weare, Berkelbach and coworkers \cite{greene2019}, and co-ordinate descent FCI of Lu and coworkers \cite{wang2019}.

Depending on the utilized features, the number of electrons and accessible
basis sizes can vary. The i-FCIQMC implementation including the
semi-stochastic version is highly scalable and has been successfully applied
to Hilbert space sizes of up to $10^{108}$ with 54
electrons \cite{shepherd2012}. Atomic basis sets up to aug-cc-pCV8Z for
first-row atoms (1138 spin orbitals) are treatable \cite{dattani2020}.
Reduced density matrices can routinely be calculated for use in accurate Stochastic-MCSCF \cite{LiManni2016,LiManni2016} for active spaces containing up to 40 electrons and 38 spatial orbitals\cite{LiManni2018,limanni2019a}. Real-time calculations are computationally more demanding but can still be performed for first-row dimers using cc-pVQZ basis sets \cite{guther2018}. For the similarity transformed FCIQMC method, the limiting factor is not the convergence of the FCIQMC, but the storage of the three-body interaction terms imposing a limit of $\sim 100$ spatial orbitals on currently available hardware \cite{Cohen2019}.
Optimized implementations for the application to lattice model systems, like the Hubbard\cite{Hubbard1963} (in a real- and momentum space formulation), $t-J$ and the Heisenberg models for a variety of lattice geometries, are implemented in \NECI. 
The applicability of FCIQMC to the Hubbard model strongly depends on the interaction strength $U/t$. 
For the very weakly correlated regime $U/t < 1$, FCIQMC is employable up to 70 lattice sites \cite{Shepherd2014}, using a momentum-space basis. 
In the interesting, yet most problematic, intermediate interaction strength regime in two dimensions, the transcorrelated (similarity-transformed) FCIQMC is necessary to obtain reliable energies in systems up to 50 sites (at and near half-filling)\cite{Dobrautz2019_b}.

The FCIQMC algorithm as implemented in \NECI is based on a sparse
representation of the wave function and a stochastic application of the
Hamiltonian. We start with the full wave function
\begin{equation}
\label{eq:fci}
\ket{\psi_\text{FCI}} = \sum_i C_i \ket{D_i}\,,
\end{equation}
with coefficients $C_i$ in a many-body basis $\ket{D_i}$. \NECI supports
Slater determinants or CSFs as a many-body basis, for simplicity, for now, the
usage of determinants is assumed, but the algorithm is analogous for CSFs, see also
section ~\ref{sec:guga_fciqmc}. The FCIQMC wave function is not
normalized. The ground state of a Hamiltonian $\hat{H}$ is now obtained by iterative imaginary time-evolution, with
the propagator expanded to first order using a discrete time-step
$\Delta \tau$ such that
\begin{equation}
\label{eq:projector}
\ket{\psi(\tau + \Delta\tau)} = \left(1 - \Delta \tau (\hat{H} - S(\tau))\right) \ket{\psi(\tau)}\,,
\end{equation}
which converges to the ground state of $\hat{H}$ for $\tau \rightarrow \infty$ for
$|\Delta \tau| < \frac{2}{W}$ where $W$ is the difference between the largest
and smallest eigenvalue of $\hat{H}$\cite{spectral-width}. Here, $S(\tau)$ is a diagonal shift applied to
$\hat{H}$, which is iteratively updated to match the ground state energy.

The full wave function is stored in a compressed manner, where only
coefficients above a given threshold value $C_\text{min}$ are
stored. Coefficients smaller than $C_\text{min}$ are stochastically rounded.
That is, in every iteration, a wave function given by coefficients $C_i(\tau)$
is stored such that
\begin{equation}
C_i(\tau) = \begin{cases} C_i & \mathrm{if}\; |C_i| > C_\text{min} \\
       \mathrm{sign}(C_i)\,C_\text{min}
       & \mathrm{else\;w.\;prob\;} \frac{|C_i|}{C_\text{min}} \\
       0 & \mathrm{otherwise}\end{cases}\,,
\end{equation}
such that $\Braket{C_i(\tau)} = C_i$. This compression is applied in every step
of the algorithm that affects the coefficients. The value $|C_i(\tau)|$ is referred to
as \textit{walker number} of the determinant $\ket{D_i}$, so $\ket{D_i}$ is
said to have $|C_i(\tau)|$ walkers assigned.

Applying the Hamiltonian to this compressed wave function is done by
separating it into a diagonal and an off-diagonal part as
\begin{align}
\label{eq:fciqmc_dyn}
\ket{\psi(\tau + \Delta \tau)} =& \underbrace{\sum_i\left(1 - \Delta \tau(H_{ii} - S(\tau))\right)
      C_i(\tau)\ket{D_i}}_\text{(b) Death step} \overarrow[-][\downarrow]{(c) Annihilation
    step}\nonumber\\&\underbrace{\sum_i
    \sum_{j\neq i} \Delta \tau H_{ji} C_i(\tau) \ket{D_j}}_\text{(a) Spawn step}\,,
\end{align}
and then performed in the three labeled steps $(a)-(c)$. First, in the \textit{spawning step},
the off-diagonal part is evaluated by stochastically sampling the sum over
$j$, storing the resulting \textit{spawned} wave-function as a separate entity
as described in the flow chart in Figure~\ref{fig:fc}. Then, in
the \textit{death step}, the diagonal contribution is evaluated
deterministically, following a stochastic rounding of the resulting
coefficients. This step is performed in-place, since the coefficients of the
previous iterations are not required anymore. Finally, the spawned wave-function
from the off-diagonal part is added in the \textit{annihilation step}, summing
up all contributions from the spawned wave-function to each determinant. \NECI
implements the initiator method, too, which labels a class of determinants
as \textit{initiators}, typically those with an associated walker number above
a given threshold, and effectively zeroes all matrix elements between
non-initiator determinants and determinants with $C_i(\tau)=0$. The
implementation thereof is also sketched in figure \ref{fig:fc}.

In the context of FCIQMC calculations, the core functionality of \NECI consists of
a highly parallelizable implementation of the initiator FCIQMC
method \cite{initiator-fciqmc} for both real and complex Hamiltonians.
There is both a generic interface for {\it ab initio} systems, specialized
implementations for the Hubbard and Heisenberg models, as well as the uniform
electron gas.
The interface for passing input information on the system to \NECI is discussed
in section~\ref{sec:interface}. To enable continuation of calculations at a
later point, \NECI can write the instantaneous wave function and
current parameters\textemdash such as the shift value\textemdash to disk, saving the current state of the
calculation.

The \NECI program~\cite{neci} itself is written in Fortran, and requires extended Fortran
2003 support, which is the default for current Fortran
compilers. Parallelization is achieved using the Message Passing Interface
(MPI)~\cite{mpi_standard}, and support for MPI 3.0 or newer is required. \NECI further requires
the BLAS~\cite{blackford2002} and LAPACK~\cite{laug} lineara algebra
libraries, which are available in numerous packages. Usage of the HDF5 library~\cite{hdf5} for parallel I/O is supported, but not
required. If used, the linked HDF5 library has to be built with Fortran
support and for parallel applications. For installation, \texttt{cmake} is
required, as well as the \texttt{fypp} Fortran
preprocessor\cite{fypp}. For pseudo random number generation, the double
precision SIMD oriented fast Mersenne Twister (dSFMT) \cite{saito2008, dSFMT}
implementation of the Mersenne Twister method \cite{matsumoto1998} is used. The stable version of the program can be obtained from github at
\url{https://github.com/ghb24/NECI_STABLE}, licensed under the GNU General
Public License 3.0. Some advanced or experimental features are only contained in the development version, for access to the
development version, please contact the corresponding authors. All features
presented here are eventually to be integrated to the stable version. Detailed
instructions on the installation can be found in the Documentation that is
available together with the code.

In the following, various important features of \NECI are explained in
detail. An overview of excitation generation, a fundamental part of every
FCIQMC calculation, is given in section~\ref{sec:excit-gen}. Then,
the semi-stochastic approach (section~\ref{sec:semi_stoch}), the
estimation of energy and use of trial wave functions
(section~\ref{sec:trial_wf}), the recently
proposed adaptive shift method to reduce the initiator error
(section~\ref{sec:ada}) and perturbative corrections to this error (section~\ref{sec:pt2}), the sampling of reduced density matrices which is crucial for interfacing the
FCIQMC method with other algorithms (section~\ref{sec:rdms}), the calculation of
excited states (section~\ref{sec:ex}), static response functions
(section~\ref{sec:response} and the real-time FCIQMC method (section~\ref{sec:rt}),
the transcorrelated approach (section~\ref{sec:moltc}) and the available symmetries,
including total spin conservation utilizing GUGA (section ~\ref{sec:guga-fciqmc}) are
discussed. Finally, the scalability of \NECI is explored (section~\ref{sec:par})
and the interfaces for usage with other code are presented
(section~\ref{sec:interface}), in particular for the Stochastic-MCSCF method (section~\ref{sec:Stochastic-MCSCF}).
 
\section{Excitation generation}
\label{sec:excit-gen}
A key component of the FCIQMC algorithm is the sampling of the Hamiltonian
matrix elements in the spawning step, where the Hamiltonian is applied
stochastically. This requires an efficient algorithm to randomly generate
connected determinants with a known probability $p_{gen}$ for any given determinant,
referred to as excitation generation. This typically means making a
symmetry constrained choice of (up to) two occupied orbitals in a determinant
and (up to) two orbitals to replace them with, such that the corresponding Hamiltonian
matrix element is non-zero. If spin-adapted functions are
used rather than determinants, the connectivity rules change but the main principles are same.

The spawning probability for a spawn from a determinant $\ket{D_i}$
to a determinant $\ket{D_j}$ is in practice given by
\begin{equation}
  p_s = \Delta \tau \frac{|H_{ij}|}{p_{gen}(j|i)}\,.
\end{equation}
This means, the purpose of $p_{gen}(j|i)$  of selecting $\ket{D_j}$ from $\ket{D_i}$ in the spawning probability $p_s$ is to allow the flexibility 
in the selection of determinants $\ket{D_j}$ from $\ket{D_i}$ so that, irrespective of how we choose 
$\ket{D_j}$ from $\ket{D_i}$, the rate at which transitions occur is not biased 
by the selection algorithm. In other words, if a
particular determinant $\ket{D_j}$ is only selected rarely from $\ket{D_i}$ (i.e. with low generation probability), then
the acceptance of the move (i.e. the spawning probability) will be with correspondingly high probability (i.e. 
proportional to the inverse of the generation probability). Conversely if a determinant $\ket{D_j}$ is 
selected with relatively high generation probability from $\ket{D_i}$, then its
acceptance probability will be correspondingly low.  In other words, from the
point of view of the exactness of the FCIQMC algorithm, the precise manner in
which excitations are made is immaterial: as long as the probability
$p_{gen}(j|i) > 0$ when $|H_{ij}|>0$, the algorithm will ensure that transitions from $D_i\rightarrow D_j$ occur at a rate proportional to $|H_{ij}|$, and hence the walker dynamics converges onto the exact ground-state 
solution of the Hamiltonian matrix. However, from the point of view of {\em efficiency}, different 
algorithms to generate excitations are by no means equivalent.

That is, events with a very large $\frac{|H_{ij}|}{p_{gen}(j|i)}$ can lead to very large spawns and
thus endanger the stability of an i-FCIQMC calculation. For time-step
optimization, \NECI offers a general histogramming method, which determines
the time-step from a histogram of $\frac{|H_{ij}|}{p_{gen}(j|i)}$ \cite{Dobrautz2019}, as well as
an optimized special case thereof, which only takes into account the maximal
ratio \cite{smart2014}. If required, internal weights of the excitation generators such
a bias towards double excitations are then optimized in the same fashion 
to maximise the time-step.

However, as a result, the time-step and thus overall efficiency of the
simulation is driven by the worst-cases of the $\frac{|H_{ij}|}{p_{gen}(j|i)}$
ratio discovered
within the explored Hilbert space. Thus an optimal excitation generator
should create excitations with a probability distribution to the Hamiltonian
matrix elements, such that
\begin{equation}
  \frac{|H_{ij}|}{p_{gen}(j|i)} \approx \mathrm{const.}
\end{equation}
This is the optimal probability distribution, since then, the acceptance rate
is solely determined by the time step \cite{holmes2016}.

\NECI supports a variety of algorithms to perform excitation generation, with
the most notable being the pre-computed heat-bath (PCHB) sampling (a variant
of the heat-bath sampling presented in \cite{holmes2016}, as described in the appendix ~\ref{app:pchb}), the on-the-fly Cauchy-Schwartz
method \cite{cs_excit} (described in the appendix ~\ref{app:nu_excit}),
the pre-computed Power-Pitzer method \cite{neufeld2019} and lattice-model
excitation generators both for real-space and momentum-space
lattices. Additionally, a three-body excitation generator and a uniform
excitation generator are available, which are essential for treating systems
with the transcorrelated ansatz when including three-body interactions.

As heat-bath excitation generation can have high memory requirements,
it might be impractical for some systems. There, the on-the-fly Cauchy-Schwartz
method can maintain very good  $\frac{|H_{ij}|}{p_{gen}(j|i)}$ ratios
without significant memory cost, albeit at $\mathcal{O}(N)$ computational
cost, $N$ being the number of orbitals, and possibly with dynamic load imbalance. The details of the Cauchy-Schwartz excitation
generation are discussed in the appendix.
 
\section{Semi-stochastic FCIQMC}
\label{sec:semi_stoch}
In many chemical systems the wave function is dominated by a relatively small number of determinants. In a stochastic algorithm, the efficiency can be improved substantially by treating these determinants in a partially deterministic manner.

Petruzielo \emph{et al.} suggested a semi-stochastic algorithm\cite{Petruzielo2012}, where the FCIQMC projection operator $\hat{P} = \sum_{ij} P_{ij} |D_i \rangle \langle D_j|$, is applied exactly within a small but important subspace, which we call the deterministic space, $\mathcal{D}$. Specifically, we write
\begin{equation}
\hat{P} = \hat{P}^{\mathcal{D}} + \hat{P}^{\mathcal{S}},
\end{equation}
where
\begin{equation}
\hat{P}^{\mathcal{D}} = \sum_{i \in \mathcal{D}, \, j \in \mathcal{D}} P_{ij} |D_i \rangle \langle D_j |.
\end{equation}
The $\hat{P}^{\mathcal{D}}$ operator therefore accounts for all spawnings which are both from and to determinants in $\mathcal{D}$. The stochastic projection operator, $\hat{P}^{\mathcal{S}}$, contains all remaining terms. The matrix elements of $\hat{P}^{\mathcal{D}}$ are calculated and stored in a fixed array, and applied exactly each iteration by a matrix-vector multiplication. The operator $\hat{P}^{\mathcal{S}}$ is then applied stochastically by the usual FCIQMC spawning algorithm.

The semi-stochastic adaptation requires storing the Hamiltonian matrix within $\mathcal{D}$, which we denote $\boldsymbol{H}^{\mathcal{D}}$. In \texttt{NECI}, $\boldsymbol{H}^{\mathcal{D}}$ is stored in a sparse format, distributed across all processes. To calculate $\boldsymbol{H}^{\mathcal{D}}$, we have implemented the fast generation scheme of Li \emph{et al.}\cite{Li2018} This approach has allowed us to use deterministic spaces containing up to $\sim 10^7$ determinants. However, a more typical size of $\mathcal{D}$ is between $10^4$ and $10^5$.

Ideally, a deterministic space of a given size ($N_{\mathcal{D}}$) should be
chosen to contain the determinants with the largest value of $|C_i|$ in the
exact FCI wave function. This optimal choice is not possible in practice, but
various approaches exist to make an approximate selection. Umrigar and
co-workers suggest using selected configuration interaction (SCI) to make the
selection.\cite{Petruzielo2012} Within \texttt{NECI}, the most common approach is to
choose the $N_{\mathcal{D}}$ determinants which have the largest weight in the
FCIQMC wave function, at a given iteration.\cite{Blunt2015} Therefore, a
typical FCIQMC simulation in \NECI will be performed until convergence (at
some iteration number $N_{\mathrm{conv.}}$) using the fully-stochastic
algorithm, at which point the semi-stochastic approach is turned on, selecting
the $N_{\mathcal{D}}$ most populated determinants in the instantaneous wave
function to form $\mathcal{D}$. The appropriate parameters ($N_{\mathcal{D}}$
and $N_{\mathrm{conv.}}$) are specified in the \NECI input file. \NECI
supports performing periodic re-evaluation of the $N_{\mathcal{D}}$ most populated
determinants, updating the deterministic space $\mathcal{D}$ with a given frequency.

Using the semi-stochastic adaptation with a moderate deterministic space (on the
order of $\sim 10^4$) can improve the efficiency of FCIQMC by multiple orders
of magnitudes. This is particularly true in weakly correlated systems. The
semi-stochastic approach can also be used in \NECI when sampling reduced
density matrices (RDMs) as described in section \ref{sec:rdms}. Here, contributions to RDMs are included exactly
between all pairs of determinants within $\mathcal{D}$. It has been shown that
this can substantially reduce the error on RDM-based
estimators.\cite{Blunt2015} Using the semi-stochastic adaptation in \NECI
disables the load-balancing unless a periodic update of $\mathcal{D}$ is performed.
 
\section{Trial wave functions}
\label{sec:trial_wf}
The most common energy estimator used in FCIQMC is the reference-based projected estimator,
\begin{equation}
E_{\mathrm{Ref}} = \frac{ \langle D_{\mathrm{Ref}} | \hat{H} | \Psi \rangle
}{ \langle D_{\mathrm{Ref}} | \Psi \rangle },
\label{eq:proje}
\end{equation}
where $| D_{\mathrm{Ref}} \rangle$ is an appropriate reference determinant
(usually the Hartree--Fock determinant). In case $\ket{\Psi}$ is an
eigenstate, this yields the exact energy, but in general it is a
non-variational estimator. This is the default estimator for the energy,
and can be obtained with minimal overhead.

\NECI has the option to use projected estimators based on more accurate trial
wave functions, which can significantly reduce statistical error in energy
estimates.
For this reason we define a trial subspace $\mathcal{T}$, which is spanned by
$N_{\mathcal{T}}$ determinants.
Similarly to the deterministic space, $\mathcal{T}$ should ideally be formed
from the determinants with the largest contribution in the FCI wave function,
or some good approximation to these determinants.
Projecting $\hat{H}$ into $\mathcal{T}$ gives us a $N_{\mathcal{T}} \times
N_{\mathcal{T}}$ matrix, which we denote $\boldsymbol{H}^{\mathcal{T}}$, whose
eigenstates can be used as trial wave functions for more accurate energy
estimators.

Let us denote an eigenstate of $\boldsymbol{H}^{\mathcal{T}}$ by $| \Psi^{\mathcal{T}} \rangle = \sum_{i \in \mathcal{T}} C_i^{\mathcal{T}} | D_i \rangle$, with eigenvalue $E^{\mathcal{T}}$. Then a trial function-based estimator can be defined as
\begin{align}
E_{\mathrm{Trial}} &= \frac{ \langle \Psi^{\mathcal{T}} | \hat{H} | \Psi \rangle }{ \langle \Psi^{\mathcal{T}} | \Psi \rangle }, \\
    &= E^{\mathcal{T}} + \frac{ \sum_{j \in \mathcal{C}} C_j V_j }{ \sum_{i \in \mathcal{T}} C_i C_i^{\mathcal{T}} }.
\end{align}
Here, $\mathcal{C}$ is the space of all determinants connected to $\mathcal{T}$ by a single application of $\hat{H}$ (not including those in $\mathcal{T}$). $C_i$ denotes walker coefficients in the FCIQMC wave function, and $V_j$ is defined within $\mathcal{C}$ as
\begin{equation}
V_j = \sum_{i \in \mathcal{T}} \bra{D_j}\hat H\ket{D_i} C_i^{\mathcal{T}},  \;\;\;\;\;\;\;\;  | D_j \rangle \in \mathcal{C},\; \ket{D_i} \in \mathcal{T}.
\end{equation}

To calculate the estimator $E_{\mathrm{Trial}}$ we therefore require several
large arrays: first, $\boldsymbol{H}^{\mathcal{T}}$, which is stored in a
sparse format, in the same manner as the deterministic Hamiltonian in the
semi-stochastic scheme; second, $\ket{\psi^{\mathcal{T}}}$, which must be calculated by the Lanczos or Davidson algorithm; third, $\boldsymbol{V}$, which is a vector in the entire $\mathcal{C}$ space. The number of coefficients to store in $\mathcal{C}$ is larger than in $\mathcal{T}$ by a significant amount, typically by several orders of magnitude. Indeed, storing $\mathcal{V}$ can become the largest memory requirement. Because of this, using trial wave functions is typically more memory intensive in \NECI than using the semi-stochastic approach, for a given space size. We therefore suggest using a smaller trial space, $\mathcal{T}$, compared to the deterministic space, $\mathcal{D}$.

Note that the initiator error on $E_{\mathrm{Trial}}$ is not the same as the
initiator error on $E_{\mathrm{Ref}}$.
For example, $E_{\mathrm{Trial}}$ becomes exact as
$| \Psi^{\mathcal{T}} \rangle$ approaches the FCI wave function.
For practical trial wave functions, however, the two energy estimates typically
give similar initiator errors for ground-state energies in our experience.
An exception occurs for excited states (see Section~\ref{sec:ex}).
In this case, the wave function is usually not well approximated by a single
reference determinant, and $E_{\mathrm{Trial}}$ with an appropriate
$\mathcal{T}$ yields a great improvement, both for the statistical and initiator error.
 
\section{Adaptive Shift}
\label{sec:ada}
The initiator criterion~\cite{initiator-fciqmc} is important in making FCIQMC a practical method allowing us to  achieve convergence at a dramatically  lower number of walkers than the full FCIQMC~\cite{BoothThomAlavi2009}.  However,  this approximation introduces a bias in the energy when an insufficient number of walkers is used. This  bias can be attributed to the fact that non-initiators are systematically undersampled due to the lack of feedback from their local Hilbert space. To correct this, we can allow each non-initiator determinant $\ket{D_i}$ to have its own \emph{local} shift $S_i(\tau)$ as an appropriate fraction of the full shift $S(\tau)$
\begin{equation}
	S_i(\tau) = f_i  \times S(\tau)\;.
\end{equation}
The fraction $f_i$ is computed by monitoring which spawns are accepted due to the initiator criterion and accumulating positive weights over the accepted and rejected ones:
\begin{equation}\label{eq:as_fval}
  f_i=\frac{ \sum_{j\in accepted} w_{ij} }{ \sum_{j\in all} w_{ij} }\;.
\end{equation}
These weights $w_{ij}$ are derived from perturbation theory~\cite{Lowdin1951} where the first-order contribution of determinant $\ket{D_i}$ to the amplitude of determinant $\ket{D_j}$ is used as a weight for spawns from $\ket{D_i}$ to $\ket{D_j}$
\begin{equation}
  w_{ij} = \frac{|H_{ij}|}{H_{jj}-E_0}\;.
\end{equation}
It is worth noting that, regardless of how the weights are chosen, expression~(\ref{eq:as_fval}) guarantees that initiators get the full shift. Also as the number of walkers increases, the local Hilbert space of a non-initiator becomes more and more populated, restoring the full method in the large  walker limit.

\begin{figure}
	\centering
	\includegraphics[width= \columnwidth]{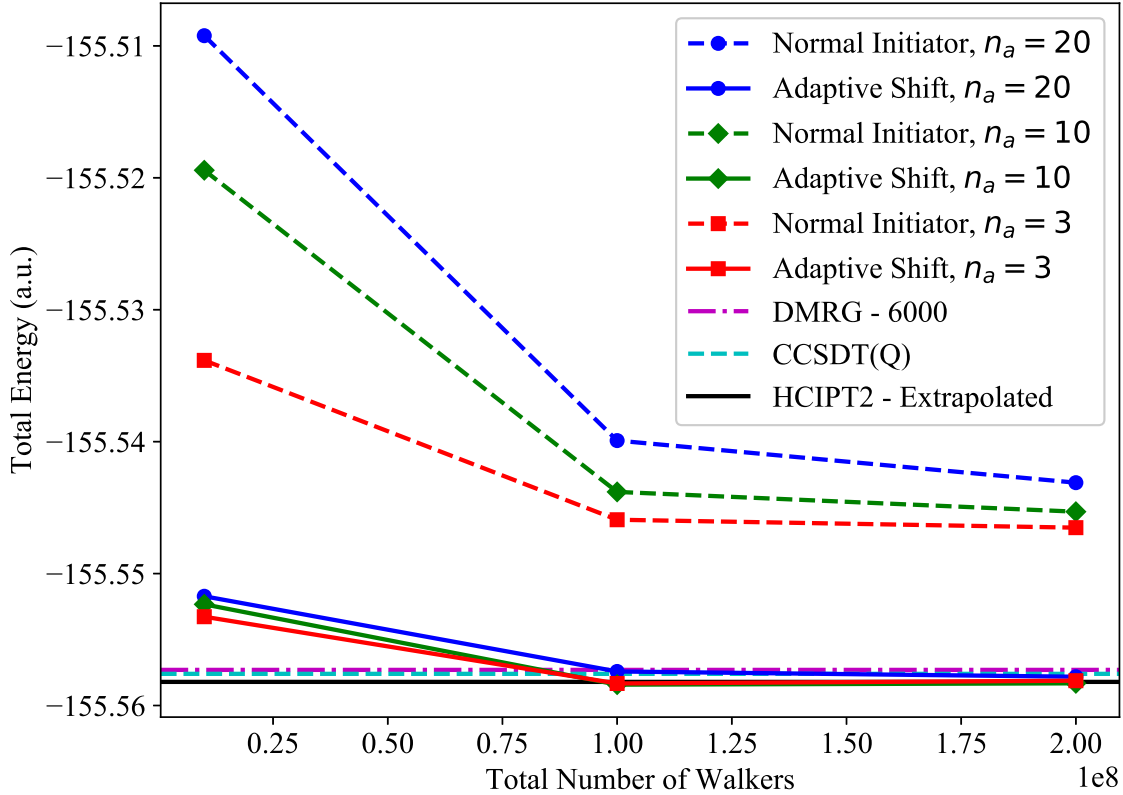}
	\caption{Example of application of the adaptive shift method: Total energies of butadiene for the normal initiator and the
	adaptive shift method, as a function of the number of walkers, for three
	values of the initiator parameter $n_a$. The adaptive shift results
	converge to: $-155.5581(2)\,\mathrm{E_h}$,
	$-155.5583(2)\,\mathrm{E_h}$ and $-155.5578(2)\,\mathrm{E_h}$ for $n_a$ of 3, 10
	and 20, respectively. The DMRG value of $-155.5573\,\mathrm{E_h}$, obtained
	with a bond dimension of 6000~\cite{Olivares2015}, the CCSDT(Q) value
	of $-155.5576\,\mathrm{E_h}$ and the extrapolated HCIPT2 value of
	$-155.5582(1)\,\mathrm{E_h}$~\cite{Chien2018} are in good agreement with
	that. Reproduced from Ghanem \textit{et al.} JCP 151, 224108 (2019)~\cite{adaptive_shift} with
	the permission of AIP Publishing.} 
	\label{fig:as_buta}
\end{figure}

We call the above approach for unbiasing the initiator approximation,
the\emph{ adaptive-shift} method \cite{adaptive_shift}. In
Fig.~\ref{fig:as_buta}, examplary results (from ~\cite{adaptive_shift}) from using the adaptive shift method
are displayed, comparing total energies of the butadiene molecule in
ANO-L-pVDZ basis (22 electrons in 82 spatial orbitals), obtained with the normal
initiator method and the adaptive shift method using three different values of
the initiator parameter $n_a$ : 3, 10 and 20. The adaptive shift results are in good agreement with
other benchmark values from DMRG, CCSDT(Q) and extrapolated HCIPT2. In contrast, the normal initiator method has a bias of over 10 mH. Also notice how by using the adaptive shift, the results become, to a large extent, independent of the initiator parameter $n_a$.

\section{Perturbative corrections to initiator error}
\label{sec:pt2}
An alternative approach to removing initiator error in \NECI is through a perturbative correction\cite{Blunt2018_2}. In the initiator approximation, spawning events from non-initiators to unoccupied determinants are typically discarded. These discarded events make up a significant fraction of all spawning attempts made, which in turn accounts for much of the total simulation time. While it is necessary to discard these spawned walkers to prevent disastrous noise from the sign problem\cite{Spencer2012}, this step is extremely wasteful.

These discarded walkers actually contain significant information which can be
used to greatly increase the accuracy of the initiator FCIQMC approach. Specifically, these walkers may sample up to double excitations from the currently-occupied determinants (a similar argument can be used to justify the above adaptive shift approach). In analogy with a comparable approach taken in selected CI methods, these discarded walkers can be used to sample a second-order correction to the energy from Epstein-Nesbet perturbation theory.

The correction is calculated by
\begin{equation}
\Delta E_2 = \frac{1}{ (\Delta \tau)^2 } \sum_{i \, \in \, \textrm{rejected}} \frac{ S_i^1 S_i^2 }{ E_0 - H_{ii} }.
\label{eq:pt2_correction}
\end{equation}
Here, $\Delta \tau$ is the time step, $E_0$ is the i-FCIQMC estimate of the
energy, and $S_i^r$ is the total spawned weight onto determinant $|
D_i \rangle$ in replica $r$ (the replica approach will be discussed in more detail in Section~\ref{sec:rdms}). This correction requires that two replica FCIQMC
simulations are being performed simultaneously, to avoid biases in this
estimator. The summation here is performed over all spawning attempts which are discarded $\emph{on both replicas}$ simultaneously.

This must only be applied to correct the variational energy estimator from i-FCIQMC. Such variational energies in \NECI can either be calculated directly\cite{Blunt2019, Blunt2019_2}, or from two-body reduced density matrices, which may be sampled in FCIQMC.

This perturbative correction is essentially free to accumulate, since all spawned walkers contributing to Eq.~(\ref{eq:pt2_correction}) are created regardless. The only significant extra cost comes from the requirement to perform two replica simulations. However, for large systems the noise on this correction can become significant, which necessitates further running time to reduce statistical errors.

This correction has proven extremely successful in practice, particularly for weakly correlated systems, where it is typical to see $80 - 90\%$ of remaining initiator error removed\cite{Blunt2018_2, Blunt2019, Blunt2019_2}.
 
\section{Density matrix sampling and pure expectation values}
\label{sec:rdms}
While the total energy is an important quantity to extract from quantum
systems, a more complete characterization of a system requires the ability to
extract information about other expectation values.
If these expectation values are derived from operators which do not commute
with the Hamiltonian of the system, then a `projected' estimate of the
expectation value akin to Eq.~\ref{eq:proje} is not possible, and alternatives
within FCIQMC are required in order to compute them.
This is the case for many key quantities such as nuclear derivatives (forces on
atoms), dipole moments and higher-order electrical moments, as well as other
observables such as pair distribution functions \cite{Thomas2015}.
They all can be obtained via the corresponding $n$-body reduced density matrix
($n$-RDM), where $n$ is the rank of the operator in question, that fully
characterizes the correlated distribution and coherence of $n$ electrons
relative to each other.
This information can also be used to calculate quantum information measures,
which are not observables but which characterize the entanglement within a
system, such as correlation entropies \cite{Overy2014}.

To characterize the strength of coupling between {\em different} states under
certain operators, e.g. the oscillator strength of optical excitations, as well
as obtaining other dynamical information requires computing {\em transition}
density matrices (tRDMs) between stochastic samples of {\em different} states,
which can be sampled within FCIQMC using the excited state feature discussed in
section \ref{sec:ex} \cite{BoothSpectra, Blunt2017}.
Furthermore, the two states considered may not sample eigenstates of the
system, but one of them can be a {\em response} state of the system, then the
resulting tRDMs characterize the response of a system to a perturbation,
corresponding to a higher derivative of the energy such as the polarizability
of the system, which will be addressed in section~\ref{sec:response} \cite{Samanta2018}.
Finally, RDMs can also be used to characterize the expectation value of an
{\em effective} Hamiltonian in a subspace of a system \cite{Blunt2015_2,Blunt2018}.
This effective Hamiltonian can include effects such as electronic correlations
coupling the space to a wider external set of states.
The plurality of electronic structure methods of this kind, such as explicitly
correlated `F12' corrections for basis set incompleteness
\cite{BoothF12,Gruneis13,Kersten2016}; multi-configurational self-consistent
field \cite{Thomas2015_2,LiManni2016}; internally-contracted multireference
perturbation theories \cite{Anderson2020}; embedding
methods\cite{Fertitta2018,Fertitta2019}; and the Multi-Configuration
Pair-Density Functional Theory (MC-PDFT) \cite{limanni2014a}, further attest the
importance of faithful and efficient sampling of RDMs in electronic structure
theory.

All expectation values of interest can be derived from contractions with a general reduced density matrix object, defined as
\begin{equation}
    \Gamma_{i_1 i_2 \dots i_n, j_1 j_2 \dots j_n}^{A,B} = \langle \Psi_A |\hat{a}^\dagger_{i_1} \hat{a}^\dagger_{i_2}... \hat{a}^\dagger_{i_n} \hat{a}_{j_n} \hat{a}_{j_{n-1}} ... \hat{a}_{j_1} | \Psi_B \rangle ,
\end{equation}
where $n$ denotes the `rank' of the RDM, and the choice of the states $A$ and $B$ define the type of RDM, as described above. In this section we focus on the sampling of the 2-RDM. This is generally the most common RDM required, as most expectation values of interest are (up to) two-body operators, including the total energy of the system. Furthermore, within FCIQMC, the fact that the rank of the RDM required is then the same as the rank of the Hamiltonian which is sampled within the stochastic dynamics, leads to a novel algorithm which ensures that the overhead to compute the 2-RDM is relatively small and manageable\cite{Overy2014}.

Expanding the expression for the 2-RDM in terms of the exact FCI wave function (Eq.~\ref{eq:fci}), we find
\begin{equation}
\Gamma_{k l, m n}^{A, B} = \sum_{i,j} C_{i}^{A*} C_{j}^{B} \langle D_{i} | \hat{a}^\dagger_{k} \hat{a}^\dagger_{l} \hat{a}_{n} \hat{a}_{m} | D_{j} \rangle , \label{eqn:FCI2RDM}
\end{equation}
where ${i,j}$ index the many-electron Slater determinants and $k,l,m,n$ denote single-particle orbitals.
We will focus on the case where we are sampling $| \Psi_A \rangle = | \Psi_B \rangle = | \Psi_0 \rangle$, the ground state of the system, 
since the same basic principles are applied to sampling the tRDMs, where the other walker distribution may represent an excited state or a response state, with more details for these cases considered in Refs.~\onlinecite{Blunt2017,Samanta2018}.
The expectation values derived from these RDMs describe `pure' expectation values, to distinguish them from the projective estimate of expectation values given in Eq.~\ref{eq:proje}.

There are some features of the form of Eq.~\ref{eqn:FCI2RDM} that should be noted. Firstly, the 2-RDM requires the sampled amplitudes on all determinants in the space connected to each other via (up to) a double electron substitutions. This means that this expectation value requires a global sampling of connections in the entire Hilbert space, in contrast to the projected energy estimate, which requires only a consideration of the determinant amplitudes which are connected directly through ${\hat H}$ to the reference determinant (or small trial wave function, see Sec.~\ref{sec:trial_wf}). Secondly, it is seen that the pairs of determinants in Eq.~\ref{eqn:FCI2RDM} are exactly the same as the pairs of determinants connected in general through the Hamiltonian operator used to sample the FCIQMC dynamics in Eq.~\ref{eq:fciqmc_dyn}, assuming that the matrix element is not zero due to (accidental) symmetry between the determinants. This allows an algorithm to sample the 2-RDM concurrently with the sampling of the Hamiltonian required for the spawning steps between occupied determinant pairs.

A final point to note, is that the $n$-RDM is a non-linear functional of the FCI amplitudes -- specifically being a quadratic form. Within the FCIQMC sampling, the $C_{i}$ amplitudes are stochastic variables represented as walkers ($C_{i}(\tau)$) which at any one iteration are in general very different from the true $C_{i}$, but when averaged over long times have an expected mean amplitude which is the same as (or a very good approximation to) $C_{i}$. However, due to this non-linearity in the form of the 2-RDM, the average of the sampled amplitude product is not equal to the product of the average amplitude, $\langle C_{i}^* (\tau) C_{j} (\tau) \rangle_{\tau} \neq \langle C_{i}^* (\tau) \rangle_{\tau} \langle C_{j} (\tau) \rangle_{\tau}$, as it neglects the (co-)variance between the sampled determinant amplitudes.
Initial applications of RDM sampling in FCIQMC neglected these correlations in the sampling of the RDMs, which significantly hampered the results, especially for the diagonal elements of the RDMs\cite{BoothF12}. The result is that even if each determinant were correctly sampled on average, the stochastic error in the sampling would manifest as systematic error in the RDMs, and thus only give correct results in the large walker limit, but not the large sampling limit, even if the wave function were correctly resolved.

The resolution to this problem came via the `replica trick'\cite{Overy2014,Blunt2014},
which changes the quadratic RDM functional into a bilinear
one\cite{Zhang1993}. This formally removes the systematic error in the RDM
sampling, at the expense of requiring a second walker distribution. The
premise is to ensure that these two walker distributions are entirely
independent and propagated in parallel, sampling the same (in this instance
ground-state) distribution. This ensures an unbiased
sampling of the desired RDM, by ensuring that each RDM contribution is derived
from the product of an uncorrelated amplitude from each replica walker
distribution. The sampling algorithm then proceeds by ensuring that during the
spawning step, the current amplitudes are packaged and communicated along with
any spawned walkers. During the annihilation stage, these amplitudes are then
multiplied by the amplitude on the child determinant from the other replica
distribution, and this product then contributes to all $n$-RDMs which are
accumulated, and equal to the rank of the excitation or higher. In this way,
the efficient and parallel annihilation algorithm is used to avoid latency of
additional communication operations, with the necessary packaging of the
amplitude and specification of the parent determinant along with each spawned
walker being the only additional overhead. The \NECI implementation
allows for up to 20 replicas to be run, which exceeds any needs arising in the
context of RDM calculation.

Full details about the ground-state 2-RDM sampling algorithm can be found in
Ref.~\onlinecite{Overy2014}, however we mention a few salient additional
details here. The RDMs are stored in fully distributed and sparse data structures, allowing the accumulation of RDMs for very large numbers of orbitals. The sampling of the RDMs is also not inherently hermitian. While the sampling within FCIQMC obeys detailed balance, the flux of walkers spawned from $\ket{D_i} \rightarrow \ket{D_j}$ is only equal to the reverse flux on average, and therefore the stochastic noise ensures that the swapping of the two states does not give identical accumulated RDM amplitudes for finite sampling (note that for transition RDMs this is not expected, with more details in Ref.~\onlinecite{Blunt2017}). Similarly, the states sampled in FCIQMC are not normalized, and therefore neither are the sampled RDMs. Both of these aspects are addressed at the end of the calculation, where the RDMs are explicitly made hermitian via averaging appropriate entries, and the normalization is constrained by ensuring that the trace of the RDMs give the appropriate number of electrons\cite{Overy2014}.

The dominant cost of RDM sampling in large systems comes from the sampling of elements defined by pairs of creation and annihilation operators with the same orbital index. These correspond to tuples of occupied orbitals common to both $|D_{i} \rangle$ and $|D_{j} \rangle$ states. We term these contributions {\em promotions}, as they contribute to a rank of a RDM greater than the excitation level between $|D_{i} \rangle$ and $|D_{j} \rangle$. For instance, single excitation spawning events need to contribute to all $N-1$ elements of the 2-RDM corresponding to common occupied orbitals in the two determinants. The most extreme case comes from the `diagonal' contributions to the RDMs, where ${i=j}$, which requires $N(N-1)/2$ contributions to the 2-RDM to be included where each index defining the RDM element corresponds to the same occupied orbital in the two determinants. To mitigate this cost, these diagonal elements are stored locally on each MPI process, and only infrequently accumulated at the end of an RDM `sampling block', or when the determinant becomes unoccupied, with the amplitude averaged over the sampling block. This substantially reduces the frequency of the $\mathcal{O}(N^2)$ operations required to sample these promoted contributions from the diagonal of Eq.~\ref{eqn:FCI2RDM}.

Other efficiency boosting modifications to the algorithm, such as the semi-stochastic adaptation\cite{Blunt2015} (detailed in Sec~\ref{sec:semi_stoch}) are also seamlessly integrated with the RDM accumulation. Within the deterministic core space the RDM contributions are exactly accumulated along with the exact propagation, with the connections from the deterministic to the stochastic spaces sampled in the standard fashion. This combination of RDM sampling with the semi-stochastic algorithm can greatly reduce the stochastic errors in the RDMs by ensuring that contributions from large weighted determinant amplitudes are explicitly and deterministically included. Furthermore, the reference determinant and its direct excitations are also exactly accumulated. This is partly because these are likely important contributions, but principally, if the reference is a Hartree--Fock determinant then the coupling to its single excitations via the Hamiltonian will be zero due to Brillouin's theorem. These single excitations will nevertheless contribute to the RDMs, and therefore are included explicitly.

The sampling of RDMs with a rank greater than two is also now possible within
the FCIQMC algorithm and \NECI code. The importance of these quantities is primarily in their use in internally-contracted multireference perturbation theories, although a number of other uses for these quantities also exist\cite{Anderson2020}. These methods allow for the FCIQMC dynamics to only consider an active orbital subspace, hugely reducing both the full Hilbert space of the stochastic dynamics as well as the required timestep, while the accumulation of up to 4-RDMs (or contracted lower-order intermediates for efficiency) allows for a rigorous coupling of the strong correlation in the low-energy active space to the dynamic correlation in the wider `external' space via post-processing of these higher-body RDMs with integrals of the external space. Sampling of higher-body RDMs cannot use the identical algorithm to the 2-RDMs, since it now requires the product of determinant amplitudes separated by up to 4-electron excitations, which are not explicitly sampled via the standard FCIQMC propagation algorithm. To allow for this sampling, we include an additional spawning step per walker of excitations with a rank between three and $n$, where $n$ is the rank of the highest RDM accumulated. This additional spawning is controlled with a variable stochastic resolution, ensuring that the frequency of these samples is relatively rare to control the cost of sampling these excitations (approximately only one higher-body spawn for every 10-20 traditional (up to two-body) spawning attempts). There is no timestep associated with these excitations, and every attempt is `successful', transferring information about higher-body correlations in the system and contributing to these higher-body excitations, but not modifying the distribution of the sampled wave function. However, the dominant cost of sampling these higher-body RDMs is not  the sampling events themselves, but rather the promotion of lower-rank excitations to these higher-body intermediates. Nevertheless, the faithful sampling of these higher-body properties has allowed for the stochastic estimate of fully internally-contracted perturbation theories in large active spaces, with similar number of walkers required to sample the 2-RDM in an active space\cite{Anderson2020}.
 
\section{Excited state calculations}
\label{sec:ex}
In many applications, besides ground state energies, the properties of excited
states are of interest. If states in different symmetry sectors are targeted, this can be
easily achieved by performing separate calculations in each sector, yielding
the ground state with a given symmetry. If, however, several eigenstates
with the same symmetry are required, then this approach is not sufficient.
The FCIQMC method is not inherently limited to ground state
calculations, and can employ a Gram-Schmidt orthogonalization technique to
calculate a set of orthogonal eigenstates \cite{Blunt2015b, Blunt2017}.
The obtained states will then be the lowest energy states with a given symmetry.

Calculating eigenstates sequentially and
orthogonalizing against all previously calculated states carries the problem
of only orthogonalizing against a single snapshot of the wave function,
which will lead to a biased estimate of the excited states. Instead, calculating all
states in parallel and orthogonalizing after each iteration gives much better results.

The required modifications to the algorithm are minimal. To calculate a
set of $m$ eigenstates, $m$ FCIQMC calculations are
run in parallel, with the additional step of performing the
instantaneous orthogonalization between the $m$ states, performed at the end of each
iteration. The orthogonalization requires $\mathcal{O}(m^2)$
operations and uses one global communication per state. To run $m$ parallel
calculations, the replica feature presented in section \ref{sec:rdms} is used
to efficiently sample a number of states in parallel. After each FCIQMC iteration,
for each state, the contributions from all states of lower energies are
projected out. The update step for the $n$-th wave function $\ket{\psi_n}$ is then modified to
\begin{equation}
  \ket{\psi_n(\tau + \Delta \tau)} = \hat{O}_n(\tau + \Delta \tau) \left(1 - \Delta \tau
  \left( \hat{H} - S_n(\tau) \right) \right) \ket{\psi_n(\tau)}\,,
\end{equation}

with the orthogonalization operator for the $n$-th state
\begin{equation}
  \hat{O}_n(\tau) = 1 - \sum_{m<n} \frac{\ket{\psi_m(\tau)}\bra{\psi_m(\tau)}}{\Braket{\psi_m(\tau)|\psi_m(\tau)}}\,.
\end{equation}
With this definition of the orthogonalization operator, the ground state
FCIQMC wave function $(n = 0)$ is left unaffected. The first excited state
$(n=1)$ is then orthogonalized against the ground state (using the
updated wave functions at $\tau + \Delta \tau$, after annihilation has
been performed). The second excited state is orthogonalized against
both the ground and first excited state, and so on.

To enforce the FCIQMC wave function discretization, after performing the
orthogonalization, all determinants with a coefficient smaller than the
minimal threshold (typically $1$) are stochastically rounded (either down
to $0$ or up to $1$, in an unbiased manner). This is required to prevent
proliferation of very small walkers, which adversely affects the wave
function compression.

\section{Response Theory  within FCIQMC to calculate static molecular
properties}
\label{sec:response}

Response theory is a well-established formalism to calculate molecular properties using quantum chemical methods\cite{Monkhorst1977,Dalgaard1983,Christiansen:1998p36,Helgaker2012}. 
It is, in general, formulated for a time-dependent field which allows to compute both static and dynamic molecular properties.
However, it is currently only implemented for a static field within \NECI \cite{Samanta2018}. 

Calculation of molecular properties using response theory relies on the evaluation of the response vectors which are the first or higher order wave functions of the system in the presence of an external perturbation $\hat{V}$.  
According to Wigner's ``(2n+1)'' rule, response vectors up to order $n$ are required to obtain response properties up to order $2n+1$\cite{Christiansen:1998p36}. 
For calculating second-order properties such as dipole polarizability, the first-order response vector, $\textbf{C}^{(1)}$ , needs to be obtained along with the zero-order wave function parameter $\textbf{C}^{(0)}$.
While  $\textbf{C}^{(0)}$ uses the original FCIQMC working equation \ref{eq:fciqmc_dyn},  $\textbf{C}^{(1)}$ is updated according to
\begin{equation}
\Delta C^{(1)}_i =  \underbrace{-\Delta\tau\sum_j(H_{ij}-S(\tau))C^{(1)}_j}_{\text{Hamiltonian dynamics}} - \underbrace{\Delta\tau \alpha V_{ij}C^{(0)}_j}_{\text{Perturbation dynamics}}	.	\label{eq:ResEq1}
\end{equation}

The response vector is discretized into signed walkers in the same way it is done for $\textbf{C}^{(0)}$. 
The dynamics of the response-state walker is simulated according to Eq.~\ref{eq:ResEq1} using an additional pair of replica and it works in parallel with the dynamics of the zero-order state.
Additional spawning and death steps are devised for the response-state walker dynamics, due to the presence of the perturbation,  alongside the original spawning and death steps in the dynamics. 
The dependence of the response state on the zero-order states comes from these two aforementioned additional steps.
A Gram-Schmidt orthogonalization is applied to the response-state walker distribution with respect to the zero-order walker distribution at each iteration using the same functionality as described in section~\ref{sec:ex}. 
This ensures orthogonality of the response vectors with respect to all lower-order wave function parameters.

The norm of the response walkers is fixed by the choice of the normalization of the zero-order walkers and it can, in principle, grow at a much faster rate than the zero-order norm. 
Therefore,  in Eq.~\ref{eq:ResEq1} we introduce the parameter $\alpha$ to control the norm of the response walkers and to reduce the computational effort expended in simulating their dynamics. 
We aim at matching the number of response-state walkers ($N_w^{(1)}$)  with the number of zero-order walkers ($N_w^{(0)}$) by updating $\alpha$ periodically as
\begin{equation}
\alpha=\frac{N_w^{(0)} }{N_w^{(1)}}	.
\end{equation}
Once the walker number stabilizes, the value of $\alpha$ is kept fixed, while accumulating statistics. 
As $\alpha$ scales the norm of the response vector, it needs to be taken into account while evaluating response properties. 

Response properties are then obtained from transition reduced density matrices (tRDMs) which are stochastically accumulated following Eq.~\ref{eqn:FCI2RDM}.
For example, dipole polarizability is obtained from the one-electron tRDMs between the zero- and first-order wave function as
\begin{equation}
\alpha_{xy}=- \frac{1}{2}\sum_{pq} \left[ \hat{x}_{pq} \gamma_{p,q}^y+ \hat{y}_{pq} \gamma_{p,q}^x  \right],	\label{eq:pol_gamma_1}
\end{equation}
with the $\gamma_{p,q}^y$ being calculated from the two-electron tRDM as
\begin{equation}
\gamma_{pq}^y=  \frac{1}{(N-1)} \sum_a \left[ \frac{1}{\alpha_1} \Gamma_{pa,qa}^{(0)(1)}[1]+\frac{1}{\alpha_2}\Gamma_{pa,qa}^{(0)(1)}[2]\right].	\label{eq:pol}
\end{equation}
Due to the use of two replica per state while sampling both zero- and first-order states, statistically independent and unbiased estimator of  tRDMs can be constructed in two alternative ways which are denoted here as `$[1]$' and `$[2]$'.
The perturbation used in the computation of  the tRDMs in Eq.~\ref{eq:pol} is the dipole operator $\hat{y}$. 
The factor $\frac{1}{\alpha}$ appears due to the re-scaling of the response vector following Eq.~\ref{eq:ResEq1}.

\section{Real-time FCIQMC}
\label{sec:rt}
\begin{figure}[t!]
\includegraphics[width=\textwidth]{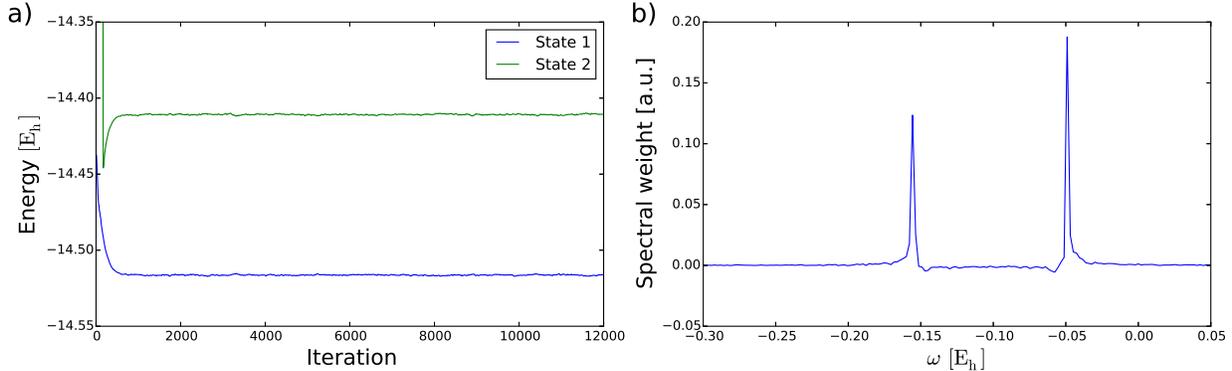}
\caption{\textbf{a)} Energy over iteration for an excited state calculation
with \NECI
for the Beryllium atom targeting two states in the $\mathrm{B_{1g}}$ irrep of
the $D_{2h}$ symmetry group (corresponding to P-states). The two states have
triplet/singlet character and the energy difference is
$105.5\,\mathrm{mH}$. \textbf{b)} Spectral decomposition of a $2s\rightarrow
2p$ excited state of the Beryllium atom created using real-time evolution with
\texttt{NECI}, containing the two lowest energy
P-states which correspond to the states targeted in a). The gap between the two
states is $106.6\,\mathrm{mH}$, agreeing with the excited state calculation
within the spectral resolution of $2.1\,\mathrm{mH}$. The zero of the energy
axis corresponds to the kation ground state energy. The output files are available in the supplementary material~\cite{excited_state_output}. In experiment, a value of
$93.8\,\mathrm{mH}$ is observed for this energy gap \cite{kramida1997}.}
\label{fig:exc_state}
\end{figure}

For the purpose of obtaining spectroscopic data or targeting highly excited
states, the calculation of orthogonal sets of eigenstates quickly becomes
unfeasible, as to obtain a certain eigenstate, all eigenstates of lower
energy with the same symmetry have to be computed as well. Spectral functions and
the resulting excitation energies can however be calculated using real-time
evolution of the wave function, yielding time-resolved Green's functions which
contain information on the full spectrum.
In addition to the stochastic imaginary time evolution of a wave function
using in the calculation of individual states, \NECI supports performing
real-time and arbitrary complex-time calculations, evolving the wave function
alongside a complex time trajectory \cite{guther2018}. As Green's functions
are quadratic in the coefficients of the wave function and averaging over
multiple iterations is not an option when evolving a wave function with a
real-time component, running multiple calculations in parallel akin to excited
state calculations discussed in section \ref{sec:ex} is mandatory, as is
running with complex coefficients. The real-time propagation can be used to
obtain energy gaps from spectral densities and thus target excited states. In
contrast to the direct calculation of excited states, these have not to be
calculated one by one and in order of ascending energy, however. In
Figure~\ref{fig:exc_state}, a simple example for applying both the
excited-state search and the real-time evolution to the Beryllium atom in a
cc-pVDZ basis set to obtain the singlet-triplet gap of the lowest P-state is given.
An issue with running real-time calculations is the difficulty of population
control, as the death step is essentially replaced by a rotation in the
complex plane. This issue can be mitigated by a rotation of the trajectory,
evolving along a trajectory in complex plane. \NECI supports an automated
trajectory selection that updates the angle $\alpha$ of the time trajectory in
the complex plane to maintain a constant population. The Green's function
obtained in the complex plane can then be used to obtain real-frequency
spectral functions using analytic continuation~\cite{silver1990,jarell1996}, with the analytic continuation
being significantly easier and more information being recoverable the closer to the real
axis the trajectory is \cite{guther2018}.
As, in contrast to the projector FCIQMC, errors arising from the expansion
of the propagator are a concern when running complex-time calculations, \NECI
uses a second order Runge-Kutta integrator here, to sufficiently reduce the time-step error.

\section{Transcorrelated Method}
\label{sec:moltc}
The computational cost of a Full CI method usually scales exponentially with respect to the size of the basis set. On the other hand, the low regularity of wave functions (characterized by the electronic cusp \cite{Kato57}) causes a very
slow convergence towards the basis set limit.  For calculations aiming at highly accurate results, it is very helpful to speed up such slow convergence. 

A Jastrow Ansatz\cite{jastrow55} offers a way to factor out the cusp from the wave function
\begin{equation}
  \ket{\Psi}=e^{\hat T} \ket{\Phi},   \label{Jastrow}
\end{equation} 
where $\hat T=\sum_{i<j} u({\bf r}_i,{\bf r}_j)$ is a symmetric function ($u({\bf r}_i,{\bf r}_j)=u({\bf r}_j,{\bf r}_i)$)  over electron pairs, and $\ket{\Phi}$ is an anti-symmetric many-body function. By including the cusp term $|{\bf r}_i-{\bf r}_j|/2$ in $u({\bf r}_i,{\bf r}_j)$,  the regularity of $\ket{\Phi}$ is improved at least by one order over $\ket{\Psi}$\cite{FHHO09}. We can also include other terms in $u({\bf r}_i,{\bf r}_j)$ to capture as much dynamic correlations as possible.
By using variational quantum Monte Carlo methods (VMC), the pair correlation function $u({\bf r}_i,{\bf r}_j)$ can be obtained for a single determinant $\ket{\Phi}$ (e.g., $\ket{\Phi_{HF}}$) or a linear combination of small number of determinants (e.g., a small CAS wave function).

The transcorrelated method of Boys and Handy \cite{Boys69} provides a simple and efficient way to treat the Jastrow Ansatz, where the original Schr\"odinger equation is transformed into a non-Hermitian eigenvalue problem
\begin{equation}
\tilde{H}\ket{\Phi}  =  E \ket{\Phi},\quad \tilde{H} = e^{-\hat T} \hat{H} e^{\hat T}. \label{TC_eq}
\end{equation}
The advantage of this form of $\hat T$ is that the similarity transformation leads to an expansion which terminates at second order
\begin{eqnarray}
\tilde{H} &=& \hat{H} + [\hat{H},\hat T] + \frac{1}{2}[[\hat{H},\hat T],\hat T] \\
          &=& \hat{H}-\sum_{i}\left(\frac{1}{2}\nabla_{i}^{2}\hat T+(\nabla_{i}\hat T)\nabla_{i}+\frac{1}{2}(\nabla_{i}\hat T)^{2}\right)\\
 &     = &     \hat{H}-\sum_{i< j}\hat{K}({\bf r}_{i},{\bf r}_{j})-\sum_{i<j<k}\hat{L}({\bf r}_{i},{\bf r}_{j},{\bf r}_{k}).
\end{eqnarray}
The similarity transformation introduces a novel two body operator $\hat{K}$ and a three-body potential $\hat{L}$ 
\begin{eqnarray}
       \hat{K} ({\bf r}_{i},{\bf r}_{j})&=&\frac{1}{2}\left(\nabla_{i}^{2}u({\bf r}_{i},{\bf r}_{j})+\nabla_{j}^{2}u({\bf r}_{i},{\bf r}_{j})+(\nabla_{i}u({\bf r}_{i},{\bf r}_{j}))^{2}+(\nabla_{j}u({\bf r}_{j},{\bf r}_{i}))^{2}\right)\nonumber\\
       & & +\left(\nabla_{i}u({\bf r}_{i},{\bf r}_{j}))\cdot\nabla_{i}+(\nabla_{j}u({\bf r}_{i},{\bf r}_{j}))\cdot\nabla_{j}\right)
     \\
 \hat L({\bf r}_{i},{\bf r}_{j},{\bf r}_{k})&=&\nabla_{i}u({\bf r}_{i},{\bf r}_{j})\cdot \nabla_{i}u({\bf r}_{i},{\bf r}_{k})+\nabla_{j}u({\bf r}_{j},{\bf r}_{i})\cdot \nabla_{j}u({\bf r}_{j},{\bf r}_{k})\nonumber\\
 & & +
 \nabla_{k}u({\bf r}_{k},{\bf r}_{i})\cdot \nabla_{k}u({\bf r}_{k},{\bf r}_{j}).
\end{eqnarray}
The whole transcorrelated Hamiltonian can be written in second quantised form as
\begin{eqnarray}
\label{eq:final-transcorr-H}
\tilde{H}&=&   \sum_{pq\sigma} h^p_q a^\dagger_{p\sigma} a_{q\sigma}+ \frac{1}{2} \sum_{pqrs } (V^{pq}_{rs}-K^{pq}_{rs}) \sum_{\sigma \tau} a^\dagger_{p\sigma} a^\dagger_{q\tau} a_{s\tau} a_{r\sigma} \nonumber\\
  && - \frac{1}{6}\sum_{pqrstu} L^{pqr}_{stu}  \sum_{\sigma \tau \lambda} a^\dagger_{p\sigma} a^\dagger_{q\tau} a^\dagger_{r\lambda} a_{u\lambda} a_{t\tau} a_{s\sigma},
\end{eqnarray} 
where $h^p_q=\langle\phi_p|h|\phi_q\rangle$ and $V^{pq}_{rs}=\langle\phi_p\phi_q|r_{12}^{-1}|\phi_r\phi_s\rangle$ are the one- and two-body terms of the molecular Hamiltonian, while $K^{pq}_{rs}=\langle\phi_p\phi_q|\hat{K}|\phi_r\phi_s\rangle$ and $L^{pqr}_{stu}=\langle\phi_p\phi_q\phi_r|\hat{L}|\phi_s\phi_t\phi_u\rangle$ originate from the $\hat{K}$ and $\hat{L}$ operators. 

This transcorrelated method has been investigated by FCIQMC using \NECI, as it can essentially speed up the convergence with respect
to basis sets. On the other hand the effective Hamiltonian is
non-hermitian and contains up to three-body potentials. Luo and Alavi have
explored a transcorrelated approach where only up to two-body potentials are
involved \cite{Luo2018}. The performance on uniform electron gases indicates
this approach could be developed into an efficient FCIQMC method for plane wave
basis sets in the future.
For general molecular systems, the full transcorrelated Hamiltonian (\ref{eq:final-transcorr-H}) is
implemented in \NECI, where $\hat T$ is fixed and treated as an input function, while $\ket{\Phi}$ is sampled by the FCIQMC algorithm. 
The lack of a lower bound of the energy due to the non-Hermiticity of the similarity transformed Hamiltonian poses a severe problem for variational approaches.
However, as a projective technique, FCIQMC does not have an inherent problem
sampling the ground-state right eigenvector by repetitive application of the projector (\ref{eq:projector}) and obtaining the corresponding ground-state eigenvalue. 

The matrix
elements $K^{pq}_{rs}$ and $L^{pqr}_{stu}$ are pre-calculated and have to be supplied as
input. The matrix elements of $K$ can be passed combined with the Coulomb
integrals, while the matrix elements of $L$ are passed in a separate input file.
This treatment is efficient for small atomic and
molecular systems, but for large systems the storage of the $L$ matrix becomes
a bottleneck. Here, efficient low rank tensor product expansion of $L$, might
in the future make it practical to treat even larger systems. \NECI supports storage of $L$ in a dense and a sparse format as well
as on-the-fly calculation of $L^{pqr}_{stu}$ from a tensor decomposition.
Additionally, major technical changes to the FCIQMC implementation are
required for sampling up to triple excitations, which generally leads to
reduced time-steps. The development of efficient excitation generation, which can alleviate the time-step 
bottleneck, is the subject of current work.

\begin{figure}
\includegraphics[width=0.7\columnwidth,angle=-90]{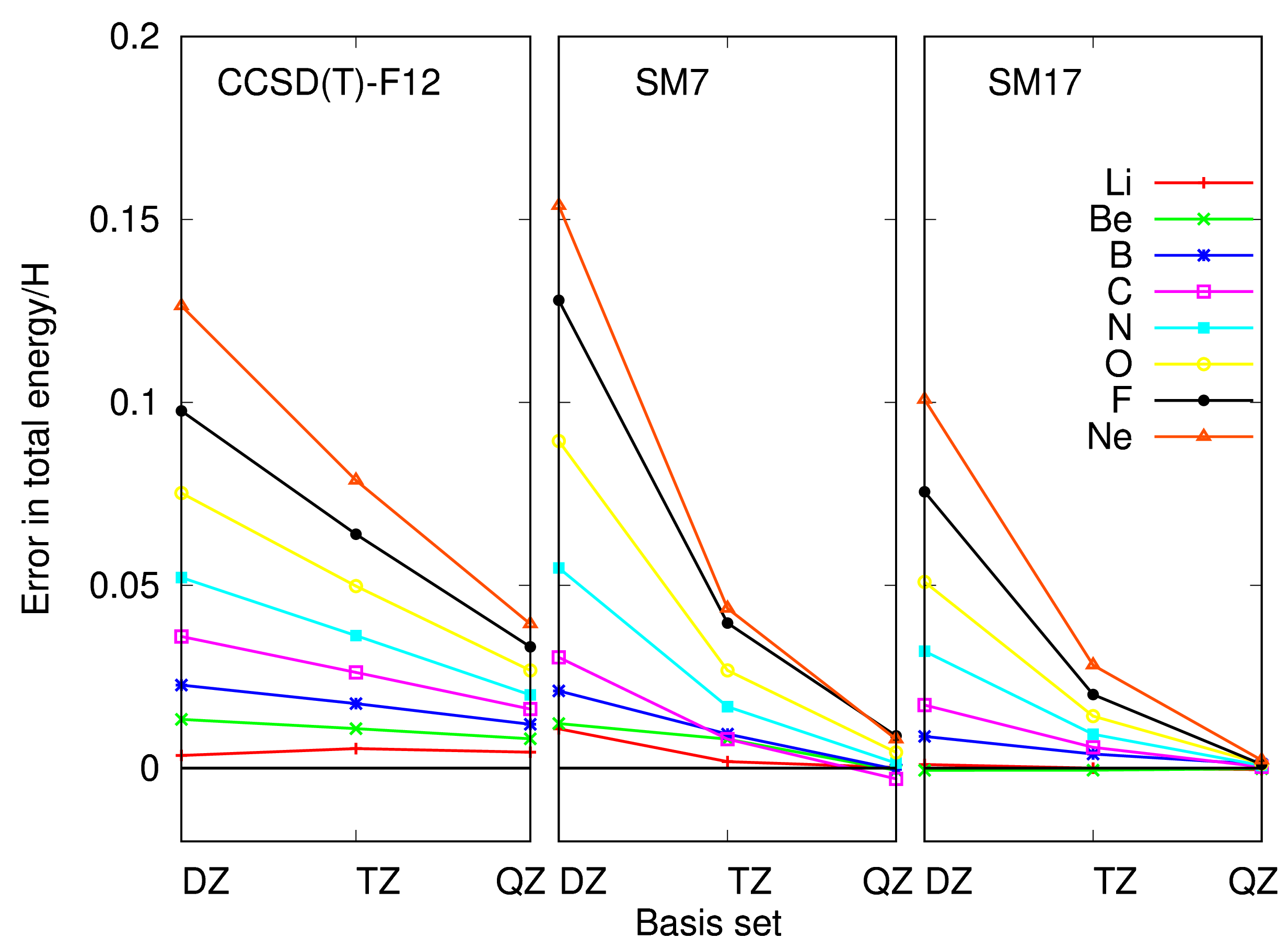}
\caption{
Exemplary application of the transcorrelated method: Errors in the total
energies of the first-row atoms, in Hartree, for the two correlation functions
and the F12 methodology. Reproduced from Cohen \textit{et al.}, JCP 151,
061101 (2019)~\cite{Cohen2019} with the permission of AIP Publishing.}
\label{fig:errors_energies}
\end{figure}

This method has been tested on the first row atoms~\cite{Cohen2019}, which
shall serve as an example here. Two
different correlation factors obtained by Schmidt and
Moskowitz\cite{Schmidt90} based on variance-minimisation VMC, which contain 7
and 17 terms of polynomial type basis functions have been employed there. 
The 7 term factor (SM7) contains mainly electron-electron correlation terms together with some electron-nuclear terms,  while the 17 term factor (SM17) uses more terms to describe also the electron-electron-nuclear correlation.
For the full CI expansion of
$\ket{\Phi}$, three different basis sets, cc-pVDZ, cc-pVTZ and
cc-pVQZ respectively have been used. In Fig.~\ref{fig:errors_energies} the convergence of the
total energies errors are displayed for the two different correlation factors,
in comparison with the the CCSD(T)-F12 method. This demonstrates that
improving the correlation factor can lead to a significant speed up of the
basis set convergence. Using the 17 term factor, the CBS
limit results can already be reached (within errors $<1$ mH) using a cc-pVQZ basis sets.

\section{Symmetries and spin-adapted FCIQMC}
\label{sec:guga-fciqmc}

Symmetry is a concept of paramount importance in the description and understanding of physical and chemical processes. 
According to Noether's theorem there is a direct connection between conserved quantities of a system and its inherent symmetries. 
Thus, identifying them allows a deeper insight in the physical mechanisms of studied systems.
Moreover, the usage of symmetries in electronic structure calculations enables a much more efficient formulation of the problem at hand. 
The Hamiltonian formulated in a basis respecting these symmetries has a block-diagonal structure, with zero overlap between states belonging to different `good' quantum numbers. 
This greatly reduces the necessary computational effort to solve these problems and allows much larger systems to be studied. 

\subsection{\label{sec:common-symmetries}Common Symmetries utilized in Electronic Structure Calculations and \texttt{NECI}} 

There are several symmetries which are commonly used in electronic structure calculations, due to the above mentioned benefits and their ease of implementation.
And our FCIQMC code \texttt{NECI} is no exception in this regard. \\

{\bf Conservation of the $\hat S_z$ spin-projection}\\
As mentioned in section~\ref{sec:intro}, FCIQMC is usually formulated in a complete basis of Slater determinants (SDs). 
SDs are eigenfunctions of the total $\hat S_z$ operator, and consequently, if the studied Hamiltonian, $\hat H$, is spin-independent (no applied magnetic field and spin-orbit interaction) it commutes with $\hat S_z$, $[\hat H, \hat S_z] = 0$.
The conservation of the $m_s$ eigenvalue in a FCIQMC calculation thus follows quite naturally: the initial chosen $m_s$ sector, determined by the starting SD used, will never be left by the random excitation generation process sketched in section~\ref{sec:excit-gen}.
No terms in the spin-conserving Hamiltonian will ever cause any state in the simulation to have a different $m_s$ value than the initial one. As a consequence the sampled wavefunction will always be an eigenfunction of $\hat S_z$ with a chosen $m_s$, determined at the start of a calculation. \\

{\bf Discrete  and Point Group Symmetries in FCIQMC}\\
\texttt{NECI} is also capable of utilizing Abelian point group symmetries, with D$_{2h}$ being the `largest' spatial group (similar to other quantum chemistry packages, e.g. \texttt{Molcas}\cite{molcas-general} and \texttt{Molpro}\cite{molpro-general-1,molpro-general-2}), momentum conservation (due to translational invariance) in the Hubbard model and uniform electron gas calculations and preservation of the $m_l$ eigenvalues of the orbital angular momentum operator $\hat L_z$ (the underlying molecular orbitals have to be constructed as eigenfunction of $\hat L_z$). This is implemented via a symmetry-conserving excitation generation step and is explained in more detail in Appendix~\ref{app:symmetries}.

\subsection{\label{sec:total-spin}Total spin conservation}

One important symmetry of \emph{spin-preserving, nonrelativistic} Hamiltonians is the global $SU(2)$ spin-rotation symmetry.
However, despite the theoretical benefits, the total $SU(2)$ spin symmetry is not as widely used as other symmetries, like translational or point group symmetries, due to their usually impractical and complicated implementation.

There are two kind of implementations of total spin conservation  in our FCIQMC code \NECI.
One approximate one is based on 
\emph{Half-Projected Hartree-Fock (HPHF) functions}\cite{hphf, helgaker,neci,hphf-fciqmc,linear-scaling-fciqmc}. 
Their rationale relies on the fact that for an \emph{even} number of electrons, every spin state $\ket{S}$ contains degenerate eigenfunctions with $m_s = 0$.  Using \emph{time-reversal symmetry} arguments a HPHF function can be constructed as
\begin{equation}
\ket {H_i} = \begin{cases}
\ket {D_i} & \text{for fully close-shell determinants} \\
\frac{1}{\sqrt 2} \left(\ket{D_i} \pm \ket{\overline{D_i}}\right) & \text{otherwise,}\\
\end{cases} 
\end{equation}
where $\overline{\ket{D_i}}$ indicates the spin-flipped version of $\ket{D_i}$. 
Depending on the sign of the open-shell coupled determinants, $\ket{H_i}$ are eigenfunctions of $\hat{\mathbf S}^2$ with odd ($-$) or even ($+$) eigenvalue $S$.
The use of HPHF is restricted to systems with an even number of electrons and can only target the lowest even- and odd-$S$ state. Thus, it can not differentiate between, e.g.\ a singlet $S = 0$ and quintet $S = 2$ state.

\subsubsection{The (graphical) Unitary Group Approach (GUGA)}

Our full implementation of total spin conservation is based on the graphical Unitary group approach (GUGA). 
It relies on the observation that the spin-free excitation operators $\hat E_{ij}$ in the spin-free formulation of the electronic Hamiltonian,
\begin{equation}
\label{eq:spin-free}
\hat H = \sum_{ij}^n t_{ij} \hat E_{ij} + \sum_{ijkl}^n V_{ijkl} \left(\hat E_{ij}\hat E_{kl} - \delta _{jk} \hat E_{il}\right),
\end{equation}
have the same commutation relations,
\begin{equation}
[\hat E_{ij},\hat E_{kl}] = \delta_{jk}\hat E_{il} - \delta_{il} \hat E_{kj},
\end{equation}
as the generators of the Unitary group $U(n)$.
This connection allows the usage of the Gel'fand-Tsetlin (GT) basis\cite{gelfand-1,gelfand-2,gelfand-3}, which is irreducible and invariant under the action of the operators $\hat E_{ij}$, in electronic structure calculations. 
The GT basis is a general basis for any irrep of $U(n)$, but Paldus\cite{Paldus1974,Paldus1975,Paldus1976} realized that only a special subset is relevant for the electronic problem~(\ref{eq:spin-free}), due to the Pauli exclusion principle. 
Based on Paldus' work, Shavitt\cite{Shavitt1977} further developed an even more compact representation by introducing the graphical extension of the UGA. This leads to the most efficient encoding of a spin-adapted GT basis state (CSF) in form of a step-vector $\ket{\mathbf{d}}$. 
This step-vector representation has the same storage cost of two bits per spatial orbital as Slater determinants. 
The entries of this step-vector encode the change of the total number of electrons $\Delta N_i$ and the change of the total spin $\Delta S_i$ of subsequent spatial orbitals $i$. This is summarized in Table~\ref{tab:step-vector}. 
\begin{table}
\centering
\caption{\label{tab:step-vector}Possible step-values $d_i$ and the corresponding change in number of electrons $\Delta N_i$ and total spin $\Delta S_i$ of subsequent spatial orbitals $i$.}
\begin{tabular}{ccc}
\botrule
$d_i$ & $\Delta N_i$ & $\Delta S_i$ \\
\hline
0 & 0 & 0 \\
1 & 1 & \phantom{-}1/2 \\
2 & 1 & -1/2 \\
3 & 2 & 0 \\
\botrule
\end{tabular}
\end{table}
All possible CSFs for a chosen number of spatial orbitals $N$, number of electrons $n$ and total spin $S$ are then given by all step-vectors $\ket{d} = \ket{d_1,d_2,\dots,d_N}$ fulfilling the restrictions
\begin{equation}
\label{eq:step-v-restrictions}
\sum_{i=1}^N \Delta n_i = n, \quad \sum_{i=1}^N \Delta S_i = S, \quad \text{and}\quad S_k = \sum_{i=1}^k \Delta S_i \geq 0.
\end{equation}
The last restriction in Eq.~\ref{eq:step-v-restrictions} corresponds to the fact that the (intermediate) total spin must never be less than 0. 

The most important finding of Paldus and Shavitt\cite{Shavitt1978,Paldus1981} was that the Hamiltonian matrix elements\textemdash more specifically the \emph{coupling coefficients} between two CSFs, e.g. $\bra{m'}\hat E_{ij} \ket{m}$\textemdash can be entirely formulated within the framework of the GUGA; without any reference and thus necessity to transform to a Slater determinant based formulation. 
Although CSFs can be expressed as a linear combination of SDs, the complexity of this transformation scales exponential with the number of open-shell orbitals of a specific CSF\cite{Shavitt1981}. Thus, it is prohibitively hard to rely on such a transformation and for already more than $\approx 15$ electrons a formulation without any reference to SDs is much more preferable.

\begin{figure}
\centering
\includegraphics[width=0.5\columnwidth]{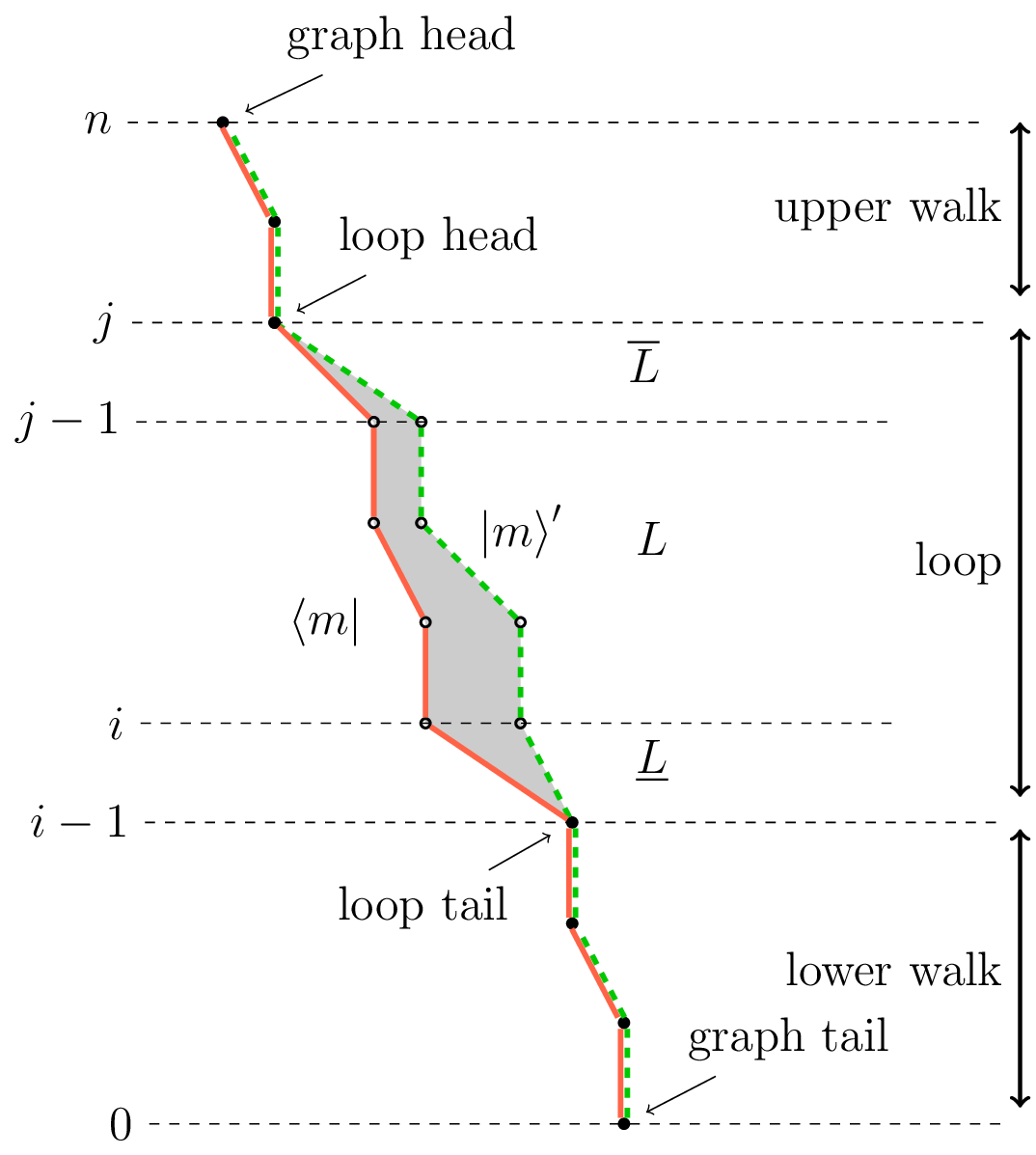}
\caption{\label{fig:single}Graphical representation of the coupling coefficient between two CSFs, $\bra{m}\hat E_{ij}\ket{m'}$.}
\end{figure}

Furthermore, Shavitt and Paldus\cite{Shavitt1978,Paldus1981} were able to find a very efficient formulation of the coupling coefficients as a product of terms, via the graphical extension of the UGA. Matrix elements between two given CSFs only depend on the shape of the loop enclosed by their graphical representation, as depicted in Fig.~\ref{fig:single}.
The coupling coefficient of the one-body operator $\hat E_{ij}$ is given by 
\begin{equation}
\label{eq:single-mat-ele}
\bra{m'}\hat E_{ij}\ket{m} = \prod_{k=i}^{j}W(Q_k; d_k',d_k,\Delta S_k, S_k),
\end{equation}
where the product terms depend on the step-values of the two CSFs, $d_k'$ and $d_k$, the difference in the current spin $\Delta S_k$ (with the restriction $S_k' - S_k = \pm 1/2$) and the intermediate spin $S_k$ of $\ket{m}$ at orbital $k$. $Q_k$ in Eq.~(\ref{eq:single-mat-ele}) depends on the \emph{shape} of the loop formed by $\ket{m}$ and $\ket{m'}$ at level $k$ and is tabulated in e.g. Ref.~[\onlinecite{Shavitt1978}].
Additionally, the two CSFs, $\ket{m}$ and $\ket{m'}$, must coincide outside the range $(i,j)$ for Eq.~(\ref{eq:single-mat-ele}) to be non-zero. 

\subsubsection{Spin-adapted excitation generation - GUGA-FCIQMC}
\label{sec:guga_fciqmc}
The compact representation of spin-adapted basis functions in form of step-vectors and the product form of the coupling coefficients (\ref{eq:single-mat-ele}) allow for a very efficient implementation in our stochastic FCIQMC code \texttt{NECI}.
As mentioned in Sec.~\ref{sec:excit-gen}, the \emph{excitation generation} step is at the heart of any FCIQMC code. 

The main difference to a SD-based implementation of FCIQMC, apart from the more involved matrix element calculation~(\ref{eq:single-mat-ele}), is the higher connectivity within a CSF basis. 
For a given excitation operator $\hat E_{ij}$, with spatial orbital indices $(i,j)$, there is usually more than one possible excited CSF $\ket{m'}$ when applied to $\ket{m}$, 
$\hat E_{ij}\ket{m} = \sum_k c_k \ket{m'_k}$.
All valid \emph{spin-recouplings} within the excitation range $(i,j)$ can have a non-zero coupling coefficient as well. 
This fact is usually the prohibiting factor in spin-adapted
approaches. However, there is a quite virtuous combination of the concepts of FCIQMC and the GUGA formalism, as one only needs to pick \textbf{one} possible excitation from $\ket{m}$ to $\ket{m'}$ in the excitation generation step of FCIQMC, see Sec.~\ref{sec:excit-gen}. 

We resolved this issue, by randomly choosing one possible valid branch in the graphical representation, depicted in Fig.~\ref{fig:single}, for randomly chosen spatial orbital indices $i,j(,k,l)$.
Additionally we weight the random moves according to the expected magnitude of the coupling coefficients\cite{Dobrautz2019,dobrautz-phd}
to ensure $p_{gen}(m'\vert m) \propto \vert H_{m'm}\vert  $.
This approach avoids the possible exponential scaling as a function of the open-shell orbitals of connected states within a CSF based approach. 

However, this comes with the price of reduced generation probabilities and consequently a lower imaginary time-step, as mentioned in Sec.~\ref{sec:excit-gen}.
Combined with an additional effort of calculating these random choices in the excitation generation and the on-the-fly matrix element computation, the GUGA-FCIQMC implementation has a worse scaling with the number of spatial orbitals $N$ compared to a Slater determinant based implementation\cite{Dobrautz2019}.

However, the benefits of using a spin-adapted basis are a \emph{reduced Hilbert space size}, \emph{elimination of spin-contamination} in the sampled wavefunction and most importantly: the spin-adapted FCIQMC implementation via the GUGA allows targeting specific spin states, which are otherwise not attainable with a SD based implementation as discussed in Ref.~\onlinecite{Dobrautz2019}. 

The unique specification of a target spin allows resolving near degenerate spin states and consequently numerical results can be interpreted more clearly. This enables more insight in the intricate interplay of nearly degenerate spin states and their effect on the chemical and physical properties of matter.  

\subsubsection{\label{sec:h-chain}Example: Hydrogen chain in a minimal basis}

\begin{figure}
\centering
\includegraphics[width=0.8\columnwidth]{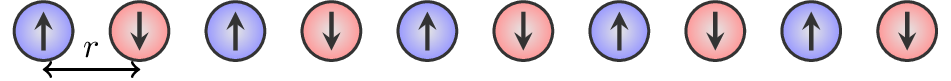}
\caption{\label{fig:hydrogen-levels}Schematic representation of a one-dimensional hydrogen chain of $L$ hydrogen atoms with equal separation $r$.}
\end{figure}

The GUGA-FCIQMC method has been benchmarked~\cite{dobrautz-phd} by applying it
to a linear chain of $L$ equidistant hydrogen atoms \cite{hydrogen-chain-2}
recently studied to test a variety of quantum chemical methods
\cite{hydrogen-chain}, which shall serve as an example here. 
Using a minimal STO-6G basis there is only one orbital per H atom and the system resembles a one-dimensional Hubbard model \cite{Hubbard1963,ppp-1,ppp-2,Gutzwiller1963} with long-range interaction.
Studying a system of hydrogen atoms removes complexities like core electrons or relativistic effects and thus is an convenient benchmark system for quantum chemical methods.

For large equidistant separation of the H atoms a localized basis, obtained with the default Boys-localization in Molpro's \texttt{LOCALI} routine, with singly occupied orbitals centred at each hydrogen is more appropriate than a HF basis. 
Thus, this is an optimal difficult benchmark system of the GUGA-FCIQMC method, since the complexity of a spin-adapted basis depends on the number of open-shell orbitals, which is maximal for this system. 
Particularly targeting the low-spin eigenstates of such highly open-shell systems poses a difficult challenge within a spin-adapted formulation. 
This situation is depicted schematically in Fig.~\ref{fig:hydrogen-levels}. 

We studied this system to show that we are able to treat systems with up to 30
open-shell orbitals with our stochastic implementation of the GUGA approach~\cite{dobrautz-phd}.
We calculated the $S = 0, 1$ and $2$ (only $S=0$ for $L=30$) energy per atom up to $L = 30$ H atoms in a minimal STO-6G basis at the stretched $r = 3.6\, a_0$ geometry\cite{hydrogen-chain} and compared it with DMRG \cite{block-dmrg-1, block-dmrg-2, chansharma2011,white1992, hydrogen-chain} reference results. 
The results are shown in Table~\ref{tab:hydrogen-chain}, where we see excellent agreement within chemical accuracy with the reference results.

An important fact is the order of the orbitals though. Similar to the DMRG method it is most beneficial to order the orbitals according to their overlap, since the number of possible spin recouplings depends on the number of open shell orbitals in the excitation range. If we make a poor choice in the ordering of orbitals, excitations between physically adjacent and thus strongly overlapping orbitals are accompanied by numerous possible spin-recouplings in the excitation range, if stored far apart in the list of orbitals. 
This behaviour is thoroughly discussed in Ref~[\onlinecite{LiManni2019}].

\begin{table}
\centering
\caption{\label{tab:hydrogen-chain}\small Example for application of
  GUGA-FCIQMC: Difference of the energy per site $E/L$ of an hydrogen chain
  for different number $L$ of H atoms and total spin $S$ in a STO-6G basis set
  at the stretched bond distance of $r = 3.6\,a_0$ compared with DMRG
  \cite{block-dmrg-1,block-dmrg-2,hydrogen-chain} reference
  results\cite{hydrogen-chain}. The GUGA-FCIQMC results were obtained without
  the initiator approximation\cite{initiator-fciqmc}. Reproduced from
  Dobrautz, Ph.D. thesis (2019)~\cite{dobrautz-phd}.}
\small
\renewcommand{\arraystretch}{1.0}
\begin{tabular}{ccccc}
\botrule
L &  S & $E_{ref}\,[E_h]$ & $E_{FCIQMC} \, [E_h]$ & $\Delta E \, [mE_h]$ \\
\hline
20 & 0 & -0.481979 	& -0.481978(1)\phantom{0} & -0.001(1)\phantom{000} \\
20 & 1 & -0.481683 	& -0.481681(11) & -0.002(11)\phantom{00} \\
20 & 2 & -0.480766 	& -0.480764(18) & -0.002(18)\phantom{00}\\
\hline
30 & 0 & -0.482020 & -0.481972(31) & -0.047(31)\phantom{00} \\
 \botrule
\end{tabular}
\end{table}
 
\section{Parallel scaling}
\label{sec:par}
When applying for access to large computing clusters, it is often necessary to
demonstrate that the software being used (in this case \NECI) is capable of
using the hardware efficiently. Ideally, the speed-up relative to using some
base number of compute cores should grow perfectly linearly with the number of
cores. In 2014, Booth \textit{et. al.}\cite{linear-scaling-fciqmc}, presented an
example with 500$\times10^6$ walkers in which no deviation from a linear
speed-up is noticeable when comparing using 512 cores to using 32, and even at
2048 cores, a speed-up by a factor of 57.5 was reported, which is 90\% of the ideal speed-up factor of 64.
In that work, the largest number of cores explored was 2048. By comparing the performance for a calculation with 100$\times10^6$ walkers and 500$\times10^6$ walkers, the same figure showed that the speed-up became closer to the ideal speed-up when the number of walkers was increased, suggesting that when using even more walkers, the efficiency comes even closer to 100\% of the ideal speed-up factor.

Since 90\% of the ideal speed-up factor was achieved in 2014 with only 500$\times10^6$
walkers on 2048 cores, and large compute clusters nowadays tend to have tens of
thousands of cores available, we report scaling data for a much larger
number of walkers on up to 24,800 cores in
Table~\ref{tab:scaling}. The calculations were done using the integrals in FCIDUMP format for the (54e,54o) active space first described in \cite{Reiher2017a} for the FeMoco molecule, and the output files are provided in the supplementary material~\cite{scaling_output}. 

The scaling analysis presented in Table \ref{tab:scaling}
was done with 32 billion walkers on each of the two replicas used
for the RDM sampling. Calculations at 32 billion walkers are expensive,
so we only completed enough iterations to determine an accurate estimate
of the average runtime per iteration for the scaling analysis, and
not enough iterations to accurately estimate the energy.\\

One may ask whether or not the scaling observed in Table \ref{tab:scaling}
was performed for a reasonable number of walkers for this active space.
To answer this question, we compare in Table \ref{tab:Best-non-extrapolated-energies}
the best (non-extrapolated) DMRG and sHCI-PT2 energies in the literature
\cite{Li2019} to energies obtained with i-FCIQMC at only 8 billion
walkers/replica, and find that the i-FCIQMC-RDM and i-FCIQMC-PT2 energies
are closer together than the sHCI-VAR and sHCI-PT2 energies, indicating
that the i-FCIQMC energies are closer to the true FCI limit where the
difference between variational and PT2 energies should vanish. The
DMRG result lies about half-way between the two i-FCIQMC results, but
fairly well below the lower of the sHCI results (a forthcoming publication specifically about the FeMoco system is planned, in which more details will be presented, but the purpose of this paper is to give an overview of the NECI code). \\

Furthermore, comparing the time per iteration between $8\times 10^9$ and
$32\times10^9$ walkers shows that a high parallel efficiency is also achieved
at the lower walker number. The determinants in \NECI are stored using a
hash table, making i-FCIQMC linearly scaling in the walker
number \cite{linear-scaling-fciqmc}, so the ideal time per iteration with
$32\times10^9$ walkers at 19960 cores according to the result for $8\times 10^9$
walkers at 16000 cores would be 23.4 seconds, which is only marginally smaller
than the reported 23.5 seconds. Note however, that this is the relative
efficiency between large scale calculations, which demonstrates performance
gain from extending parallelization at large scales, not from parallelization
over the entire range of scales, which is addressed to some extent by the
Chromium dimer example below.

\begin{table}
\centering
\caption{Best non-extrapolated energies obtained for the CAS(54,54) of the
FeMoco molecule, with three different methods. DMRG and sHCI energies
were calculated in Ref. \citep{Li2019}, and i-FCIQMC results were obtained
in this work with 8 billion walkers on each of the two replicas for
the RDM sampling. \label{tab:Best-non-extrapolated-energies}}
\small
\renewcommand{\arraystretch}{1.0}
\begin{tabular}{@{\extracolsep{\fill}}lr@{\extracolsep{0pt}.}l}
\botrule
\addlinespace[2mm]
\multirow{1}{*}{~~~~~~~~~~~Method} & \multicolumn{2}{c}{Total Energy~~~~~~~~}\tabularnewline\addlinespace[2mm]
\hline
\addlinespace[2mm]
~~~~~~~~~~~i-FCIQMC-RDM & -13\,482&174\,95(4)~~~~~~~~\tabularnewline
~~~~~~~~~~~i-FCIQMC-PT2\,\,\, & -13\,482&178\,45(40)~~~~~~~~\tabularnewline
~~~~~~~~~~~sHCI-VAR\,\,\, & -13\,482&160\,43~~~~~~~~\tabularnewline
~~~~~~~~~~~sHCI-PT2\,\,\, & -13\,482&173\,38~~~~~~~~\tabularnewline
~~~~~~~~~~~DMRG & -13\,482&176\,81~~~~~~~~\tabularnewline\addlinespace[2mm]
\botrule
\end{tabular}
\end{table}

In the case of the Chromium dimer (cc-pVDZ, 28 electrons correlated in 76 spatial orbitals) 
considered in figure \ref{fig:scaling}, the average time per iteration per walker
ranges from $3.18\times 10^{-9}\,\mathrm{s}$ at 640 cores to $2.51\times
10^{-10}\,\mathrm{s}$ at 10240 cores and  $1.53\times 10^{-10}\,\mathrm{s}$ at
20480 cores, corresponding to a parallel speed-up of 82.1\% from 10240 to 20480
cores and an overall speed-up of 65.2\% over the full range. The deviation from
ideal scaling almost exclusively stems from the communication of the spawns,
at lower walker numbers, the communicative overhead is more significant, reducing the
parallel efficiency compared to the FeMoco example. Nevertheless, a very high yield can be obtained from
scaling up the number of cores, even for already large scales.

\subsection{Load balancing}

\begin{figure}[t!]
  \includegraphics[width=\columnwidth]{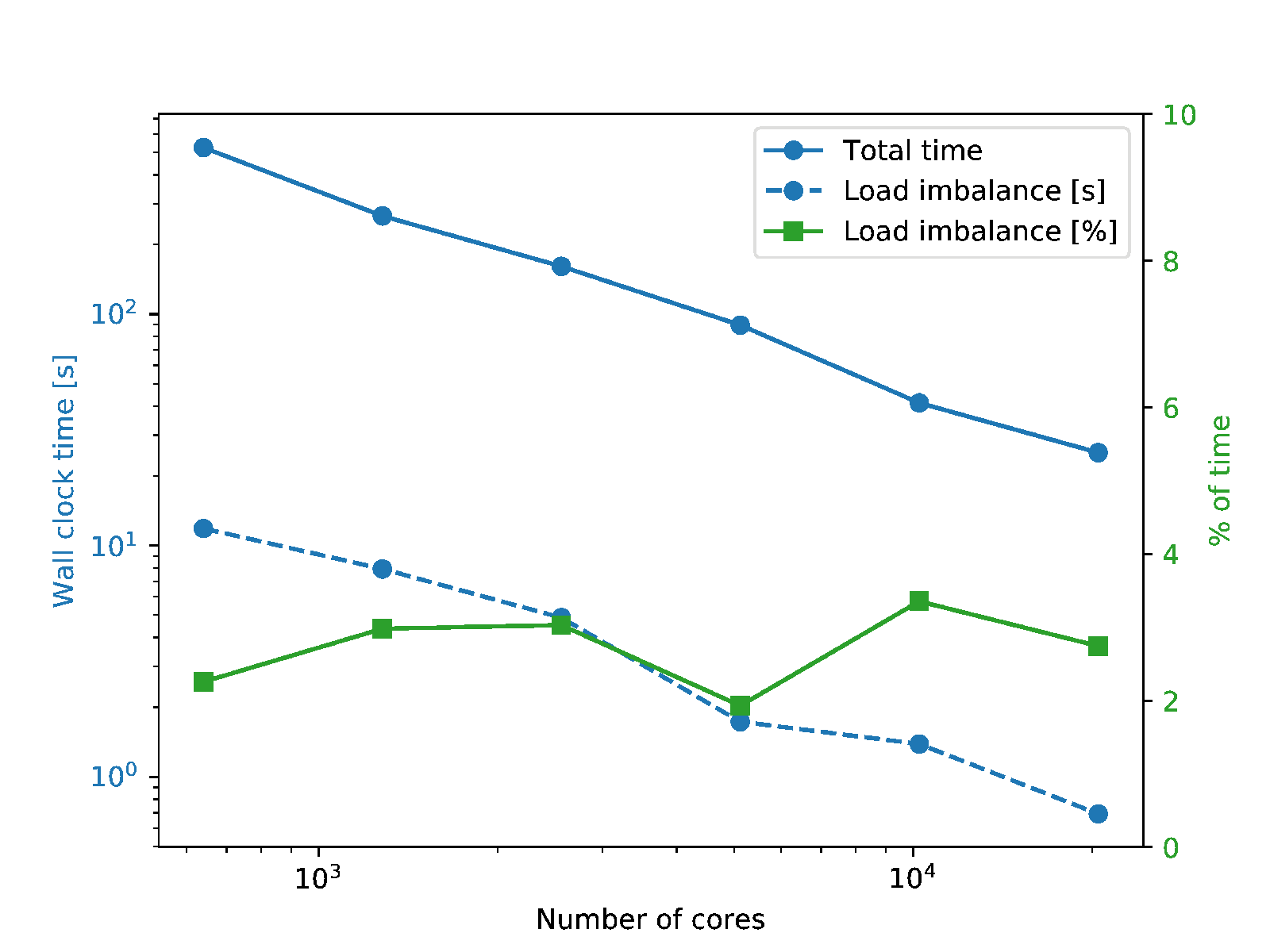}
  \caption{Total time and time lost due to
    load imbalance for running 100 iterations with 1.6 Billion walkers for the
  Cr$_2$/cc-pVDZ (28e in 76o) on 640 to 20480 cores (not counting initialisation). The calculations were run on Intel
  Xeon Gold 6148 Skylake processors, with a 100 Gb/s OmniPath node
  interconnect. The code was compiled using the Intel Fortran compiler,
  version 19.0.4. A semi-stochastic core-space of 50000 determinants was used,
  and PCHB excitation generation. For the largest number of cores, the
    time step is $3.68\times 10^{-4}$ with an average acceptance ratio of
    12.51\%, which is representative for all numbers of cores. The load imbalance time is
  measured as the accumulated difference between the maximum and average time per
  iteration across MPI tasks. Figure generated using {\tt Matplotlib}~\cite{matplotlib}.}
  \label{fig:scaling}
\end{figure}

The parallel efficiency of NECI is made possible by treating static load imbalance. \NECI contains a load-balancing feature\cite{hande}, which
is enabled by default and periodically re-assigns some determinants to other
processors in order to maintain a constant number of walkers per processor. As
can be seen in figure ~\ref{fig:scaling}, for the given benchmarks, no
significant load imbalance occurs up to (including) 20480
cores~\cite{load_imbalance_output, femoco_lb}.
The initialization of a simulation does not feature the same speed-up due to I/O operations and
initial communication such as trial wave function setup and core space
generation. However, since it does not play a significant role for extended
calculations, we consider only the time spent in the actual
iterations.

\begin{table}[t!]
\vspace{4mm}

\begin{tabular*}{1\textwidth}{@{\extracolsep{\fill}}cccccc}
\hline
\hline
\addlinespace[2mm]
\multirow{2}{*}{\# of walkers} & \multirow{2}{*}{\# of cores} & average time & ratio of & ratio of  & efficiency of \tabularnewline\addlinespace[-2mm]
 &  & per iteration & \# of cores & average time/iteration & parallelisation\tabularnewline\addlinespace[2mm]
\hline
\addlinespace[2mm]
\multirow{1}{*}{$32\times 10^{9}$} & 19960 & 23.5 seconds & \multirow{2}{*}{1.242} & \multirow{2}{*}{1.246} & \multirow{2}{*}{99.68\%}\tabularnewline
$32\times10^9$ & 24800 & 18.8 seconds &  &  & \tabularnewline\addlinespace[2mm] 
\hline
$8\times 10^9$ & 16000 & 7.3 seconds &- &- &-\\
\hline
\hline
\end{tabular*}
\caption{Efficiency of parallelisation for a CAS(54,54) of the FeMoco molecule.
In both $32\times 10^9$ walkers cases, the time per iteration is averaged over more than 250
iterations and in both cases the unbiased sample variance over the 250+ iterations is less
than 0.5 seconds. For comparison, the time per iteration for $8\times 10^9$
walkers which was used to obtain the energy reported in table \ref{tab:Best-non-extrapolated-energies}, is given. Calculations were run on 512, 620 and 400 nodes with Intel Xeon Gold 6148 Skylake processors
with 20 cores at 2.4 GHz and 96 GB of DDR4 RAM, and all nodes were in a single
island with a 100 Gb/s OmniPath interconnect between the nodes. Hyperthreading
was not used.}\label{tab:scaling}
\end{table}

\section{Interfacing NECI}
\label{sec:interface}
The ongoing development of \NECI is focused on an efficiently scaling
  solver for the CI-problem.
It is not desirable to reimplement functionality that is already available
  in existing quantum chemistry codes.
Since the CI-problem is defined by the electronic
  integrals and subsequent methods depend on the results
  of the CI-step,
    namely the reduced density matrices,
  it is easily possible to replace a CI-solver
  of existing quantum chemistry code with \NECI.

To use \NECI only an input file and a FCIDUMP file \cite{Knowles1989},
  which is the widely understood file format for the electronic integrals,
  are required.
After running \NECI the stochastically sampled reduced density matrices
  are available as input for further calculations in other codes.
It is possible to link \NECI as library and call it directly
  or to run it as external process and do the communication with
  explicit copying of files.
The first alternative will be referred to as embedded,
  the second is the decoupled form.

Due to the stochastic nature of the Monte Carlo algorithm,
  it is not yet possible to use \NECI as a black box CI-solver
  for larger systems.
In this case it is recommended to use the decoupled form for a better
  manual control of the convergence.
Another advantage of the decoupled form is the combination of \NECI with
  different quantum chemical algorithms or implementations that do not benefit
  from massive parallelisation as much as \NECI.
This way it is possible to switch from serial or single node execution
  to multiple nodes in the CI-step.
So far \NECI has been coupled with
 \Molpro~\cite{MOLPRO-WIREs,MOLPRO_brief}, {\tt Molcas 8}\cite{molcas-general}, \OMolcas~\cite{OpenMolcas},
 \PySCF~\cite{pyscf}, and \VASP~\cite{Kresse1996}.
 
\section{Stochastic-MCSCF}
\label{sec:Stochastic-MCSCF}
The Stochastic multi-configurational self-consistent field (MCSCF) procedure emerges from the combination of
conventional MCSCF methodologies with FCIQMC as the CI-eigensolver.
Stochastic-MCSCF approaches greatly enlarge the applicability of FCIQMC to strongly correlated
molecular systems of practical interest in chemical science.

To date two implementations of Stochastic-MCSCF have been made available,
based on the interface of \NECI with \OMolcas~\cite{LiManni2016, OpenMolcas} (and {\tt Molcas 8} \cite{molcas-general}) and \PySCF~\cite{Thomas2015_2, pyscf}.
As they are both based on the complete active space (CAS) concept, they are often also referred to as
Stochastic-CASSCF methods.

The Stochastic-CASSCF implemented in PySCF is based on a second order CASSCF
algorithm~\cite{Sun2017} which decouples the orbital optimization problem from
the active space CI problem, allowing for easy interfacing with \NECI.

At each \emph{macro-iteration}, a FCIQMC simulation is performed at the current point of orbital expansion,
and density matrices are stochastically sampled (see section~\ref{sec:rdms}). These are then passed back to PySCF,
which updates the orbital coefficients accordingly, using either a 1-step\cite{Sun2017} or 2-step approach\cite{Yanai2009}.

The Stochastic-CASSCF implemented in \OMolcas is based on the quasi-second order Super-CI orbital optimization.
Optimal orbitals (in the variational sense) are found by solving the Super-CI secular equations in the $\ket{\text{Super}-CI}$ basis,
defined by the CAS wave function at the point of expansion, $|0\rangle$, and all its possible single excitations
\begin{equation}
|\text{Super}-CI\rangle = |0\rangle + \sum_{p>q}{\chi_{pq} (\hat{E}_{pq} - \hat{E}_{qp}) | 0 \rangle}
\end{equation}
The wave function is improved by mixing single excitations to the $|0\rangle$ wave function.
As the CASSCF optimization proceeds, the $\chi_{pq}$ coefficients decrease until they vanish,
and $|0\rangle$ will reveal the variational stationary point.
Third-order density matrix elements of the exact Super-CI approach are avoided by utilizing an effective one-electron Hamiltonian,
as discussed in greater details in Reference~\citenum{LiManni2016}.

\begin{figure}
	\centering
	\includegraphics[width= 6cm]{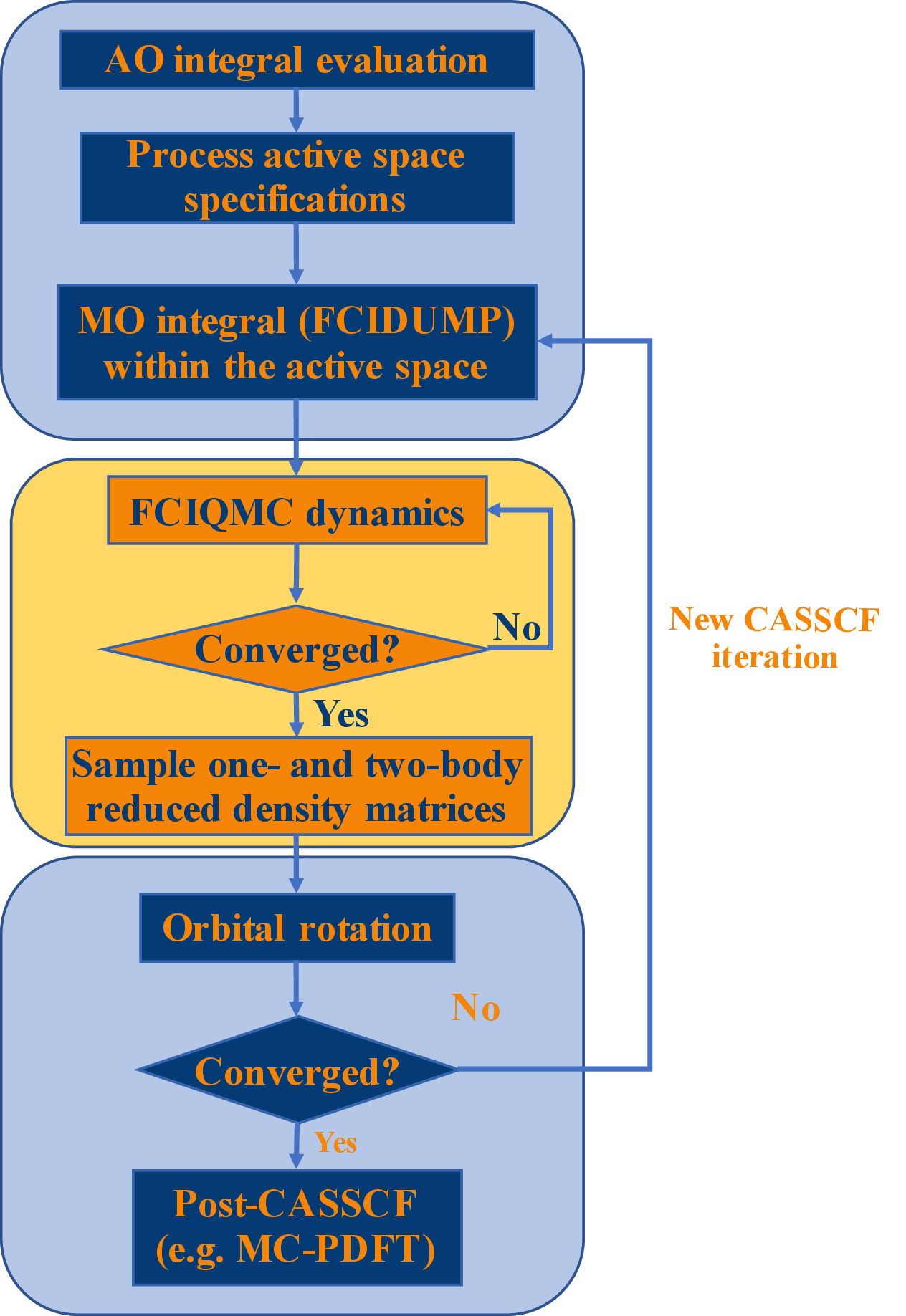}
	\caption{
	Flow chart summarizing the Stochastic-CASSCF steps. The blue boxes represent parts of the algorithm performed at
	the \OMolcas or \PySCF interfacing software. The center yellow box shows the two crucial FCIQMC steps, stochastic optimization
	of the CAS-CI wave function and sampling of one- and two-body reduced density matrices.
	When embedded schemes are employed, additional external potentials are added within the interfacing software when
	generating the FCIDUMP file. Post-CASSCF procedures, such as the MC-PDFT methodology, follow the Stochastic-CASSCF approach
	within the interfacing software.
	}
	\label{Fig:Stochastic-MCSCF}
\end{figure}

A flow chart of Stochastic-CASSCF describing the various steps of the CASSCF wave function optimization
is given in Figure~\ref{Fig:Stochastic-MCSCF}.

The Stochastic-CASSCF approach has successfully been applied to a number of challenging chemical problems.
The accuracy of the method has been demonstrated on simple test cases, such as benzene and naphthalene~\cite{Thomas2015_2}
and more complex molecular systems, namely coronene~\cite{Thomas2015_2}, free-base porphyrin and Mg-porphyrin~\cite{LiManni2016}.
More recently the method has also been applied to understand the mechanism stabilizing intermediate spin states in
Fe(II)-porphyrin~\cite{LiManni2018,limanni2019a}, the study of a
  $[Fe(III)_{2}S_{2}(SCH_{3})_{2}]^{2-}$ iron-sulfur model system in its
  oxidized form \cite{LiManni2019}, and new superexchange paths in corner-sharing cuprates~\cite{Bogdanov2018}.

To date, only state specific Stochastic-CASSCF optimizations have been reported.
However, state-average Stochastic-CASSCF optimizations are a straightforward extension that can be reached by taking advantage of
the \NECI capability to optimize excited states wave functions, as discussed in section~\ref{sec:ex}.
The Stochastic-CASSCF method can also be coupled to the \emph{adaptive shift} approach discussed in section~\ref{sec:ada}
with a great enhancement in performance.
 
\section{Conclusion}
With \NECI we present a state of the art FCIQMC program capable of running a large
variety of versions of the FCIQMC algorithm. This includes the semi-stochastic
FCIQMC feature, energy estimation using trial wave functions, the
stochastic sampling of reduced density matrices, and excited state
calculations. Further features of \texttt{NECI}'s FCIQMC implementation discussed are
the real-time FCIQMC method and the adaptive shift method, as well as a
spin-adapted formulation of the algorithm and support for
transcorrelated Hamiltonians. We demonstrated the scalability of
the program to up to 24800 cores, showing that the code can run
efficiently on large-scale machines.

Finally, we highlighted the interoperability of \NECI with other quantum
chemistry software, in particular \OMolcas and \PySCF, which can be used to run
Stochastic-CASSCF calculations.
 
\section{Supplementary Materials}
Example FCIQMC output files for excited state calculations (\nolinkurl{
  output\_file\_excited\_state\_be2\_b1g.txt} and
\nolinkurl{stats\_file\_excited\_state\_be2\_b1g.txt}) and real-time
calculations including the resulting spectrum
(\nolinkurl{output\_file\_real\_time\_be2\_b1g.txt}, and
\nolinkurl{fft\_spectrum\_be2\_b1g.txt}) for the examples presented in
section \ref{fig:exc_state} are available in the supplemental
material. Furthermore, the supplement contains the output files for scaling
(\nolinkurl{output\_file\_scaling\_with\_*\_cores.txt} and
\nolinkurl{output\_file\_energy\_with\_8b\_walkers.txt}) and load imbalance
analysis
(\nolinkurl{output\_file\_load\_imbalance\_n*.txt}). Exemplary output and
integral files for
a similarity transformed FCIQMC calculation of the Neon atom in a cc-pVDZ
basis set are also supplied in the supplement
(\nolinkurl{tcdump\_Ne\_st\_pVDZ.h5} and \nolinkurl{ FCIDUMP\_Ne\_st\_pVDZ}
integral files and \nolinkurl{output\_file\_Ne\_st\_pVDZ.txt} output
\nolinkurl{
  stats\_file\_Ne\_st\_pVDZ.txt} files). All output files contain the
corresponding FCIQMC input.

\section{Data availability statement}
The data that supports the findings of this study are available within the
article and its supplementary material~\cite{supplement}. The \NECI program
can be obtained at \url{https://github.com/ghb24/NECI_STABLE}, the development
version can be obtained from the corresponding author upon reasonable request.

\begin{acknowledgments}
  The early development of NECI was supported by the EPSRC under grant numbers
  EP/J003867/1 and EP/I014624/1.
  
  We would like to thank Olle Gunnarsson, David Tew, Daniel Kats, Aron Cohen,
  and Vamshi Katakuri for insightful discussions.
  
  The high performance benchmarks discussed in section ~\ref{sec:par},
  were ran on the MPCDF (Max Planck Computing \& Data Facility) system Cobra.
\end{acknowledgments}

\appendix

\section{Stochastic excitation generation and $p_{gen}$}
In the following appendices we will consider in some detail the process of (random) excitation generation in FCIQMC - a crucial yet rather flexible aspect of the algorithm. We will consider some general aspects, such as implementation of Abelian symmetries in the excitation process, as well as non-uniform excitation generation, as is often desirable in quantum chemical Hamiltonians. There are other classes of systems (such as Hubbard models, Transcorrelated Hamiltonians, spin models such as Heisenberg systems, etc) for which more specialised considerations are necessary for efficient excitation generation but we will not consider them here. 

The first general point about excitation generation, (by which we mean starting from a given determinant $\ket{D_i}$ we randomly pick either one or two electrons, and a corresponding number of holes to substitute them with, to create a second determinant $\ket{D_j}$), is that if $|H_{ij}|>0$, then the probability ($p_{gen}(j|i)$) to select $\ket{D_j}$ and $\ket{D_i}$, must also be greater than 0.  Furthermore, $p_{gen}(j|i)$ must be {\em computable}, and in general the effort to do so will depend on the algorithm chosen to execute the excitation process.   

Let us discuss in more detail the process of stochastic excitation generation, and its
impact on $p_{gen}$.
Suppose we are simulating a system of $n$ electrons in $2N$ spin orbitals $\{\phi_1,...,\phi_{2N}\}$.
A given determinant $\ket{D_i}$ can be defined by its occupation number representation, 
$I = \ket{n_1,...,n_{2N}}$, which is 
a binary string such that $n_i=1$ if orbital $i$ is occupied (`an electron in $\ket{D_i}$'), and $n_i=0$ if it is unoccupied 
(`a hole in $\ket{D_i}$').
Each orbital carries a spin quantum number $\sigma(\phi_i)$, and may also carry a symmetry label, $\Gamma(\phi_i)$. These are both discrete symmetries, with $\sigma=\pm 1/2$, and $\Gamma=\Gamma_1,...,\Gamma_G$, 
where $G$ is the number of irreducible representations available in the point-group of system 
under consideration. 
We will only consider Abelian groups, so that the product of symmetry labels 
uniquely specifies another symmetry label. This simplifies the task of selecting excitations, 
although it does not necessarily exploit the full symmetry of the problem. 

\subsection{Uniform excitation generation} 
Now we wish to perform a stochastic excitation generation, which we will initially 
consider without the use of any symmetry/spin information. 
For example, we can select a 
pair of electrons, $i,j$ (with $i<j$) in $\ket{D_i}$, at random, and a pair of holes
$a,b$ (with $a<b$), and perform the transition $ij\rightarrow ab$. 
The corresponding matrix element is
\begin{equation}
H_{ij}^{ab} = \braket{ij | ab } - \braket{ij | ba} \equiv \braket{ij || ab }
\end{equation}
We will denote the electron pair simply as $ij$ and the hole pair $ab$. 

For this simple procedure, it is clear that the probability to choose
$\ket{D_j}$ from $\ket{D_i}$
is simply:
\begin{equation}
 p_{gen}(j|i) = {n\choose 2}^{-1}{2N-n \choose 2}^{-1}
\end{equation}
from which it follows that $p_{gen}(j|i)\sim (nN)^{-2}$.
This procedure does not take symmetry or spin quantum numbers into account, and it is quite 
possible that the corresponding Hamiltonian matrix element is zero. To ensure that we do not generate
such excitations, we need to select the hole pairs so that following two conditions are met:
\begin{eqnarray}
 \sigma(\phi_i)+\sigma(\phi_j)&=&\sigma(\phi_a)+\sigma(\phi_b), \\
 \Gamma(\phi_i)\times\Gamma(\phi_j)&=&\Gamma(\phi_a)\times\Gamma(\phi_b).
\end{eqnarray}
These restriction greatly impact the way in which we will select $i,j$ and $a,b$, 
and the resulting generation probability. 

\subsubsection{Imposing symmetries via conditional probabilities\label{app:symmetries}} 
One way to impose symmetries in excitation generation while keeping track 
of the generation probabilities is via
the notion of conditional probabilities. 
For example, rather than drawing $(ij)$ 
and $(ab)$ independently, with probability $p(ab,ij)=p(ab)p(ij)$,
one can instead draw $(ab)$ given that one has already drawn $(ij)$; the probability 
for this process is given by 
\begin{equation}
p(ab,ij)=p(ab|ij)p(ij),
\end{equation} 
where $p(ij)$ is the probability to select $(ij)$ in the first place. 
If $(ij)$ has a particular characteristic that confers 
a physical (e.g. symmetry-related) constraint on $(ab)$, this can be implemented at the stage in which we select $(ab)$: 
$(ab)$ need only be selected from among those hole-pairs for which the constraint 
is satisfied. For example, if the electrons $(ij)$ have opposite spins then the holes $(ab)$ must also 
have opposite spins. The smaller number of possibilities in choosing the $ab$ pair then leads to 
a larger $p(ab|ij)$ compared to $p(ab)$, which can be thought of as a renormalisation
of the latter probability to take into account the constraint. 

The concept of conditional probabilities can be further extended so that the pair $(ij)$ itself
is made to satisfy a particular condition. Suppose we introduce a set of 
of conditions $\{\calC_1,\calC_2,...\}$ such that the union of 
all such conditions is exhaustive. It is possible to draw conditional probabilities with respect to 
such conditions.  For example,
\begin{eqnarray}
\calC_1 & = & \mbox{`electron pair have the same spin'} \\
\calC_2 & = & \mbox{`electron pair have opposite spins'}
\end{eqnarray}
then one can write:
\begin{eqnarray}
p(ab,ij)=p(ab,ij|\calC_1)p(\calC_1)+p(ab,ij|\calC_2)p(\calC_2)
\end{eqnarray}
with
\begin{equation}
p(\calC_1)+p(\calC_2)=1.
\end{equation} 
$p(\calC_1)$, the probability to select same-spin excitations, can be chosen arbitrarily, which 
then fixes $p(\calC_2)$ according to the above.

The advantage of this formulation is that we can skew the selection of electron pairs, for example, towards 
opposite spin  excitations if that proves advantageous, and to be able to compute the 
resulting probabilities. Furthermore, we can write:
\begin{eqnarray}
p(ab,ij|\calC_1)=p(ab|ij)p(ij|\calC_1)\\
p(ab,ij|\calC_2)=p(ab|ij)p(ij|\calC_2)
\end{eqnarray}
which allows us to select a pair of electrons satisfying condition $\calC_1$, and subsequently
draw a pair of holes given one has selected an electron pair with the same spin (which 
implies that hole-pair must be chosen to have the same spin as the electron-pair). 

\subsection{Cauchy-Schwartz excitation generation}\label{app:nu_excit}
Let us now consider how to generate the hole pairs in a non-uniform manner, to reflect the fact that, 
in {\em ab initio} Hamiltonians, the matrix elements vary strongly in magnitude. 
Since the spawning probability is proportional to the ratio
$\abs{H_{ij}}/p_\mathrm{gen}(j|i)$, it is clearly desirable to generate excitations
which make this ratio as uniform as possible, ideally with $p_\mathrm{gen}(j|i)\propto|H_{ij}|$.
In this way, one would ensure a relatively uniform probability of successful spawning, which ideally would be 
close to one, implying a low rejection rate.
Keeping the discussion focussed on double excitations (the generalisation to single excitations being straightforward) 
the question that arises is: how best can one select $ij$ and $ab$ such that $p_\mathrm{gen}(j|i)\propto\abs{H_{ij}}$ to a good
approximation, and $p_{gen}$ remains exactly computable without excessive cost. We will 
see there is a compromise to be made. One can ensure precise proportionality between
$p_\mathrm{gen}(j|i)$ and $\abs{H_{ij}}$ but only at prohibitive cost. Alternatively, one might be able to
select $ij$ and $ab$ to effect the transition $\ket{D_i}\rightarrow \ket{D_j}$ based on computationally inexpensive heuristics, to
provide approximate proportionality, which will nevertheless allow for a large overall improvement in 
efficiency. 

To ensure exact proportionality between $p_\mathrm{gen}(j|i)$ and $\abs{H_{ij}}$ it is 
necessary to enumerate all electron-pairs and hole-pairs which are possible 
from $\ket{D_i}$, and to construct the {\em cumulative probability function} (CPF), from which the
desired distribution can be straightforwardly sampled. The (unnormalised) CPF is: 
\begin{equation}
	F_{ab,ij}[D] = \sum_{ee^\prime\in D  }^{ij} \sum_{hh^\prime\in D  }^{ab} \abs{\braket{ee^\prime || hh^\prime}} \label{4indcpf}
\end{equation}
In this expression, the sum over $ee^\prime$ runs over all enumerated electron pairs in $D$ up to $ij$, 
and similarly for the hole-pairs (up to $ab$). The CPF is a non-decreasing function of its
discrete  arguments,  and its inverse transform enables one to select $ab$ and $ij$ with probability 
proportional to $\abs{\braket{ij || ab}}$. From the point of generation probabilities, this
is the ideal excitation generator, allowing for
a uniform spawning probability (which can be made to equal unity, implying zero rejection rates.)
Unfortunately the CPF costs $\mathcal{O}(n^2N^2)$ to set up (for each determinant $\ket{D_i}$), making it
prohibitive in practice.

To make practical progress, we need an approximate distribution function which is 
much cheaper to calculate. Two observations can be made in this relation. First, if the two electrons
have different spin, then the Hamiltonian matrix element consists of only one rather 
than two terms. This is because upon excitation $ij\rightarrow ab$, the two holes 
must match the spins of the two electrons. For example $\sigma(a)=\sigma(i)$ and 
$\sigma(b)=\sigma(j)$. In this case, the Hamiltonian matrix element reduces to:
\begin{equation}
	H_{ij} = \braket{ij | ab}
\end{equation} 
with the exchange term $\braket{ij | ba} = 0$.

With this simpler matrix element, we now ask: given that we have chosen an electron pair $ij$,
how can we select the hole pair $ab$ so that, with high probability, the resulting 
matrix element $\braket{ij | ab}$ is large? At this point we can appeal to the Cauchy-Schwarz
inequality, which provides a strict upper bound:
\begin{equation}
\braket{ij | ab}\le \sqrt{     \braket{ii| aa } \braket{ jj |bb }} 
\end{equation}
This suggests that, as long as $\braket{ij | ab}$ is non-zero by symmetry, it may be 
advantageous to select the hole $a$ so that $\braket{ii| aa }$ is large, and 
the hole $b$ so that $\braket{ jj |bb } $ large. 
Because $i$ and $j$ have different spins, the selection of $a$ and $b$ will be {\em independent}
of each other, with $a$ for example being chosen from the $\alpha$-spin holes available, and $b$ from the $\beta$-spin holes.
To do this, we set up two CPFs:
\begin{eqnarray}
	F_{a}[i\in \alpha D] & = & \sum_{h\in \alpha D}^{a} \sqrt{     \braket{hh|ii}},  \\
	F_{b}[j\in \beta D] &=   & \sum_{h\in \beta D}^{b} \sqrt{     \braket{hh|jj}},
\end{eqnarray}
where the sums over $h$ runs over the $\alpha$ or $\beta$ {\em holes} in $D$. (The notation $i\in \alpha D$ means an 
$\alpha$-electron in $D$, and $h\in \alpha D$ means an $\alpha$-hole in $D$). 
Unlike Eq.(\ref{4indcpf}), these CPFs cost only $\mathcal{O}(N)$ to set up, and allow (via their inverse transforms) the selection of $a$ and $b$ 
with probabilities proportional to $\sqrt{\braket{aa | ii}}$ and $\sqrt{\braket{jj | bb}}$ 
respectively. 

The Cauchy-Schwarz bound on an individual 4-index integral provides a very
useful factorised approximation for the 
purposes of excitation generation, especially for opposite-spin excitations. 
The case for same-spin excitations is less favourable because it involves
the difference between two 4-index integrals, and in this case we must obtain an upper bound for this difference expressed 
in a factorised form. We use the following much less tight upper bound:
\begin{align}
	\abs{\braket{ij|ab}-\braket{ij|ba}} 
		&\le  [\sqrt{     \braket{aa| ii}}  + \sqrt{     \braket{ aa |jj }} ] \\
		&\times [ \sqrt{     \braket{bb| ii }} + \sqrt{     \braket{ bb |jj }} ] 
\end{align} 
In practice, we must draw two holes $a$ and $b$ from the {\em same}
set of holes, avoiding the possibility of drawing 
the same hole twice. Because we would like to avoid setting
up a two-dimensional CPF (which would cost $\mathcal{O}(N^2)$), 
we create one one-dimension CPF in order to draw hole $a$,
and then remove this hole in the CPF before drawing 
the second hole. In other words we set up two related CPFs
\begin{eqnarray}
F_{a}[ij \in D] & = & \sum_h^a \sqrt{     \braket{ii|hh}} +\sqrt{     \braket{jj|hh}}, \\
F^\prime_{b}[ij \in D] & = & \begin{cases}
	F_{b}[ij \in D | a] & \text{if}\; b < a,\\
	\begin{aligned}
		&F_{b}[ij \in D] - \sqrt{\braket{ii|aa}} \\ &\;\;\;- \sqrt{\braket{jj|aa}}
	\end{aligned} &\mathrm{if}\;b \ge a,
\end{cases}
\end{eqnarray} 
drawing hole $a$ from $F_{a}$ and hole $b$ from $F^\prime_{b}$. 

Our exploration of excitation generation has led us to discover
many highly performing schemes. The Cauchy-Schwarz (CS) scheme
presented above is a good starting point, but it has a number of
weaknesses that can be further addressed. In particular, as noted
above, the upper bound obtained is particularly poor for double
excitations with the same spin, and in general the specified bound
can be too loose. Fortunately, the selection of the second hole, $b$,
is made once the first hole, $a$, has already been chosen, and as
such the exact double excitation Hamiltonian matrix elements can be
used at this stage, such that an updated CPF for selecting the second
electron is given by
\begin{equation}
	F_{b}[ij \in D | a] = \sum_{\substack{h\in D\\h\ne a}}^b \sqrt{\abs{\braket{ij|ab} - \braket{ij|ba}}}.
\end{equation}
This Part-Exact (PE) scheme no longer provides a strict bound,
but by better representing the cancellation of terms present
in these matrix elements, it provides a substantially better
approximation. More crucially, it improves the prediction of the
elements that were previously handled least effectively, and thus
relaxes the time-step constraints on the overall calculation.

Due to the increase in computational cost involved in constructing
two lists, and the additional normalisation of the probabilities 
required by causing the two selections not to be made in the same
manner, this update to the scheme increases computational cost per
iteration. In almost all systems examined this is far outweighed by
the time-step changes, especially in systems with large basis sets
or with translational symmetry. However, it is possible to find
systems where the pure CS scheme is more optimal.

\subsubsection{Preparing for excitation generation} 
For determinant $D$, to pick an excited determinant, first construct a table of hole occupancies for each spin and irreducible representation, so that 
$n_{\sigma\Gamma}[D] $ is the number of holes with spin $\sigma$ in irrep $\Gamma$ available in $D$. This is an $\mathcal{O}(n)$ process. 

We next decide whether we wish to make a single excitation or a double excitation from $D$. A single excitation is 
chosen with probability $p_{sing}$, a parameter which can be optimised as the simulation proceeds to maximise the 
acceptance ratio and time-step of the simulation. The probability to create a double excitation is chosen such that the maximal
ratios $\frac{|H_{ij}|}{p_{gen}(j|i)}$ for single and double excitations are
equal, which for \emph{ab initio} systems typically means double excitations
dominate. To a first approximation $p_{sing}=nN/(nN+n^2N^2)$, which is in
general a small number on the order of $(nN)^{-1}$. The probability of attempting a double excitation is then $p_{doub}=1-p_{sing}$. 

\subsubsection{Single Excitations}
\label{app:weighted_signles}
If a single excitation is being attempted, first select an electron (say $i$) at random, with probability $n^{-1}$. 
The spin $\sigma=\sigma(i)$ and irrep $\Gamma=\Gamma(\phi_i)$ of the electron determines the spin and 
irrep of the hole. 

To select the hole $a$, run over all $n_{\sigma\Gamma}$ holes available in $D$ 
with spin and symmetry $\sigma\Gamma$, and compute the (unnormalised) cumulative probability function,
\begin{equation}
	F^{(1)}_{a}[i\in D]= \sum_{h\in \sigma\Gamma D}^a|\braket{D_i^h |\hat{H}| D}|,
\end{equation}
where $\ket{D_i^h}$ is a single-excitation $i\rightarrow h$ from $\ket{D}$, and $\braket{D_i^h |\hat{H}| D}$
is the Hamiltonian matrix element between them. 
The normalisation of the CPF is give by the last element in the array:
\begin{equation}
 \Sigma_i = F^{(1)}_{n_{\sigma\Gamma}}[i\in D],
\end{equation}
where $n_{\sigma\Gamma}$ is the number of holes available with spin $\sigma$ in irrep $\Gamma$ in $D$. 
Using $F^{(1)}_{a}$, select hole $a$ (with probability $|\langle D_i^a | H | D \rangle|$), by inverting the 
CPF. This is selected by generating a
uniform random number $\xi$ in the interval $[0,\Sigma_i)$, and determining the index of $a$ such that the condition 
\begin{equation}
F^{(1)}_{a-1} < \xi \le F^{(1)}_{a}
\end{equation}
is met. The overall generation probability for this 
excitation is:
\begin{equation}
p_{gen}(a,i) = p(a|i)\times p(i)\times p_{sing},
\end{equation}
where 
\begin{align}
p(a|i)&= \frac{\theta_{a }}{\Sigma_i }, \\
\theta_{a} &= F^{(1)}_{a }-F^{(1)}_{a-1  }, \\
p(i)&= n^{-1}.
\end{align}
This completes the selection of a singly excited determinant. The computation of $F^{(1)}_{a}$ is an order $\mathcal{O}(nN)$ 
operation  (with $\mathcal{O}(N)$ holes being summed over, and each Hamiltonian  matrix element being $\mathcal{O}(n)$ to compute). 
Although this is expensive, 
the generation of single excitations turns out overall to be a small fraction of the total cost, largely because 
the relatively small number of times such excitations are attempted.

\subsubsection{Double Excitations with opposite spin electron-pairs} 
If a double excitation is being attempted, then firstly a pair of electrons
needs to be selected. The first electron, $i$, should be selected uniformly at
random. Following this, the CPF
\begin{equation}
	F_j[i \in D] = 
		\sum_{\substack{k \in D\\k \ne i}}^j\braket{ik|ik} \times
		\begin{cases}
			p_{opp} & \text{if $i,k$ opp},\\
			1 - p_{opp} & \text{otherwise.}
		\end{cases}
\end{equation}
should be constructed, where $p_{opp}$ is a optimisable biasing factor
towards excitations with electrons having opposite spins. The second
electron is selected through inversion of the CPF.

If the two selected electrons  have opposite spins, then the first hole to
be chosen is, by convention, always a $\beta$ electron, and the second hole always $\alpha$.
This choice is entirely arbitrary, and in some high-spin systems it may
make sense to reverse this selection. Considering all available orbitals
of this spin, the CPF
\begin{equation}
	F^{(\beta)}_a[i\in D] = \sum_{h\in \beta D}^a \sqrt{\braket{hh | ii}},
\end{equation}
is constructed, where $i$ is taken to be the electron from the selected pair
with $\beta$ spin, and the hole selected by inverting the CPF.

Once this first electron has been chosen, the symmetry of the target orbital
is now fixed by the constraint that
$\Gamma_a\otimes\Gamma^\prime = \Gamma_i \otimes \Gamma_j$. This greatly restricts
the number of holes that must be considered when constructing the final CPF,
\begin{equation}
	F^{(\alpha\Gamma^\prime a)}_b[ij \in D] =
		\sum_{h\in \alpha\Gamma^\prime D}^b\sqrt{\abs{\braket{ij|ab}}}.
\end{equation}
Note that with the conventional choice of orbital $i$ above, $\braket{ji|ah} = 0$,
and can thus be excluded. The second hole is then also obtained by inverting the
CPF. The generation probability is then given by:
\begin{align}
	p_{gen}(ab,ij) &= p(ij)p(a|i)p(b|ija)p_{doub}, \\ 
	p(ij) &= \frac{1}{N}\left(\frac{\theta_j^{(i)}}{\Sigma_i} + \frac{\theta^{(j)}_i}{\Sigma_j}\right), \\
	\theta_j^{(i)} &= F_j[i\in D] - F_{j-i}[i\in D], \\
	p(a|i) &= \frac{\theta_a}{\Sigma^{(\beta)}(i)}, \\
	\theta_a & = F^{(\beta)}_a - F^{(\beta)}_{a-1}, \\ 
	p(b|ija)&= \frac{\theta_b}{\Sigma^{(\alpha \Gamma^\prime)}(ija)},  \\ 
	\theta_{b} &= F^{(\alpha \Gamma^\prime a)}_b - F^{(\alpha \Gamma^\prime a)}_{b-1},
\end{align}
where $\Sigma_i,\Sigma^{(\beta)}(i),\Sigma^{(\alpha \Gamma^\prime)}(ija)$
are the normalisations of $F_j, F^{(\beta)}_a, F^{(\alpha\Gamma^\prime a)}_b$
respectively, and are given by the final entries of the corresponding arrays.
	
The asymetric selection of $\alpha$ and $\beta$ holes is somewhat peculiar.
It should be noted that it is possible to make this selection symmetrically,
considering \emph{all} available holes in the selection of the first hole,
and then renormalising the probabilities to account for the possibility of
selecting $b$ first. The symmetric scheme increases computational cost
substantially (twice as many holes need to be considered in the CPF, and
a further CPF must be calculated for the renormalisation). It also makes the
overall time-step behaviour worse as, although it improves the general
smoothness, for the worst-case scenario with a very rarely selected excitation
with very different $a,b$ and $b,a$ probabilities, the denominator is
increased substantially by considering more orbitals, whilst leaving the
numerator essentially unchanged.

\subsubsection{Double excitations with same-spin electron pairs}

If the pair of electrons, selected as described above, have the same
spin the process needs to account for the fact that the holes can
be selected in either order, and the probabilities need to be
adjusted to compensate.

Now, considering only holes with the same spin as the two electrons,
construct the CPF
\begin{equation}
	F_a^{(\sigma)}[ij\in D] =
		\sum_{h\in\sigma D}^a
		\sqrt{\braket{hh|ii}} + \sqrt{\braket{hh|jj}}.
\end{equation}
Hole $a$ can then be selected through inversion of this CPF, which
fixes the symmetry of hole $b$ such that
$\Gamma_a \otimes \Gamma_b = \Gamma_i \otimes \Gamma_j$. The
CPF for selecting the second hole can then be constructed from the
(much smaller) set of holes with the appropriate symmetry, such
that
\begin{equation}
	F^{(\sigma\Gamma_b a)}_b[ij \in D] =
		\sum_{\substack{h\in \alpha\Gamma_b D\\h\ne a}}^b
		\sqrt{\abs{\braket{ij|ab} - \braket{ij|ba}}}.
\end{equation}

The second hole, $b$, can then be selected through inversion of this
CPF. It is important to note that as the selection of the first
hole includes all holes of the hole with the given spin, the
selection of the holes could have been made in the reverse order, and
this needs to be taken into account in the generation probability,
which is given by:
\begin{equation}
	p_{gen}(ab,ij) = [p(a|ijb)p(b|ij) + p(b|ija)p(a|ij)]p(ij)p_{double},
\end{equation}
where
\begin{align}
	p(ij) &= \frac{1}{N}\left(\frac{\theta_j^{(i)}}{\Sigma_i} + \frac{\theta^{(j)}_i}{\Sigma_j}\right), \\
	\theta_j^{(i)} &= F_j[i\in D] - F_{j-1}[i\in D], \\
	p(a|ij) &= \frac{\theta_a}{\Sigma_a} \label{eqn:pab-start}, \\
	\theta_a &= F^{(\sigma)}_a -F^{(\sigma)}_{a-1}, \\
	p(b|ija) &= \frac{\theta^{(a)}_b}{\Sigma^{(a)}_b}, \\
	\theta^{(a)}_b &= F^{(\sigma\Gamma_b a)}_b - F^{(\sigma\Gamma_b a)}_{b-1}
\end{align}
and $\Sigma_i, \Sigma_a, \Sigma^{(a)}_b$ are the normalisations of the three
CPFs, given by their final elements. Note that in the implementation, the
normalisations of four CPFs must be calculated to be able to calculate
$p(a|ijb)$ as well as $p(b|ija)$.

\subsection{Pre-computed heat-bath sampling}
\label{app:pchb}
While the Cauchy-Schwartz excitation generator has negligible memory cost,
picking an excitation requires $\mathcal{O}(N)$ steps, each involving
Hamiltonian matrix elements, making the procedure expensive. The pre-computed heat-bath algorithm employed in \NECI is a simple
approximation derived from the heat-bath sampling \cite{holmes2016} and offers
a much faster excitation generation, at the cost of increased memory
requirement. The heat-bath probability distribution can also used to determine a cutoff
in a deterministic scheme, leading to the heat-bath CI (HCI) method \cite{holmes2016_b}. The sampling
can either use uniform single excitations or the weighted scheme outlined in section ~\ref{app:weighted_signles}, and
approximates the exact heat-bath sampling of double-excitations by uniformly picking the occupied
orbitals, and then picking two target orbitals simultaneously weighted with
the Hamiltonian matrix element. Since the double excitations play the largest
role in excitation generation, and the singles' matrix elements depend on the
determinants in addition to the excitation, it is typically most efficient to
generate only double excitations in a weighted fashion, resulting in an excellent tradeoff between
optimal weights and the cost of excitation generation.

To create a double excitation using pre-computed heat-bath generation, first,
two occupied orbitals $i$, $j$ are chosen uniformly at random, using a bias towards
spin-opposite exctitations, which is determined similar to the bias towards
double excitations. This works analogously to the Cauchy-Schwartz excitation
generation outlined in section ~\ref{app:nu_excit}. Then, a pair $a$, $b$ of
orbitals is chosen using pre-computed weights

\begin{equation}
p(ab|ij) = \frac{|H_{ij}^{ab}|}{\sum_{a'b'} |H_{ij}^{a'b'}|}\,,
\label{eq:hbs}
\end{equation}
where $H_{ij}^{ab}$ is the matrix element for a double excitation from
orbitals $i,\,j$ to orbitals $a,\,b$. These are independent of the determinant
and thus can be pre-computed at memory cost $\mathcal{O}(M^4)$. Then, pairs of
orbitals can be picked using these weights via alias sampling
\cite{Walker1977} in $\mathcal{O}(1)$ time. If one of the picked orbitals
$a$, $b$ is occupied, or all matrix elements $H_{ij}^{ab}$ are zero, the
excitation is immediately rejected, otherwise, we continue with the FCIQMC
scheme.

As it is desirable to use spatial orbital indices to save memory, but the
matrix element depends on the relative spin of the orbitals in the case of a
spin-opposite excitation since it determines if an exchange integral is used,
for each pair of spatial orbitals $i$, $j$, three probability distributions
are generated, one for the spin-parallel case, one for the spin-opposite case
without exchange and one for the spin-opposite case with exchange. Between the
latter two, we then choose the exchange case with probability
\begin{equation}
p_{exch}(ij) = \frac{\sum_{ab} |H_{i\alpha j\beta}^{a\beta
b\alpha}|}{\sum_{ab}|H_{i\alpha j\beta}^{a\beta
b\alpha}| + |H_{i\alpha j\beta}^{a\alpha
b\beta}|}\,.
\end{equation}
The denominator is the same as the denominator in Eq. (\ref{eq:hbs}) for
spin orbitals, while the numerator is the denominator in Eq. (\ref{eq:hbs})
for spatial orbitals in the exchange case. The bias $p_{exch}$ hence relates
the spatial orbital distributions to the original distribution (\ref{eq:hbs}).

This approach is tailored for rapid excitation
generation, as the process is in principle $\mathcal{O}(1)$, while yielding
acceptance rates comparable to the on-the-fly Cauchy-Schwartz generation. Due
to implementational details of \texttt{NECI}, the uniform selection of electrons
scales linearly with the number of electrons, which, however, does not
constitute a bottleneck in practical application. The
rapid excitation generation has important consequences for the scalability of
the algorithm, since the stochastic nature of the algorithm can give rise to
dynamic load imbalance if the time taken for excitation generation can vary significantly
depending on determinant and electron/orbital selection.


\begin{thebibliography}{132}%
\makeatletter
\providecommand \@ifxundefined [1]{%
 \@ifx{#1\undefined}
}%
\providecommand \@ifnum [1]{%
 \ifnum #1\expandafter \@firstoftwo
 \else \expandafter \@secondoftwo
 \fi
}%
\providecommand \@ifx [1]{%
 \ifx #1\expandafter \@firstoftwo
 \else \expandafter \@secondoftwo
 \fi
}%
\providecommand \natexlab [1]{#1}%
\providecommand \enquote  [1]{``#1''}%
\providecommand \bibnamefont  [1]{#1}%
\providecommand \bibfnamefont [1]{#1}%
\providecommand \citenamefont [1]{#1}%
\providecommand \href@noop [0]{\@secondoftwo}%
\providecommand \href [0]{\begingroup \@sanitize@url \@href}%
\providecommand \@href[1]{\@@startlink{#1}\@@href}%
\providecommand \@@href[1]{\endgroup#1\@@endlink}%
\providecommand \@sanitize@url [0]{\catcode `\\12\catcode `\$12\catcode
  `\&12\catcode `\#12\catcode `\^12\catcode `\_12\catcode `\%12\relax}%
\providecommand \@@startlink[1]{}%
\providecommand \@@endlink[0]{}%
\providecommand \url  [0]{\begingroup\@sanitize@url \@url }%
\providecommand \@url [1]{\endgroup\@href {#1}{\urlprefix }}%
\providecommand \urlprefix  [0]{URL }%
\providecommand \Eprint [0]{\href }%
\providecommand \doibase [0]{http://dx.doi.org/}%
\providecommand \selectlanguage [0]{\@gobble}%
\providecommand \bibinfo  [0]{\@secondoftwo}%
\providecommand \bibfield  [0]{\@secondoftwo}%
\providecommand \translation [1]{[#1]}%
\providecommand \BibitemOpen [0]{}%
\providecommand \bibitemStop [0]{}%
\providecommand \bibitemNoStop [0]{.\EOS\space}%
\providecommand \EOS [0]{\spacefactor3000\relax}%
\providecommand \BibitemShut  [1]{\csname bibitem#1\endcsname}%
\let\auto@bib@innerbib\@empty
%</preamble>
\bibitem [{\citenamefont {Alavi}(2000)}]{Alavi2000}%
  \BibitemOpen
  \bibfield  {author} {\bibinfo {author} {\bibfnamefont {A.}~\bibnamefont
  {Alavi}},\ }\bibfield  {title} {\enquote {\bibinfo {title} {Two interacting
  electrons in a box: An exact diagonalization study},}\ }\href {\doibase
  10.1063/1.1316045} {\bibfield  {journal} {\bibinfo  {journal} {The Journal of
  Chemical Physics}\ }\textbf {\bibinfo {volume} {113}},\ \bibinfo {pages}
  {7735--7745} (\bibinfo {year} {2000})},\ \Eprint
  {http://arxiv.org/abs/https://doi.org/10.1063/1.1316045}
  {https://doi.org/10.1063/1.1316045} \BibitemShut {NoStop}%
\bibitem [{\citenamefont {Thompson}\ and\ \citenamefont
  {Alavi}(2002)}]{ThompsonAlavi2002}%
  \BibitemOpen
  \bibfield  {author} {\bibinfo {author} {\bibfnamefont {D.~C.}\ \bibnamefont
  {Thompson}}\ and\ \bibinfo {author} {\bibfnamefont {A.}~\bibnamefont
  {Alavi}},\ }\bibfield  {title} {\enquote {\bibinfo {title} {Two interacting
  electrons in a spherical box: An exact diagonalization study},}\ }\href
  {\doibase 10.1103/PhysRevB.66.235118} {\bibfield  {journal} {\bibinfo
  {journal} {Phys. Rev. B}\ }\textbf {\bibinfo {volume} {66}},\ \bibinfo
  {pages} {235118} (\bibinfo {year} {2002})}\BibitemShut {NoStop}%
\bibitem [{\citenamefont {Booth}, \citenamefont {Thom},\ and\ \citenamefont
  {Alavi}(2009)}]{BoothThomAlavi2009}%
  \BibitemOpen
  \bibfield  {author} {\bibinfo {author} {\bibfnamefont {G.~H.}\ \bibnamefont
  {Booth}}, \bibinfo {author} {\bibfnamefont {A.~J.~W.}\ \bibnamefont {Thom}},
  \ and\ \bibinfo {author} {\bibfnamefont {A.}~\bibnamefont {Alavi}},\
  }\bibfield  {title} {\enquote {\bibinfo {title} {Fermion monte carlo without
  fixed nodes: A game of life, death, and annihilation in slater determinant
  space},}\ }\href {\doibase 10.1063/1.3193710} {\bibfield  {journal} {\bibinfo
   {journal} {The Journal of Chemical Physics}\ }\textbf {\bibinfo {volume}
  {131}},\ \bibinfo {pages} {054106} (\bibinfo {year} {2009})}\BibitemShut
  {NoStop}%
\bibitem [{\citenamefont {Blankenbecler}, \citenamefont {Scalapino},\ and\
  \citenamefont {Sugar}(1981)}]{sugar1981}%
  \BibitemOpen
  \bibfield  {author} {\bibinfo {author} {\bibfnamefont {R.}~\bibnamefont
  {Blankenbecler}}, \bibinfo {author} {\bibfnamefont {D.~J.}\ \bibnamefont
  {Scalapino}}, \ and\ \bibinfo {author} {\bibfnamefont {R.~L.}\ \bibnamefont
  {Sugar}},\ }\bibfield  {title} {\enquote {\bibinfo {title} {Monte carlo
  calculations of coupled boson-fermion systems. i},}\ }\href {\doibase
  10.1103/PhysRevD.24.2278} {\bibfield  {journal} {\bibinfo  {journal} {Phys.
  Rev. D}\ }\textbf {\bibinfo {volume} {24}},\ \bibinfo {pages} {2278--2286}
  (\bibinfo {year} {1981})}\BibitemShut {NoStop}%
\bibitem [{\citenamefont {Sugiyama}\ and\ \citenamefont
  {Koonin}(1986)}]{koonin1986}%
  \BibitemOpen
  \bibfield  {author} {\bibinfo {author} {\bibfnamefont {G.}~\bibnamefont
  {Sugiyama}}\ and\ \bibinfo {author} {\bibfnamefont {S.}~\bibnamefont
  {Koonin}},\ }\bibfield  {title} {\enquote {\bibinfo {title} {Auxiliary field
  monte-carlo for quantum many-body ground states},}\ }\href {\doibase
  https://doi.org/10.1016/0003-4916(86)90107-7} {\bibfield  {journal} {\bibinfo
   {journal} {Annals of Physics}\ }\textbf {\bibinfo {volume} {168}},\ \bibinfo
  {pages} {1 -- 26} (\bibinfo {year} {1986})}\BibitemShut {NoStop}%
\bibitem [{\citenamefont {Zhang}\ and\ \citenamefont
  {Krakauer}(2003)}]{Zhang2003}%
  \BibitemOpen
  \bibfield  {author} {\bibinfo {author} {\bibfnamefont {S.}~\bibnamefont
  {Zhang}}\ and\ \bibinfo {author} {\bibfnamefont {H.}~\bibnamefont
  {Krakauer}},\ }\bibfield  {title} {\enquote {\bibinfo {title} {Quantum monte
  carlo method using phase-free random walks with slater determinants},}\
  }\href {\doibase 10.1103/PhysRevLett.90.136401} {\bibfield  {journal}
  {\bibinfo  {journal} {Phys. Rev. Lett.}\ }\textbf {\bibinfo {volume} {90}},\
  \bibinfo {pages} {136401} (\bibinfo {year} {2003})}\BibitemShut {NoStop}%
\bibitem [{\citenamefont {Zhang}, \citenamefont {Carlson},\ and\ \citenamefont
  {Gubernatis}(1997)}]{zhang1997}%
  \BibitemOpen
  \bibfield  {author} {\bibinfo {author} {\bibfnamefont {S.}~\bibnamefont
  {Zhang}}, \bibinfo {author} {\bibfnamefont {J.}~\bibnamefont {Carlson}}, \
  and\ \bibinfo {author} {\bibfnamefont {J.~E.}\ \bibnamefont {Gubernatis}},\
  }\bibfield  {title} {\enquote {\bibinfo {title} {Constrained path monte carlo
  method for fermion ground states},}\ }\href@noop {} {\bibfield  {journal}
  {\bibinfo  {journal} {Physical Review B}\ }\textbf {\bibinfo {volume} {55}},\
  \bibinfo {pages} {7464} (\bibinfo {year} {1997})}\BibitemShut {NoStop}%
\bibitem [{\citenamefont {Dobrautz}, \citenamefont {Smart},\ and\ \citenamefont
  {Alavi}(2019)}]{Dobrautz2019}%
  \BibitemOpen
  \bibfield  {author} {\bibinfo {author} {\bibfnamefont {W.}~\bibnamefont
  {Dobrautz}}, \bibinfo {author} {\bibfnamefont {S.~D.}\ \bibnamefont {Smart}},
  \ and\ \bibinfo {author} {\bibfnamefont {A.}~\bibnamefont {Alavi}},\
  }\bibfield  {title} {\enquote {\bibinfo {title} {Efficient formulation of
  full configuration interaction quantum monte carlo in a spin eigenbasis via
  the graphical unitary group approach},}\ }\href {\doibase 10.1063/1.5108908}
  {\bibfield  {journal} {\bibinfo  {journal} {The Journal of Chemical Physics}\
  }\textbf {\bibinfo {volume} {151}},\ \bibinfo {pages} {094104} (\bibinfo
  {year} {2019})},\ \Eprint
  {http://arxiv.org/abs/https://doi.org/10.1063/1.5108908}
  {https://doi.org/10.1063/1.5108908} \BibitemShut {NoStop}%
\bibitem [{\citenamefont {Cleland}, \citenamefont {Booth},\ and\ \citenamefont
  {Alavi}(2010)}]{initiator-fciqmc}%
  \BibitemOpen
  \bibfield  {author} {\bibinfo {author} {\bibfnamefont {D.}~\bibnamefont
  {Cleland}}, \bibinfo {author} {\bibfnamefont {G.~H.}\ \bibnamefont {Booth}},
  \ and\ \bibinfo {author} {\bibfnamefont {A.}~\bibnamefont {Alavi}},\
  }\bibfield  {title} {\enquote {\bibinfo {title} {Communications: Survival of
  the fittest: Accelerating convergence in full configuration-interaction
  quantum monte carlo},}\ }\href {\doibase 10.1063/1.3302277} {\bibfield
  {journal} {\bibinfo  {journal} {The Journal of Chemical Physics}\ }\textbf
  {\bibinfo {volume} {132}},\ \bibinfo {pages} {041103} (\bibinfo {year}
  {2010})}\BibitemShut {NoStop}%
\bibitem [{\citenamefont {Ghanem}, \citenamefont {Lozovoi},\ and\ \citenamefont
  {Alavi}(2019)}]{adaptive_shift}%
  \BibitemOpen
  \bibfield  {author} {\bibinfo {author} {\bibfnamefont {K.}~\bibnamefont
  {Ghanem}}, \bibinfo {author} {\bibfnamefont {A.~Y.}\ \bibnamefont {Lozovoi}},
  \ and\ \bibinfo {author} {\bibfnamefont {A.}~\bibnamefont {Alavi}},\
  }\bibfield  {title} {\enquote {\bibinfo {title} {Unbiasing the initiator
  approximation in full configuration interaction quantum monte carlo},}\
  }\href {\doibase 10.1063/1.5134006} {\bibfield  {journal} {\bibinfo
  {journal} {The Journal of Chemical Physics}\ }\textbf {\bibinfo {volume}
  {151}},\ \bibinfo {pages} {224108} (\bibinfo {year} {2019})},\ \Eprint
  {http://arxiv.org/abs/https://doi.org/10.1063/1.5134006}
  {https://doi.org/10.1063/1.5134006} \BibitemShut {NoStop}%
\bibitem [{\citenamefont {Petruzielo}\ \emph {et~al.}(2012)\citenamefont
  {Petruzielo}, \citenamefont {Holmes}, \citenamefont {Changlani},
  \citenamefont {Nightingale},\ and\ \citenamefont {Umrigar}}]{Petruzielo2012}%
  \BibitemOpen
  \bibfield  {author} {\bibinfo {author} {\bibfnamefont {F.~R.}\ \bibnamefont
  {Petruzielo}}, \bibinfo {author} {\bibfnamefont {A.~A.}\ \bibnamefont
  {Holmes}}, \bibinfo {author} {\bibfnamefont {H.~J.}\ \bibnamefont
  {Changlani}}, \bibinfo {author} {\bibfnamefont {M.~P.}\ \bibnamefont
  {Nightingale}}, \ and\ \bibinfo {author} {\bibfnamefont {C.~J.}\ \bibnamefont
  {Umrigar}},\ }\bibfield  {title} {\enquote {\bibinfo {title} {Semistochastic
  projector monte carlo method},}\ }\href@noop {} {\bibfield  {journal}
  {\bibinfo  {journal} {Phys. Rev. Lett.}\ }\textbf {\bibinfo {volume} {109}},\
  \bibinfo {pages} {230201} (\bibinfo {year} {2012})}\BibitemShut {NoStop}%
\bibitem [{\citenamefont {Blunt}\ \emph
  {et~al.}(2015{\natexlab{a}})\citenamefont {Blunt}, \citenamefont {Smart},
  \citenamefont {Kersten}, \citenamefont {Spencer}, \citenamefont {Booth},\
  and\ \citenamefont {Alavi}}]{Blunt2015}%
  \BibitemOpen
  \bibfield  {author} {\bibinfo {author} {\bibfnamefont {N.~S.}\ \bibnamefont
  {Blunt}}, \bibinfo {author} {\bibfnamefont {S.~D.}\ \bibnamefont {Smart}},
  \bibinfo {author} {\bibfnamefont {J.~A.~F.}\ \bibnamefont {Kersten}},
  \bibinfo {author} {\bibfnamefont {J.~S.}\ \bibnamefont {Spencer}}, \bibinfo
  {author} {\bibfnamefont {G.~H.}\ \bibnamefont {Booth}}, \ and\ \bibinfo
  {author} {\bibfnamefont {A.}~\bibnamefont {Alavi}},\ }\bibfield  {title}
  {\enquote {\bibinfo {title} {{Semi-stochastic full configuration interaction
  quantum Monte Carlo: Developments and application}},}\ }\href {\doibase
  10.1063/1.4920975} {\bibfield  {journal} {\bibinfo  {journal} {The Journal of
  Chemical Physics}\ }\textbf {\bibinfo {volume} {142}},\ \bibinfo {pages}
  {184107} (\bibinfo {year} {2015}{\natexlab{a}})}\BibitemShut {NoStop}%
\bibitem [{\citenamefont {Blunt}\ \emph
  {et~al.}(2015{\natexlab{b}})\citenamefont {Blunt}, \citenamefont {Smart},
  \citenamefont {Booth},\ and\ \citenamefont {Alavi}}]{Blunt2015b}%
  \BibitemOpen
  \bibfield  {author} {\bibinfo {author} {\bibfnamefont {N.~S.}\ \bibnamefont
  {Blunt}}, \bibinfo {author} {\bibfnamefont {S.~D.}\ \bibnamefont {Smart}},
  \bibinfo {author} {\bibfnamefont {G.~H.}\ \bibnamefont {Booth}}, \ and\
  \bibinfo {author} {\bibfnamefont {A.}~\bibnamefont {Alavi}},\ }\bibfield
  {title} {\enquote {\bibinfo {title} {An excited-state approach within full
  configuration interaction quantum monte carlo},}\ }\href {\doibase
  10.1063/1.4932595} {\bibfield  {journal} {\bibinfo  {journal} {The Journal of
  Chemical Physics}\ }\textbf {\bibinfo {volume} {143}},\ \bibinfo {pages}
  {134117} (\bibinfo {year} {2015}{\natexlab{b}})}\BibitemShut {NoStop}%
\bibitem [{\citenamefont {Zhang}\ and\ \citenamefont
  {Kalos}(1993)}]{Zhang1993}%
  \BibitemOpen
  \bibfield  {author} {\bibinfo {author} {\bibfnamefont {S.}~\bibnamefont
  {Zhang}}\ and\ \bibinfo {author} {\bibfnamefont {M.~H.}\ \bibnamefont
  {Kalos}},\ }\bibfield  {title} {\enquote {\bibinfo {title} {Bilinear quantum
  monte carlo: Expectations and energy differences},}\ }\href {\doibase
  10.1007/BF01053583} {\bibfield  {journal} {\bibinfo  {journal} {Journal of
  Statistical Physics}\ }\textbf {\bibinfo {volume} {70}},\ \bibinfo {pages}
  {515--533} (\bibinfo {year} {1993})}\BibitemShut {NoStop}%
\bibitem [{\citenamefont {Overy}\ \emph {et~al.}(2014)\citenamefont {Overy},
  \citenamefont {Booth}, \citenamefont {Blunt}, \citenamefont {Shepherd},
  \citenamefont {Cleland},\ and\ \citenamefont {Alavi}}]{Overy2014}%
  \BibitemOpen
  \bibfield  {author} {\bibinfo {author} {\bibfnamefont {C.}~\bibnamefont
  {Overy}}, \bibinfo {author} {\bibfnamefont {G.~H.}\ \bibnamefont {Booth}},
  \bibinfo {author} {\bibfnamefont {N.~S.}\ \bibnamefont {Blunt}}, \bibinfo
  {author} {\bibfnamefont {J.~J.}\ \bibnamefont {Shepherd}}, \bibinfo {author}
  {\bibfnamefont {D.}~\bibnamefont {Cleland}}, \ and\ \bibinfo {author}
  {\bibfnamefont {A.}~\bibnamefont {Alavi}},\ }\bibfield  {title} {\enquote
  {\bibinfo {title} {Unbiased reduced density matrices and electronic
  properties from full configuration interaction quantum monte carlo},}\ }\href
  {\doibase 10.1063/1.4904313} {\bibfield  {journal} {\bibinfo  {journal} {The
  Journal of Chemical Physics}\ }\textbf {\bibinfo {volume} {141}},\ \bibinfo
  {pages} {244117} (\bibinfo {year} {2014})}\BibitemShut {NoStop}%
\bibitem [{\citenamefont {Blunt}\ \emph {et~al.}(2014)\citenamefont {Blunt},
  \citenamefont {Rogers}, \citenamefont {Spencer},\ and\ \citenamefont
  {Foulkes}}]{Blunt2014}%
  \BibitemOpen
  \bibfield  {author} {\bibinfo {author} {\bibfnamefont {N.~S.}\ \bibnamefont
  {Blunt}}, \bibinfo {author} {\bibfnamefont {T.~W.}\ \bibnamefont {Rogers}},
  \bibinfo {author} {\bibfnamefont {J.~S.}\ \bibnamefont {Spencer}}, \ and\
  \bibinfo {author} {\bibfnamefont {W.~M.~C.}\ \bibnamefont {Foulkes}},\
  }\bibfield  {title} {\enquote {\bibinfo {title} {Density-matrix quantum monte
  carlo method},}\ }\href@noop {} {\bibfield  {journal} {\bibinfo  {journal}
  {Phys. Rev. B}\ }\textbf {\bibinfo {volume} {89}},\ \bibinfo {pages} {245124}
  (\bibinfo {year} {2014})}\BibitemShut {NoStop}%
\bibitem [{\citenamefont {Blunt}, \citenamefont {Booth},\ and\ \citenamefont
  {Alavi}(2017)}]{Blunt2017}%
  \BibitemOpen
  \bibfield  {author} {\bibinfo {author} {\bibfnamefont {N.~S.}\ \bibnamefont
  {Blunt}}, \bibinfo {author} {\bibfnamefont {G.~H.}\ \bibnamefont {Booth}}, \
  and\ \bibinfo {author} {\bibfnamefont {A.}~\bibnamefont {Alavi}},\ }\bibfield
   {title} {\enquote {\bibinfo {title} {Density matrices in full configuration
  interaction quantum monte carlo: Excited states, transition dipole moments,
  and parallel distribution},}\ }\href {\doibase 10.1063/1.4986963} {\bibfield
  {journal} {\bibinfo  {journal} {The Journal of Chemical Physics}\ }\textbf
  {\bibinfo {volume} {146}},\ \bibinfo {pages} {244105} (\bibinfo {year}
  {2017})}\BibitemShut {NoStop}%
\bibitem [{\citenamefont {Anderson}, \citenamefont {Shiozaki},\ and\
  \citenamefont {Booth}(2020)}]{Anderson2020}%
  \BibitemOpen
  \bibfield  {author} {\bibinfo {author} {\bibfnamefont {R.~J.}\ \bibnamefont
  {Anderson}}, \bibinfo {author} {\bibfnamefont {T.}~\bibnamefont {Shiozaki}},
  \ and\ \bibinfo {author} {\bibfnamefont {G.~H.}\ \bibnamefont {Booth}},\
  }\bibfield  {title} {\enquote {\bibinfo {title} {Efficient and stochastic
  multireference perturbation theory for large active spaces within a full
  configuration interaction quantum monte carlo framework},}\ }\href {\doibase
  10.1063/1.5140086} {\bibfield  {journal} {\bibinfo  {journal} {The Journal of
  Chemical Physics}\ }\textbf {\bibinfo {volume} {152}},\ \bibinfo {pages}
  {054101} (\bibinfo {year} {2020})},\ \Eprint
  {http://arxiv.org/abs/https://doi.org/10.1063/1.5140086}
  {https://doi.org/10.1063/1.5140086} \BibitemShut {NoStop}%
\bibitem [{\citenamefont {Li~Manni}, \citenamefont {Smart},\ and\ \citenamefont
  {Alavi}(2016)}]{LiManni2016}%
  \BibitemOpen
  \bibfield  {author} {\bibinfo {author} {\bibfnamefont {G.}~\bibnamefont
  {Li~Manni}}, \bibinfo {author} {\bibfnamefont {S.~D.}\ \bibnamefont {Smart}},
  \ and\ \bibinfo {author} {\bibfnamefont {A.}~\bibnamefont {Alavi}},\
  }\bibfield  {title} {\enquote {\bibinfo {title} {Combining the complete
  active space self-consistent field method and the full configuration
  interaction quantum monte carlo within a super-ci framework, with application
  to challenging metal-porphyrins},}\ }\href {\doibase
  10.1021/acs.jctc.5b01190} {\bibfield  {journal} {\bibinfo  {journal} {Journal
  of Chemical Theory and Computation}\ }\textbf {\bibinfo {volume} {12}},\
  \bibinfo {pages} {1245--1258} (\bibinfo {year} {2016})}\BibitemShut {NoStop}%
\bibitem [{\citenamefont {Thomas}\ \emph
  {et~al.}(2015{\natexlab{a}})\citenamefont {Thomas}, \citenamefont {Sun},
  \citenamefont {Alavi},\ and\ \citenamefont {Booth}}]{Thomas2015_2}%
  \BibitemOpen
  \bibfield  {author} {\bibinfo {author} {\bibfnamefont {R.~E.}\ \bibnamefont
  {Thomas}}, \bibinfo {author} {\bibfnamefont {Q.}~\bibnamefont {Sun}},
  \bibinfo {author} {\bibfnamefont {A.}~\bibnamefont {Alavi}}, \ and\ \bibinfo
  {author} {\bibfnamefont {G.~H.}\ \bibnamefont {Booth}},\ }\bibfield  {title}
  {\enquote {\bibinfo {title} {Stochastic multiconfigurational self-consistent
  field theory},}\ }\href {\doibase 10.1021/acs.jctc.5b00917} {\bibfield
  {journal} {\bibinfo  {journal} {Journal of Chemical Theory and Computation}\
  }\textbf {\bibinfo {volume} {11}},\ \bibinfo {pages} {5316--5325} (\bibinfo
  {year} {2015}{\natexlab{a}})}\BibitemShut {NoStop}%
\bibitem [{\citenamefont {Guther}\ \emph {et~al.}(2018)\citenamefont {Guther},
  \citenamefont {Dobrautz}, \citenamefont {Gunnarsson},\ and\ \citenamefont
  {Alavi}}]{guther2018}%
  \BibitemOpen
  \bibfield  {author} {\bibinfo {author} {\bibfnamefont {K.}~\bibnamefont
  {Guther}}, \bibinfo {author} {\bibfnamefont {W.}~\bibnamefont {Dobrautz}},
  \bibinfo {author} {\bibfnamefont {O.}~\bibnamefont {Gunnarsson}}, \ and\
  \bibinfo {author} {\bibfnamefont {A.}~\bibnamefont {Alavi}},\ }\bibfield
  {title} {\enquote {\bibinfo {title} {Time propagation and spectroscopy of
  fermionic systems using a stochastic technique},}\ }\href {\doibase
  10.1103/PhysRevLett.121.056401} {\bibfield  {journal} {\bibinfo  {journal}
  {Phys. Rev. Lett.}\ }\textbf {\bibinfo {volume} {121}},\ \bibinfo {pages}
  {056401} (\bibinfo {year} {2018})}\BibitemShut {NoStop}%
\bibitem [{\citenamefont {Luo}\ and\ \citenamefont {Alavi}(2018)}]{Luo2018}%
  \BibitemOpen
  \bibfield  {author} {\bibinfo {author} {\bibfnamefont {H.}~\bibnamefont
  {Luo}}\ and\ \bibinfo {author} {\bibfnamefont {A.}~\bibnamefont {Alavi}},\
  }\bibfield  {title} {\enquote {\bibinfo {title} {Combining the
  transcorrelated method with full configuration interaction quantum monte
  carlo: Application to the homogeneous electron gas},}\ }\href {\doibase
  10.1021/acs.jctc.7b01257} {\bibfield  {journal} {\bibinfo  {journal} {Journal
  of Chemical Theory and Computation}\ }\textbf {\bibinfo {volume} {14}},\
  \bibinfo {pages} {1403--1411} (\bibinfo {year} {2018})},\ \bibinfo {note}
  {pMID: 29431996},\ \Eprint
  {http://arxiv.org/abs/https://doi.org/10.1021/acs.jctc.7b01257}
  {https://doi.org/10.1021/acs.jctc.7b01257} \BibitemShut {NoStop}%
\bibitem [{\citenamefont {Dobrautz}, \citenamefont {Luo},\ and\ \citenamefont
  {Alavi}(2019)}]{Dobrautz2019_b}%
  \BibitemOpen
  \bibfield  {author} {\bibinfo {author} {\bibfnamefont {W.}~\bibnamefont
  {Dobrautz}}, \bibinfo {author} {\bibfnamefont {H.}~\bibnamefont {Luo}}, \
  and\ \bibinfo {author} {\bibfnamefont {A.}~\bibnamefont {Alavi}},\ }\bibfield
   {title} {\enquote {\bibinfo {title} {Compact numerical solutions to the
  two-dimensional repulsive hubbard model obtained via nonunitary similarity
  transformations},}\ }\href {\doibase 10.1103/PhysRevB.99.075119} {\bibfield
  {journal} {\bibinfo  {journal} {Phys. Rev. B}\ }\textbf {\bibinfo {volume}
  {99}},\ \bibinfo {pages} {075119} (\bibinfo {year} {2019})}\BibitemShut
  {NoStop}%
\bibitem [{\citenamefont {Cohen}\ \emph {et~al.}(2019)\citenamefont {Cohen},
  \citenamefont {Luo}, \citenamefont {Guther}, \citenamefont {Dobrautz},
  \citenamefont {Tew},\ and\ \citenamefont {Alavi}}]{Cohen2019}%
  \BibitemOpen
  \bibfield  {author} {\bibinfo {author} {\bibfnamefont {A.~J.}\ \bibnamefont
  {Cohen}}, \bibinfo {author} {\bibfnamefont {H.}~\bibnamefont {Luo}}, \bibinfo
  {author} {\bibfnamefont {K.}~\bibnamefont {Guther}}, \bibinfo {author}
  {\bibfnamefont {W.}~\bibnamefont {Dobrautz}}, \bibinfo {author}
  {\bibfnamefont {D.~P.}\ \bibnamefont {Tew}}, \ and\ \bibinfo {author}
  {\bibfnamefont {A.}~\bibnamefont {Alavi}},\ }\bibfield  {title} {\enquote
  {\bibinfo {title} {Similarity transformation of the electronic
  schr{\"o}dinger equation via jastrow factorization},}\ }\href {\doibase
  10.1063/1.5116024} {\bibfield  {journal} {\bibinfo  {journal} {The Journal of
  Chemical Physics}\ }\textbf {\bibinfo {volume} {151}},\ \bibinfo {pages}
  {061101} (\bibinfo {year} {2019})},\ \Eprint
  {http://arxiv.org/abs/https://doi.org/10.1063/1.5116024}
  {https://doi.org/10.1063/1.5116024} \BibitemShut {NoStop}%
\bibitem [{\citenamefont {Jeszenszki}, \citenamefont {Alavi},\ and\
  \citenamefont {Brand}(2019)}]{Jezenski2019}%
  \BibitemOpen
  \bibfield  {author} {\bibinfo {author} {\bibfnamefont {P.}~\bibnamefont
  {Jeszenszki}}, \bibinfo {author} {\bibfnamefont {A.}~\bibnamefont {Alavi}}, \
  and\ \bibinfo {author} {\bibfnamefont {J.}~\bibnamefont {Brand}},\ }\bibfield
   {title} {\enquote {\bibinfo {title} {Are smooth pseudopotentials a good
  choice for representing short-range interactions?}}\ }\href {\doibase
  10.1103/PhysRevA.99.033608} {\bibfield  {journal} {\bibinfo  {journal} {Phys.
  Rev. A}\ }\textbf {\bibinfo {volume} {99}},\ \bibinfo {pages} {033608}
  (\bibinfo {year} {2019})}\BibitemShut {NoStop}%
\bibitem [{\citenamefont {Malone}\ \emph {et~al.}(2015)\citenamefont {Malone},
  \citenamefont {Blunt}, \citenamefont {Shepherd}, \citenamefont {Lee},
  \citenamefont {Spencer},\ and\ \citenamefont {Foulkes}}]{Malone2015}%
  \BibitemOpen
  \bibfield  {author} {\bibinfo {author} {\bibfnamefont {F.~D.}\ \bibnamefont
  {Malone}}, \bibinfo {author} {\bibfnamefont {N.~S.}\ \bibnamefont {Blunt}},
  \bibinfo {author} {\bibfnamefont {J.~J.}\ \bibnamefont {Shepherd}}, \bibinfo
  {author} {\bibfnamefont {D.~K.~K.}\ \bibnamefont {Lee}}, \bibinfo {author}
  {\bibfnamefont {J.~S.}\ \bibnamefont {Spencer}}, \ and\ \bibinfo {author}
  {\bibfnamefont {W.~M.~C.}\ \bibnamefont {Foulkes}},\ }\bibfield  {title}
  {\enquote {\bibinfo {title} {Interaction picture density matrix quantum monte
  carlo},}\ }\href@noop {} {\bibfield  {journal} {\bibinfo  {journal} {J. Chem.
  Phys.}\ }\textbf {\bibinfo {volume} {143}},\ \bibinfo {pages} {044116}
  (\bibinfo {year} {2015})}\BibitemShut {NoStop}%
\bibitem [{\citenamefont {Malone}\ \emph {et~al.}(2016)\citenamefont {Malone},
  \citenamefont {Blunt}, \citenamefont {Brown}, \citenamefont {Lee},
  \citenamefont {Spencer}, \citenamefont {Foulkes},\ and\ \citenamefont
  {Shepherd}}]{Malone2016}%
  \BibitemOpen
  \bibfield  {author} {\bibinfo {author} {\bibfnamefont {F.~D.}\ \bibnamefont
  {Malone}}, \bibinfo {author} {\bibfnamefont {N.~S.}\ \bibnamefont {Blunt}},
  \bibinfo {author} {\bibfnamefont {E.~W.}\ \bibnamefont {Brown}}, \bibinfo
  {author} {\bibfnamefont {D.~K.~K.}\ \bibnamefont {Lee}}, \bibinfo {author}
  {\bibfnamefont {J.~S.}\ \bibnamefont {Spencer}}, \bibinfo {author}
  {\bibfnamefont {W.~M.~C.}\ \bibnamefont {Foulkes}}, \ and\ \bibinfo {author}
  {\bibfnamefont {J.~J.}\ \bibnamefont {Shepherd}},\ }\bibfield  {title}
  {\enquote {\bibinfo {title} {Accurate exchange-correlation energies for the
  warm dense electron gas},}\ }\href@noop {} {\bibfield  {journal} {\bibinfo
  {journal} {Phys. Rev. Lett.}\ }\textbf {\bibinfo {volume} {117}},\ \bibinfo
  {pages} {115701} (\bibinfo {year} {2016})}\BibitemShut {NoStop}%
\bibitem [{\citenamefont {Spencer}\ \emph {et~al.}(2019)\citenamefont
  {Spencer}, \citenamefont {Blunt}, \citenamefont {Choi}, \citenamefont
  {Etrych}, \citenamefont {Filip}, \citenamefont {Foulkes}, \citenamefont
  {Franklin}, \citenamefont {Handley}, \citenamefont {Malone}, \citenamefont
  {Neufeld}, \citenamefont {Di~Remigio}, \citenamefont {Rogers}, \citenamefont
  {Scott}, \citenamefont {Shepherd}, \citenamefont {Vigor}, \citenamefont
  {Weston}, \citenamefont {Xu},\ and\ \citenamefont {Thom}}]{hande}%
  \BibitemOpen
  \bibfield  {author} {\bibinfo {author} {\bibfnamefont {J.~S.}\ \bibnamefont
  {Spencer}}, \bibinfo {author} {\bibfnamefont {N.~S.}\ \bibnamefont {Blunt}},
  \bibinfo {author} {\bibfnamefont {S.}~\bibnamefont {Choi}}, \bibinfo {author}
  {\bibfnamefont {J.}~\bibnamefont {Etrych}}, \bibinfo {author} {\bibfnamefont
  {M.-A.}\ \bibnamefont {Filip}}, \bibinfo {author} {\bibfnamefont {W.~M.~C.}\
  \bibnamefont {Foulkes}}, \bibinfo {author} {\bibfnamefont {R.~S.~T.}\
  \bibnamefont {Franklin}}, \bibinfo {author} {\bibfnamefont {W.~J.}\
  \bibnamefont {Handley}}, \bibinfo {author} {\bibfnamefont {F.~D.}\
  \bibnamefont {Malone}}, \bibinfo {author} {\bibfnamefont {V.~A.}\
  \bibnamefont {Neufeld}}, \bibinfo {author} {\bibfnamefont {R.}~\bibnamefont
  {Di~Remigio}}, \bibinfo {author} {\bibfnamefont {T.~W.}\ \bibnamefont
  {Rogers}}, \bibinfo {author} {\bibfnamefont {C.~J.~C.}\ \bibnamefont
  {Scott}}, \bibinfo {author} {\bibfnamefont {J.~J.}\ \bibnamefont {Shepherd}},
  \bibinfo {author} {\bibfnamefont {W.~A.}\ \bibnamefont {Vigor}}, \bibinfo
  {author} {\bibfnamefont {J.}~\bibnamefont {Weston}}, \bibinfo {author}
  {\bibfnamefont {R.}~\bibnamefont {Xu}}, \ and\ \bibinfo {author}
  {\bibfnamefont {A.~J.~W.}\ \bibnamefont {Thom}},\ }\bibfield  {title}
  {\enquote {\bibinfo {title} {The hande-qmc project: Open-source stochastic
  quantum chemistry from the ground state up},}\ }\href@noop {} {\bibfield
  {journal} {\bibinfo  {journal} {Journal of Chemical Theory and Computation}\
  }\textbf {\bibinfo {volume} {15}},\ \bibinfo {pages} {1728--1742} (\bibinfo
  {year} {2019})}\BibitemShut {NoStop}%
\bibitem [{\citenamefont {Tubman}\ \emph {et~al.}(2016)\citenamefont {Tubman},
  \citenamefont {Lee}, \citenamefont {Takeshita}, \citenamefont {Head-Gordon},\
  and\ \citenamefont {Whaley}}]{tubman2016}%
  \BibitemOpen
  \bibfield  {author} {\bibinfo {author} {\bibfnamefont {N.~M.}\ \bibnamefont
  {Tubman}}, \bibinfo {author} {\bibfnamefont {J.}~\bibnamefont {Lee}},
  \bibinfo {author} {\bibfnamefont {T.~Y.}\ \bibnamefont {Takeshita}}, \bibinfo
  {author} {\bibfnamefont {M.}~\bibnamefont {Head-Gordon}}, \ and\ \bibinfo
  {author} {\bibfnamefont {K.~B.}\ \bibnamefont {Whaley}},\ }\bibfield  {title}
  {\enquote {\bibinfo {title} {A deterministic alternative to the full
  configuration interaction quantum monte carlo method},}\ }\href@noop {}
  {\bibfield  {journal} {\bibinfo  {journal} {The Journal of chemical physics}\
  }\textbf {\bibinfo {volume} {145}},\ \bibinfo {pages} {044112} (\bibinfo
  {year} {2016})}\BibitemShut {NoStop}%
\bibitem [{\citenamefont {Huron}, \citenamefont {Malrieu},\ and\ \citenamefont
  {Rancurel}(1973)}]{Huron1973}%
  \BibitemOpen
  \bibfield  {author} {\bibinfo {author} {\bibfnamefont {B.}~\bibnamefont
  {Huron}}, \bibinfo {author} {\bibfnamefont {J.~P.}\ \bibnamefont {Malrieu}},
  \ and\ \bibinfo {author} {\bibfnamefont {P.}~\bibnamefont {Rancurel}},\
  }\bibfield  {title} {\enquote {\bibinfo {title} {Iterative perturbation
  calculations of ground and excited state energies from multiconfigurational
  zeroth-order wavefunctions},}\ }\href {\doibase 10.1063/1.1679199} {\bibfield
   {journal} {\bibinfo  {journal} {The Journal of Chemical Physics}\ }\textbf
  {\bibinfo {volume} {58}},\ \bibinfo {pages} {5745--5759} (\bibinfo {year}
  {1973})},\ \Eprint {http://arxiv.org/abs/https://doi.org/10.1063/1.1679199}
  {https://doi.org/10.1063/1.1679199} \BibitemShut {NoStop}%
\bibitem [{\citenamefont {Holmes}, \citenamefont {Tubman},\ and\ \citenamefont
  {Umrigar}(2016)}]{holmes2016_b}%
  \BibitemOpen
  \bibfield  {author} {\bibinfo {author} {\bibfnamefont {A.~A.}\ \bibnamefont
  {Holmes}}, \bibinfo {author} {\bibfnamefont {N.~M.}\ \bibnamefont {Tubman}},
  \ and\ \bibinfo {author} {\bibfnamefont {C.~J.}\ \bibnamefont {Umrigar}},\
  }\bibfield  {title} {\enquote {\bibinfo {title} {Heat-bath configuration
  interaction: An efficient selected configuration interaction algorithm
  inspired by heat-bath sampling},}\ }\href {\doibase 10.1021/acs.jctc.6b00407}
  {\bibfield  {journal} {\bibinfo  {journal} {Journal of Chemical Theory and
  Computation}\ }\textbf {\bibinfo {volume} {12}},\ \bibinfo {pages}
  {3674--3680} (\bibinfo {year} {2016})},\ \bibinfo {note} {pMID: 27428771},\
  \Eprint {http://arxiv.org/abs/https://doi.org/10.1021/acs.jctc.6b00407}
  {https://doi.org/10.1021/acs.jctc.6b00407} \BibitemShut {NoStop}%
\bibitem [{\citenamefont {Holmes}, \citenamefont {Changlani},\ and\
  \citenamefont {Umrigar}(2016)}]{holmes2016}%
  \BibitemOpen
  \bibfield  {author} {\bibinfo {author} {\bibfnamefont {A.~A.}\ \bibnamefont
  {Holmes}}, \bibinfo {author} {\bibfnamefont {H.~J.}\ \bibnamefont
  {Changlani}}, \ and\ \bibinfo {author} {\bibfnamefont {C.~J.}\ \bibnamefont
  {Umrigar}},\ }\bibfield  {title} {\enquote {\bibinfo {title} {Efficient
  heat-bath sampling in fock space},}\ }\href {\doibase
  10.1021/acs.jctc.5b01170} {\bibfield  {journal} {\bibinfo  {journal} {Journal
  of Chemical Theory and Computation}\ }\textbf {\bibinfo {volume} {12}},\
  \bibinfo {pages} {1561--1571} (\bibinfo {year} {2016})},\ \bibinfo {note}
  {pMID: 26959242}\BibitemShut {NoStop}%
\bibitem [{\citenamefont {Sharma}\ \emph {et~al.}(2017)\citenamefont {Sharma},
  \citenamefont {Holmes}, \citenamefont {Jeanmairet}, \citenamefont {Alavi},\
  and\ \citenamefont {Umrigar}}]{sharma2017}%
  \BibitemOpen
  \bibfield  {author} {\bibinfo {author} {\bibfnamefont {S.}~\bibnamefont
  {Sharma}}, \bibinfo {author} {\bibfnamefont {A.~A.}\ \bibnamefont {Holmes}},
  \bibinfo {author} {\bibfnamefont {G.}~\bibnamefont {Jeanmairet}}, \bibinfo
  {author} {\bibfnamefont {A.}~\bibnamefont {Alavi}}, \ and\ \bibinfo {author}
  {\bibfnamefont {C.~J.}\ \bibnamefont {Umrigar}},\ }\bibfield  {title}
  {\enquote {\bibinfo {title} {Semistochastic heat-bath configuration
  interaction method: Selected configuration interaction with semistochastic
  perturbation theory},}\ }\href@noop {} {\bibfield  {journal} {\bibinfo
  {journal} {Journal of chemical theory and computation}\ }\textbf {\bibinfo
  {volume} {13}},\ \bibinfo {pages} {1595--1604} (\bibinfo {year}
  {2017})}\BibitemShut {NoStop}%
\bibitem [{\citenamefont {Lim}\ and\ \citenamefont {Weare}(2017)}]{lim2017}%
  \BibitemOpen
  \bibfield  {author} {\bibinfo {author} {\bibfnamefont {L.-H.}\ \bibnamefont
  {Lim}}\ and\ \bibinfo {author} {\bibfnamefont {J.}~\bibnamefont {Weare}},\
  }\bibfield  {title} {\enquote {\bibinfo {title} {Fast randomized iteration:
  Diffusion monte carlo through the lens of numerical linear algebra},}\
  }\href@noop {} {\bibfield  {journal} {\bibinfo  {journal} {SIAM Review}\
  }\textbf {\bibinfo {volume} {59}},\ \bibinfo {pages} {547--587} (\bibinfo
  {year} {2017})}\BibitemShut {NoStop}%
\bibitem [{\citenamefont {Greene}\ \emph {et~al.}(2019)\citenamefont {Greene},
  \citenamefont {Webber}, \citenamefont {Weare},\ and\ \citenamefont
  {Berkelbach}}]{greene2019}%
  \BibitemOpen
  \bibfield  {author} {\bibinfo {author} {\bibfnamefont {S.~M.}\ \bibnamefont
  {Greene}}, \bibinfo {author} {\bibfnamefont {R.~J.}\ \bibnamefont {Webber}},
  \bibinfo {author} {\bibfnamefont {J.}~\bibnamefont {Weare}}, \ and\ \bibinfo
  {author} {\bibfnamefont {T.~C.}\ \bibnamefont {Berkelbach}},\ }\bibfield
  {title} {\enquote {\bibinfo {title} {Beyond walkers in stochastic quantum
  chemistry: Reducing error using fast randomized iteration},}\ }\href@noop {}
  {\bibfield  {journal} {\bibinfo  {journal} {Journal of chemical theory and
  computation}\ }\textbf {\bibinfo {volume} {15}},\ \bibinfo {pages}
  {4834--4850} (\bibinfo {year} {2019})}\BibitemShut {NoStop}%
\bibitem [{\citenamefont {Wang}, \citenamefont {Li},\ and\ \citenamefont
  {Lu}(2019)}]{wang2019}%
  \BibitemOpen
  \bibfield  {author} {\bibinfo {author} {\bibfnamefont {Z.}~\bibnamefont
  {Wang}}, \bibinfo {author} {\bibfnamefont {Y.}~\bibnamefont {Li}}, \ and\
  \bibinfo {author} {\bibfnamefont {J.}~\bibnamefont {Lu}},\ }\bibfield
  {title} {\enquote {\bibinfo {title} {Coordinate descent full configuration
  interaction},}\ }\href@noop {} {\bibfield  {journal} {\bibinfo  {journal}
  {Journal of chemical theory and computation}\ }\textbf {\bibinfo {volume}
  {15}},\ \bibinfo {pages} {3558--3569} (\bibinfo {year} {2019})}\BibitemShut
  {NoStop}%
\bibitem [{\citenamefont {Shepherd}\ \emph {et~al.}(2012)\citenamefont
  {Shepherd}, \citenamefont {Booth}, \citenamefont {Gr\"uneis},\ and\
  \citenamefont {Alavi}}]{shepherd2012}%
  \BibitemOpen
  \bibfield  {author} {\bibinfo {author} {\bibfnamefont {J.~J.}\ \bibnamefont
  {Shepherd}}, \bibinfo {author} {\bibfnamefont {G.}~\bibnamefont {Booth}},
  \bibinfo {author} {\bibfnamefont {A.}~\bibnamefont {Gr\"uneis}}, \ and\
  \bibinfo {author} {\bibfnamefont {A.}~\bibnamefont {Alavi}},\ }\bibfield
  {title} {\enquote {\bibinfo {title} {Full configuration interaction
  perspective on the homogeneous electron gas},}\ }\href {\doibase
  10.1103/PhysRevB.85.081103} {\bibfield  {journal} {\bibinfo  {journal} {Phys.
  Rev. B}\ }\textbf {\bibinfo {volume} {85}},\ \bibinfo {pages} {081103}
  (\bibinfo {year} {2012})}\BibitemShut {NoStop}%
\bibitem [{\citenamefont {Dattani}\ \emph {et~al.}(2020)\citenamefont
  {Dattani}, \citenamefont {Li~Manni}, \citenamefont {Feller},\ and\
  \citenamefont {Koput}}]{dattani2020}%
  \BibitemOpen
  \bibfield  {author} {\bibinfo {author} {\bibfnamefont {N.}~\bibnamefont
  {Dattani}}, \bibinfo {author} {\bibfnamefont {G.}~\bibnamefont {Li~Manni}},
  \bibinfo {author} {\bibfnamefont {D.}~\bibnamefont {Feller}}, \ and\ \bibinfo
  {author} {\bibfnamefont {J.}~\bibnamefont {Koput}},\ }\href@noop {} {\enquote
  {\bibinfo {title} {Computer-predicted ionization energy of carbon within
  $1\,\mathrm{cm}^{-1}$ of the best experiment},}\ } (\bibinfo {year} {2020}),\
  \bibinfo {note} {under review}\BibitemShut {NoStop}%
\bibitem [{\citenamefont {Li~Manni}\ and\ \citenamefont
  {Alavi}(2018)}]{LiManni2018}%
  \BibitemOpen
  \bibfield  {author} {\bibinfo {author} {\bibfnamefont {G.}~\bibnamefont
  {Li~Manni}}\ and\ \bibinfo {author} {\bibfnamefont {A.}~\bibnamefont
  {Alavi}},\ }\bibfield  {title} {\enquote {\bibinfo {title} {Understanding the
  mechanism stabilizing intermediate spin states in fe(ii)-porphyrin},}\ }\href
  {\doibase 10.1021/acs.jpca.7b12710} {\bibfield  {journal} {\bibinfo
  {journal} {Journal of Physical Chemistry A}\ }\textbf {\bibinfo {volume}
  {122}},\ \bibinfo {pages} {4935--4947} (\bibinfo {year} {2018})}\BibitemShut
  {NoStop}%
\bibitem [{\citenamefont {Li~Manni}\ \emph {et~al.}(2019)\citenamefont
  {Li~Manni}, \citenamefont {Kats}, \citenamefont {Tew},\ and\ \citenamefont
  {Alavi}}]{limanni2019a}%
  \BibitemOpen
  \bibfield  {author} {\bibinfo {author} {\bibfnamefont {G.}~\bibnamefont
  {Li~Manni}}, \bibinfo {author} {\bibfnamefont {D.}~\bibnamefont {Kats}},
  \bibinfo {author} {\bibfnamefont {D.~P.}\ \bibnamefont {Tew}}, \ and\
  \bibinfo {author} {\bibfnamefont {A.}~\bibnamefont {Alavi}},\ }\bibfield
  {title} {\enquote {\bibinfo {title} {Role of valence and semicore electron
  correlation on spin gaps in fe(ii)-porphyrins},}\ }\href {\doibase
  10.1021/acs.jctc.8b01277} {\bibfield  {journal} {\bibinfo  {journal} {Journal
  of Chemical Theory and Computation}\ }\textbf {\bibinfo {volume} {15}},\
  \bibinfo {pages} {1492--1497} (\bibinfo {year} {2019})}\BibitemShut {NoStop}%
\bibitem [{\citenamefont {Hubbard}(1963)}]{Hubbard1963}%
  \BibitemOpen
  \bibfield  {author} {\bibinfo {author} {\bibfnamefont {J.}~\bibnamefont
  {Hubbard}},\ }\bibfield  {title} {\enquote {\bibinfo {title} {{Electron
  Correlations in Narrow Energy Bands}},}\ }\href {\doibase
  10.1098/rspa.1963.0204} {\bibfield  {journal} {\bibinfo  {journal} {Proc.
  Royal Soc. A}\ }\textbf {\bibinfo {volume} {276}},\ \bibinfo {pages} {238}
  (\bibinfo {year} {1963})}\BibitemShut {NoStop}%
\bibitem [{\citenamefont {Shepherd}, \citenamefont {Scuseria},\ and\
  \citenamefont {Spencer}(2014)}]{Shepherd2014}%
  \BibitemOpen
  \bibfield  {author} {\bibinfo {author} {\bibfnamefont {J.~J.}\ \bibnamefont
  {Shepherd}}, \bibinfo {author} {\bibfnamefont {G.~E.}\ \bibnamefont
  {Scuseria}}, \ and\ \bibinfo {author} {\bibfnamefont {J.~S.}\ \bibnamefont
  {Spencer}},\ }\bibfield  {title} {\enquote {\bibinfo {title} {Sign problem in
  full configuration interaction quantum monte carlo: Linear and sublinear
  representation regimes for the exact wave function},}\ }\href {\doibase
  10.1103/PhysRevB.90.155130} {\bibfield  {journal} {\bibinfo  {journal} {Phys.
  Rev. B}\ }\textbf {\bibinfo {volume} {90}},\ \bibinfo {pages} {155130}
  (\bibinfo {year} {2014})}\BibitemShut {NoStop}%
\bibitem [{\citenamefont {Trivedi}\ and\ \citenamefont
  {Ceperley}(1990)}]{spectral-width}%
  \BibitemOpen
  \bibfield  {author} {\bibinfo {author} {\bibfnamefont {N.}~\bibnamefont
  {Trivedi}}\ and\ \bibinfo {author} {\bibfnamefont {D.~M.}\ \bibnamefont
  {Ceperley}},\ }\bibfield  {title} {\enquote {\bibinfo {title} {{Ground-state
  correlations of quantum antiferromagnets: A Green-function Monte Carlo
  study}},}\ }\href {\doibase 10.1103/PhysRevB.41.4552} {\bibfield  {journal}
  {\bibinfo  {journal} {Phys. Rev. B}\ }\textbf {\bibinfo {volume} {41}},\
  \bibinfo {pages} {4552} (\bibinfo {year} {1990})}\BibitemShut {NoStop}%
\bibitem [{\citenamefont {Booth}\ and\ \citenamefont {Alavi~et.
  al.}(2013)}]{neci}%
  \BibitemOpen
  \bibfield  {author} {\bibinfo {author} {\bibfnamefont {G.~H.}\ \bibnamefont
  {Booth}}\ and\ \bibinfo {author} {\bibfnamefont {A.}~\bibnamefont {Alavi~et.
  al.}},\ }\href@noop {} {\enquote {\bibinfo {title} {{Standalone \texttt{NECI}
  codebase designed for FCIQMC and other stochastic quantum chemistry
  methods.}}}\ }\bibinfo {howpublished}
  {\url{https://github.com/ghb24/NECI_STABLE}} (\bibinfo {year}
  {2013})\BibitemShut {NoStop}%
\bibitem [{\citenamefont {Clarke}, \citenamefont {Glendinning},\ and\
  \citenamefont {Hempel}(1994)}]{mpi_standard}%
  \BibitemOpen
  \bibfield  {author} {\bibinfo {author} {\bibfnamefont {L.}~\bibnamefont
  {Clarke}}, \bibinfo {author} {\bibfnamefont {I.}~\bibnamefont {Glendinning}},
  \ and\ \bibinfo {author} {\bibfnamefont {R.}~\bibnamefont {Hempel}},\
  }\bibfield  {title} {\enquote {\bibinfo {title} {The mpi message passing
  interface standard},}\ }in\ \href@noop {} {\emph {\bibinfo {booktitle}
  {Programming Environments for Massively Parallel Distributed Systems}}},\
  \bibinfo {editor} {edited by\ \bibinfo {editor} {\bibfnamefont {K.~M.}\
  \bibnamefont {Decker}}\ and\ \bibinfo {editor} {\bibfnamefont {R.~M.}\
  \bibnamefont {Rehmann}}}\ (\bibinfo  {publisher} {Birkh{\"a}user Basel},\
  \bibinfo {address} {Basel},\ \bibinfo {year} {1994})\ pp.\ \bibinfo {pages}
  {213--218}\BibitemShut {NoStop}%
\bibitem [{\citenamefont {Blackford}\ \emph {et~al.}(2002)\citenamefont
  {Blackford}, \citenamefont {Petitet}, \citenamefont {Pozo}, \citenamefont
  {Remington}, \citenamefont {Whaley}, \citenamefont {Demmel}, \citenamefont
  {Dongarra}, \citenamefont {Duff}, \citenamefont {Hammarling}, \citenamefont
  {Henry} \emph {et~al.}}]{blackford2002}%
  \BibitemOpen
  \bibfield  {author} {\bibinfo {author} {\bibfnamefont {L.~S.}\ \bibnamefont
  {Blackford}}, \bibinfo {author} {\bibfnamefont {A.}~\bibnamefont {Petitet}},
  \bibinfo {author} {\bibfnamefont {R.}~\bibnamefont {Pozo}}, \bibinfo {author}
  {\bibfnamefont {K.}~\bibnamefont {Remington}}, \bibinfo {author}
  {\bibfnamefont {R.~C.}\ \bibnamefont {Whaley}}, \bibinfo {author}
  {\bibfnamefont {J.}~\bibnamefont {Demmel}}, \bibinfo {author} {\bibfnamefont
  {J.}~\bibnamefont {Dongarra}}, \bibinfo {author} {\bibfnamefont
  {I.}~\bibnamefont {Duff}}, \bibinfo {author} {\bibfnamefont {S.}~\bibnamefont
  {Hammarling}}, \bibinfo {author} {\bibfnamefont {G.}~\bibnamefont {Henry}},
  \emph {et~al.},\ }\bibfield  {title} {\enquote {\bibinfo {title} {An updated
  set of basic linear algebra subprograms (blas)},}\ }\href@noop {} {\bibfield
  {journal} {\bibinfo  {journal} {ACM Transactions on Mathematical Software}\
  }\textbf {\bibinfo {volume} {28}},\ \bibinfo {pages} {135--151} (\bibinfo
  {year} {2002})}\BibitemShut {NoStop}%
\bibitem [{\citenamefont {Anderson}\ \emph {et~al.}(1999)\citenamefont
  {Anderson}, \citenamefont {Bai}, \citenamefont {Bischof}, \citenamefont
  {Blackford}, \citenamefont {Demmel}, \citenamefont {Dongarra}, \citenamefont
  {Du~Croz}, \citenamefont {Greenbaum}, \citenamefont {Hammarling},
  \citenamefont {McKenney},\ and\ \citenamefont {Sorensen}}]{laug}%
  \BibitemOpen
  \bibfield  {author} {\bibinfo {author} {\bibfnamefont {E.}~\bibnamefont
  {Anderson}}, \bibinfo {author} {\bibfnamefont {Z.}~\bibnamefont {Bai}},
  \bibinfo {author} {\bibfnamefont {C.}~\bibnamefont {Bischof}}, \bibinfo
  {author} {\bibfnamefont {S.}~\bibnamefont {Blackford}}, \bibinfo {author}
  {\bibfnamefont {J.}~\bibnamefont {Demmel}}, \bibinfo {author} {\bibfnamefont
  {J.}~\bibnamefont {Dongarra}}, \bibinfo {author} {\bibfnamefont
  {J.}~\bibnamefont {Du~Croz}}, \bibinfo {author} {\bibfnamefont
  {A.}~\bibnamefont {Greenbaum}}, \bibinfo {author} {\bibfnamefont
  {S.}~\bibnamefont {Hammarling}}, \bibinfo {author} {\bibfnamefont
  {A.}~\bibnamefont {McKenney}}, \ and\ \bibinfo {author} {\bibfnamefont
  {D.}~\bibnamefont {Sorensen}},\ }\href@noop {} {\emph {\bibinfo {title}
  {{LAPACK} Users' Guide}}},\ \bibinfo {edition} {3rd}\ ed.\ (\bibinfo
  {publisher} {Society for Industrial and Applied Mathematics},\ \bibinfo
  {address} {Philadelphia, PA},\ \bibinfo {year} {1999})\BibitemShut {NoStop}%
\bibitem [{\citenamefont {{The HDF Group}}(NNNN)}]{hdf5}%
  \BibitemOpen
  \bibfield  {author} {\bibinfo {author} {\bibnamefont {{The HDF Group}}},\
  }\href@noop {} {\enquote {\bibinfo {title} {{Hierarchical Data Format,
  version 5}},}\ } (\bibinfo {year} {1997-NNNN}),\ \bibinfo {note}
  {http://www.hdfgroup.org/HDF5/}\BibitemShut {NoStop}%
\bibitem [{\citenamefont {Aradi}()}]{fypp}%
  \BibitemOpen
  \bibfield  {author} {\bibinfo {author} {\bibfnamefont {B.}~\bibnamefont
  {Aradi}},\ }\href@noop {} {\enquote {\bibinfo {title} {fypp fortran
  preprocessor},}\ }\bibinfo {howpublished}
  {\url{https://github.com/aradi/fypp}}\BibitemShut {NoStop}%
\bibitem [{\citenamefont {Saito}\ and\ \citenamefont
  {Matsumoto}(2008{\natexlab{a}})}]{saito2008}%
  \BibitemOpen
  \bibfield  {author} {\bibinfo {author} {\bibfnamefont {M.}~\bibnamefont
  {Saito}}\ and\ \bibinfo {author} {\bibfnamefont {M.}~\bibnamefont
  {Matsumoto}},\ }\bibfield  {title} {\enquote {\bibinfo {title} {Simd-oriented
  fast mersenne twister: a 128-bit pseudorandom number generator},}\ }in\
  \href@noop {} {\emph {\bibinfo {booktitle} {Monte Carlo and Quasi-Monte Carlo
  Methods 2006}}},\ \bibinfo {editor} {edited by\ \bibinfo {editor}
  {\bibfnamefont {A.}~\bibnamefont {Keller}}, \bibinfo {editor} {\bibfnamefont
  {S.}~\bibnamefont {Heinrich}}, \ and\ \bibinfo {editor} {\bibfnamefont
  {H.}~\bibnamefont {Niederreiter}}}\ (\bibinfo  {publisher} {Springer Berlin
  Heidelberg},\ \bibinfo {address} {Berlin, Heidelberg},\ \bibinfo {year}
  {2008})\ pp.\ \bibinfo {pages} {607--622}\BibitemShut {NoStop}%
\bibitem [{\citenamefont {Saito}\ and\ \citenamefont
  {Matsumoto}(2008{\natexlab{b}})}]{dSFMT}%
  \BibitemOpen
  \bibfield  {author} {\bibinfo {author} {\bibfnamefont {M.}~\bibnamefont
  {Saito}}\ and\ \bibinfo {author} {\bibfnamefont {M.}~\bibnamefont
  {Matsumoto}},\ }\href@noop {} {\enquote {\bibinfo {title} {double precision
  simd oriented fast mersenne twister},}\ }\bibinfo {howpublished}
  {\url{https://github.com/MersenneTwister-Lab/dSFMT}} (\bibinfo {year}
  {2008}{\natexlab{b}})\BibitemShut {NoStop}%
\bibitem [{\citenamefont {Matsumoto}\ and\ \citenamefont
  {Nishimura}(1998)}]{matsumoto1998}%
  \BibitemOpen
  \bibfield  {author} {\bibinfo {author} {\bibfnamefont {M.}~\bibnamefont
  {Matsumoto}}\ and\ \bibinfo {author} {\bibfnamefont {T.}~\bibnamefont
  {Nishimura}},\ }\bibfield  {title} {\enquote {\bibinfo {title} {Mersenne
  twister: a 623-dimensionally equidistributed uniform pseudo-random number
  generator},}\ }\href@noop {} {\bibfield  {journal} {\bibinfo  {journal} {ACM
  Transactions on Modeling and Computer Simulation (TOMACS)}\ }\textbf
  {\bibinfo {volume} {8}},\ \bibinfo {pages} {3--30} (\bibinfo {year}
  {1998})}\BibitemShut {NoStop}%
\bibitem [{\citenamefont {Smart}(2014)}]{smart2014}%
  \BibitemOpen
  \bibfield  {author} {\bibinfo {author} {\bibfnamefont {S.~D.}\ \bibnamefont
  {Smart}},\ }\emph {\bibinfo {title} {The use of spin-pure and non-orthogonal
  Hilbert spaces in full configuration interaction quantum monte-carlo}},\
  \href@noop {} {Ph.D. thesis},\ \bibinfo  {school} {University of Cambridge}
  (\bibinfo {year} {2014})\BibitemShut {NoStop}%
\bibitem [{\citenamefont {Smart}, \citenamefont {Booth},\ and\ \citenamefont
  {Alavi}()}]{cs_excit}%
  \BibitemOpen
  \bibfield  {author} {\bibinfo {author} {\bibfnamefont {S.}~\bibnamefont
  {Smart}}, \bibinfo {author} {\bibfnamefont {G.}~\bibnamefont {Booth}}, \ and\
  \bibinfo {author} {\bibfnamefont {A.}~\bibnamefont {Alavi}},\ }\href@noop {}
  {\enquote {\bibinfo {title} {Excitation generation in full configuration
  interaction quantum monte carlo based on cauchy-schwarz distributions},}\
  }\BibitemShut {NoStop}%
\bibitem [{\citenamefont {Neufeld}\ and\ \citenamefont
  {Thom}(2019)}]{neufeld2019}%
  \BibitemOpen
  \bibfield  {author} {\bibinfo {author} {\bibfnamefont {V.~A.}\ \bibnamefont
  {Neufeld}}\ and\ \bibinfo {author} {\bibfnamefont {A.~J.~W.}\ \bibnamefont
  {Thom}},\ }\bibfield  {title} {\enquote {\bibinfo {title} {Exciting
  determinants in quantum monte carlo: Loading the dice with fast, low-memory
  weights},}\ }\href {\doibase 10.1021/acs.jctc.8b00844} {\bibfield  {journal}
  {\bibinfo  {journal} {Journal of Chemical Theory and Computation}\ }\textbf
  {\bibinfo {volume} {15}},\ \bibinfo {pages} {127--140} (\bibinfo {year}
  {2019})}\BibitemShut {NoStop}%
\bibitem [{\citenamefont {Li}\ \emph {et~al.}(2018)\citenamefont {Li},
  \citenamefont {Otten}, \citenamefont {Holmes}, \citenamefont {Sharma},\ and\
  \citenamefont {Umrigar}}]{Li2018}%
  \BibitemOpen
  \bibfield  {author} {\bibinfo {author} {\bibfnamefont {J.}~\bibnamefont
  {Li}}, \bibinfo {author} {\bibfnamefont {M.}~\bibnamefont {Otten}}, \bibinfo
  {author} {\bibfnamefont {A.~A.}\ \bibnamefont {Holmes}}, \bibinfo {author}
  {\bibfnamefont {S.}~\bibnamefont {Sharma}}, \ and\ \bibinfo {author}
  {\bibfnamefont {C.~J.}\ \bibnamefont {Umrigar}},\ }\bibfield  {title}
  {\enquote {\bibinfo {title} {Fast semistochastic heat-bath configuration
  interaction},}\ }\href@noop {} {\bibfield  {journal} {\bibinfo  {journal} {J.
  Comp. Phys.}\ }\textbf {\bibinfo {volume} {149}},\ \bibinfo {pages} {214110}
  (\bibinfo {year} {2018})}\BibitemShut {NoStop}%
\bibitem [{\citenamefont {Lowdin}(1951)}]{Lowdin1951}%
  \BibitemOpen
  \bibfield  {author} {\bibinfo {author} {\bibfnamefont {P.}~\bibnamefont
  {Lowdin}},\ }\bibfield  {title} {\enquote {\bibinfo {title} {A note on the
  quantum-mechanical perturbation theory},}\ }\href@noop {} {\bibfield
  {journal} {\bibinfo  {journal} {J Chem Phys}\ }\textbf {\bibinfo {volume}
  {19}},\ \bibinfo {pages} {1396--1401} (\bibinfo {year} {1951})}\BibitemShut
  {NoStop}%
\bibitem [{\citenamefont {Olivares-Amaya}\ \emph {et~al.}(2015)\citenamefont
  {Olivares-Amaya}, \citenamefont {Hu}, \citenamefont {Nakatani}, \citenamefont
  {Sharma}, \citenamefont {Yang},\ and\ \citenamefont {Chan}}]{Olivares2015}%
  \BibitemOpen
  \bibfield  {author} {\bibinfo {author} {\bibfnamefont {R.}~\bibnamefont
  {Olivares-Amaya}}, \bibinfo {author} {\bibfnamefont {W.}~\bibnamefont {Hu}},
  \bibinfo {author} {\bibfnamefont {N.}~\bibnamefont {Nakatani}}, \bibinfo
  {author} {\bibfnamefont {S.}~\bibnamefont {Sharma}}, \bibinfo {author}
  {\bibfnamefont {J.}~\bibnamefont {Yang}}, \ and\ \bibinfo {author}
  {\bibfnamefont {G.~K.-L.}\ \bibnamefont {Chan}},\ }\bibfield  {title}
  {\enquote {\bibinfo {title} {The ab-initio density matrix renormalization
  group in practice},}\ }\href {\doibase 10.1063/1.4905329} {\bibfield
  {journal} {\bibinfo  {journal} {The Journal of Chemical Physics}\ }\textbf
  {\bibinfo {volume} {142}},\ \bibinfo {pages} {034102} (\bibinfo {year}
  {2015})},\ \Eprint {http://arxiv.org/abs/https://doi.org/10.1063/1.4905329}
  {https://doi.org/10.1063/1.4905329} \BibitemShut {NoStop}%
\bibitem [{\citenamefont {Chien}\ \emph {et~al.}(2018)\citenamefont {Chien},
  \citenamefont {Holmes}, \citenamefont {Otten}, \citenamefont {Umrigar},
  \citenamefont {Sharma},\ and\ \citenamefont {Zimmerman}}]{Chien2018}%
  \BibitemOpen
  \bibfield  {author} {\bibinfo {author} {\bibfnamefont {A.~D.}\ \bibnamefont
  {Chien}}, \bibinfo {author} {\bibfnamefont {A.~A.}\ \bibnamefont {Holmes}},
  \bibinfo {author} {\bibfnamefont {M.}~\bibnamefont {Otten}}, \bibinfo
  {author} {\bibfnamefont {C.~J.}\ \bibnamefont {Umrigar}}, \bibinfo {author}
  {\bibfnamefont {S.}~\bibnamefont {Sharma}}, \ and\ \bibinfo {author}
  {\bibfnamefont {P.~M.}\ \bibnamefont {Zimmerman}},\ }\bibfield  {title}
  {\enquote {\bibinfo {title} {Excited states of methylene, polyenes, and ozone
  from heat-bath configuration interaction},}\ }\href {\doibase
  10.1021/acs.jpca.8b01554} {\bibfield  {journal} {\bibinfo  {journal} {The
  Journal of Physical Chemistry A}\ }\textbf {\bibinfo {volume} {122}},\
  \bibinfo {pages} {2714--2722} (\bibinfo {year} {2018})},\ \bibinfo {note}
  {pMID: 29473750},\ \Eprint
  {http://arxiv.org/abs/https://doi.org/10.1021/acs.jpca.8b01554}
  {https://doi.org/10.1021/acs.jpca.8b01554} \BibitemShut {NoStop}%
\bibitem [{\citenamefont {Blunt}(2018)}]{Blunt2018_2}%
  \BibitemOpen
  \bibfield  {author} {\bibinfo {author} {\bibfnamefont {N.~S.}\ \bibnamefont
  {Blunt}},\ }\bibfield  {title} {\enquote {\bibinfo {title} {Communication: An
  efficient and accurate perturbative correction to initiator full
  configuration interaction quantum monte carlo},}\ }\href@noop {} {\bibfield
  {journal} {\bibinfo  {journal} {The Journal of Chemical Physics}\ }\textbf
  {\bibinfo {volume} {148}},\ \bibinfo {pages} {221101} (\bibinfo {year}
  {2018})}\BibitemShut {NoStop}%
\bibitem [{\citenamefont {Spencer}, \citenamefont {Blunt},\ and\ \citenamefont
  {Foulkes}(2012)}]{Spencer2012}%
  \BibitemOpen
  \bibfield  {author} {\bibinfo {author} {\bibfnamefont {J.~S.}\ \bibnamefont
  {Spencer}}, \bibinfo {author} {\bibfnamefont {N.~S.}\ \bibnamefont {Blunt}},
  \ and\ \bibinfo {author} {\bibfnamefont {W.~M.~C.}\ \bibnamefont {Foulkes}},\
  }\bibfield  {title} {\enquote {\bibinfo {title} {{The sign problem and
  population dynamics in the full configuration interaction quantum Monte Carlo
  method}},}\ }\href@noop {} {\bibfield  {journal} {\bibinfo  {journal} {The
  Journal of Chemical Physics}\ }\textbf {\bibinfo {volume} {136}},\ \bibinfo
  {pages} {054110} (\bibinfo {year} {2012})}\BibitemShut {NoStop}%
\bibitem [{\citenamefont {Blunt}, \citenamefont {Thom},\ and\ \citenamefont
  {Scott}(2019)}]{Blunt2019}%
  \BibitemOpen
  \bibfield  {author} {\bibinfo {author} {\bibfnamefont {N.~S.}\ \bibnamefont
  {Blunt}}, \bibinfo {author} {\bibfnamefont {A.~J.~W.}\ \bibnamefont {Thom}},
  \ and\ \bibinfo {author} {\bibfnamefont {C.~J.~C.}\ \bibnamefont {Scott}},\
  }\bibfield  {title} {\enquote {\bibinfo {title} {Preconditioning and
  perturbative estimators in full configuration interaction quantum monte
  carlo},}\ }\href@noop {} {\bibfield  {journal} {\bibinfo  {journal} {Journal
  of Chemical Theory and Computation}\ }\textbf {\bibinfo {volume} {15}},\
  \bibinfo {pages} {3537} (\bibinfo {year} {2019})}\BibitemShut {NoStop}%
\bibitem [{\citenamefont {Blunt}(2019)}]{Blunt2019_2}%
  \BibitemOpen
  \bibfield  {author} {\bibinfo {author} {\bibfnamefont {N.~S.}\ \bibnamefont
  {Blunt}},\ }\bibfield  {title} {\enquote {\bibinfo {title} {A hybrid approach
  to extending selected configuration interaction and full configuration
  interaction quantum monte carlo},}\ }\href@noop {} {\bibfield  {journal}
  {\bibinfo  {journal} {The Journal of Chemical Physics}\ }\textbf {\bibinfo
  {volume} {151}},\ \bibinfo {pages} {174103} (\bibinfo {year}
  {2019})}\BibitemShut {NoStop}%
\bibitem [{\citenamefont {Thomas}\ \emph
  {et~al.}(2015{\natexlab{b}})\citenamefont {Thomas}, \citenamefont {Opalka},
  \citenamefont {Overy}, \citenamefont {Knowles}, \citenamefont {Alavi},\ and\
  \citenamefont {Booth}}]{Thomas2015}%
  \BibitemOpen
  \bibfield  {author} {\bibinfo {author} {\bibfnamefont {R.~E.}\ \bibnamefont
  {Thomas}}, \bibinfo {author} {\bibfnamefont {D.}~\bibnamefont {Opalka}},
  \bibinfo {author} {\bibfnamefont {C.}~\bibnamefont {Overy}}, \bibinfo
  {author} {\bibfnamefont {P.~J.}\ \bibnamefont {Knowles}}, \bibinfo {author}
  {\bibfnamefont {A.}~\bibnamefont {Alavi}}, \ and\ \bibinfo {author}
  {\bibfnamefont {G.~H.}\ \bibnamefont {Booth}},\ }\bibfield  {title} {\enquote
  {\bibinfo {title} {Analytic nuclear forces and molecular properties from full
  configuration interaction quantum monte carlo},}\ }\href {\doibase
  10.1063/1.4927594} {\bibfield  {journal} {\bibinfo  {journal} {The Journal of
  Chemical Physics}\ }\textbf {\bibinfo {volume} {143}},\ \bibinfo {pages}
  {054108} (\bibinfo {year} {2015}{\natexlab{b}})}\BibitemShut {NoStop}%
\bibitem [{\citenamefont {Booth}\ and\ \citenamefont
  {Chan}(2012)}]{BoothSpectra}%
  \BibitemOpen
  \bibfield  {author} {\bibinfo {author} {\bibfnamefont {G.~H.}\ \bibnamefont
  {Booth}}\ and\ \bibinfo {author} {\bibfnamefont {G.~K.-L.}\ \bibnamefont
  {Chan}},\ }\bibfield  {title} {\enquote {\bibinfo {title} {Communbication:
  Excited states, dynamic correlation functions and spectral properties from
  full configuration interaction quantum monte carlo},}\ }\href {\doibase
  10.1063/1.4766327} {\bibfield  {journal} {\bibinfo  {journal} {The Journal of
  Chemical Physics}\ }\textbf {\bibinfo {volume} {137}},\ \bibinfo {pages}
  {191102} (\bibinfo {year} {2012})}\BibitemShut {NoStop}%
\bibitem [{\citenamefont {Samanta}, \citenamefont {Blunt},\ and\ \citenamefont
  {Booth}(2018)}]{Samanta2018}%
  \BibitemOpen
  \bibfield  {author} {\bibinfo {author} {\bibfnamefont {P.~K.}\ \bibnamefont
  {Samanta}}, \bibinfo {author} {\bibfnamefont {N.~S.}\ \bibnamefont {Blunt}},
  \ and\ \bibinfo {author} {\bibfnamefont {G.~H.}\ \bibnamefont {Booth}},\
  }\bibfield  {title} {\enquote {\bibinfo {title} {Response formalism within
  full configuration interaction quantum monte carlo: Static properties and
  electrical response},}\ }\href {\doibase 10.1021/acs.jctc.8b00454} {\bibfield
   {journal} {\bibinfo  {journal} {Journal of Chemical Theory and Computation}\
  }\textbf {\bibinfo {volume} {14}},\ \bibinfo {pages} {3532--3546} (\bibinfo
  {year} {2018})}\BibitemShut {NoStop}%
\bibitem [{\citenamefont {Blunt}, \citenamefont {Alavi},\ and\ \citenamefont
  {Booth}(2015)}]{Blunt2015_2}%
  \BibitemOpen
  \bibfield  {author} {\bibinfo {author} {\bibfnamefont {N.~S.}\ \bibnamefont
  {Blunt}}, \bibinfo {author} {\bibfnamefont {A.}~\bibnamefont {Alavi}}, \ and\
  \bibinfo {author} {\bibfnamefont {G.~H.}\ \bibnamefont {Booth}},\ }\bibfield
  {title} {\enquote {\bibinfo {title} {Krylov-projected quantum monte carlo
  method},}\ }\href {\doibase 10.1103/PhysRevLett.115.050603} {\bibfield
  {journal} {\bibinfo  {journal} {Phys. Rev. Lett.}\ }\textbf {\bibinfo
  {volume} {115}},\ \bibinfo {pages} {050603} (\bibinfo {year}
  {2015})}\BibitemShut {NoStop}%
\bibitem [{\citenamefont {Blunt}, \citenamefont {Alavi},\ and\ \citenamefont
  {Booth}(2018)}]{Blunt2018}%
  \BibitemOpen
  \bibfield  {author} {\bibinfo {author} {\bibfnamefont {N.~S.}\ \bibnamefont
  {Blunt}}, \bibinfo {author} {\bibfnamefont {A.}~\bibnamefont {Alavi}}, \ and\
  \bibinfo {author} {\bibfnamefont {G.~H.}\ \bibnamefont {Booth}},\ }\bibfield
  {title} {\enquote {\bibinfo {title} {Nonlinear biases, stochastically sampled
  effective hamiltonians, and spectral functions in quantum monte carlo
  methods},}\ }\href {\doibase 10.1103/PhysRevB.98.085118} {\bibfield
  {journal} {\bibinfo  {journal} {Phys. Rev. B}\ }\textbf {\bibinfo {volume}
  {98}},\ \bibinfo {pages} {085118} (\bibinfo {year} {2018})}\BibitemShut
  {NoStop}%
\bibitem [{\citenamefont {Booth}\ \emph {et~al.}(2012)\citenamefont {Booth},
  \citenamefont {Cleland}, \citenamefont {Alavi},\ and\ \citenamefont
  {Tew}}]{BoothF12}%
  \BibitemOpen
  \bibfield  {author} {\bibinfo {author} {\bibfnamefont {G.~H.}\ \bibnamefont
  {Booth}}, \bibinfo {author} {\bibfnamefont {D.}~\bibnamefont {Cleland}},
  \bibinfo {author} {\bibfnamefont {A.}~\bibnamefont {Alavi}}, \ and\ \bibinfo
  {author} {\bibfnamefont {D.~P.}\ \bibnamefont {Tew}},\ }\bibfield  {title}
  {\enquote {\bibinfo {title} {An explicitly correlated approach to basis set
  incompleteness in full configuration interaction quantum monte carlo},}\
  }\href {\doibase 10.1063/1.4762445} {\bibfield  {journal} {\bibinfo
  {journal} {The Journal of Chemical Physics}\ }\textbf {\bibinfo {volume}
  {137}},\ \bibinfo {pages} {164112} (\bibinfo {year} {2012})}\BibitemShut
  {NoStop}%
\bibitem [{\citenamefont {Gr{\"u}neis}\ \emph {et~al.}(2013)\citenamefont
  {Gr{\"u}neis}, \citenamefont {Shepherd}, \citenamefont {Alavi}, \citenamefont
  {Tew},\ and\ \citenamefont {Booth}}]{Gruneis13}%
  \BibitemOpen
  \bibfield  {author} {\bibinfo {author} {\bibfnamefont {A.}~\bibnamefont
  {Gr{\"u}neis}}, \bibinfo {author} {\bibfnamefont {J.~J.}\ \bibnamefont
  {Shepherd}}, \bibinfo {author} {\bibfnamefont {A.}~\bibnamefont {Alavi}},
  \bibinfo {author} {\bibfnamefont {D.~P.}\ \bibnamefont {Tew}}, \ and\
  \bibinfo {author} {\bibfnamefont {G.~H.}\ \bibnamefont {Booth}},\ }\bibfield
  {title} {\enquote {\bibinfo {title} {Explicitly correlated plane waves:
  Accelerating convergence in periodic wavefunction expansions},}\ }\href
  {\doibase 10.1063/1.4818753} {\bibfield  {journal} {\bibinfo  {journal} {The
  Journal of Chemical Physics}\ }\textbf {\bibinfo {volume} {139}},\ \bibinfo
  {pages} {084112} (\bibinfo {year} {2013})}\BibitemShut {NoStop}%
\bibitem [{\citenamefont {Kersten}, \citenamefont {Booth},\ and\ \citenamefont
  {Alavi}(2016)}]{Kersten2016}%
  \BibitemOpen
  \bibfield  {author} {\bibinfo {author} {\bibfnamefont {J.~A.~F.}\
  \bibnamefont {Kersten}}, \bibinfo {author} {\bibfnamefont {G.~H.}\
  \bibnamefont {Booth}}, \ and\ \bibinfo {author} {\bibfnamefont
  {A.}~\bibnamefont {Alavi}},\ }\bibfield  {title} {\enquote {\bibinfo {title}
  {Assessment of multireference approaches to explicitly correlated full
  configuration interaction quantum monte carlo},}\ }\href {\doibase
  10.1063/1.4959245} {\bibfield  {journal} {\bibinfo  {journal} {The Journal of
  Chemical Physics}\ }\textbf {\bibinfo {volume} {145}},\ \bibinfo {pages}
  {054117} (\bibinfo {year} {2016})},\ \Eprint
  {http://arxiv.org/abs/https://doi.org/10.1063/1.4959245}
  {https://doi.org/10.1063/1.4959245} \BibitemShut {NoStop}%
\bibitem [{\citenamefont {Fertitta}\ and\ \citenamefont
  {Booth}(2018)}]{Fertitta2018}%
  \BibitemOpen
  \bibfield  {author} {\bibinfo {author} {\bibfnamefont {E.}~\bibnamefont
  {Fertitta}}\ and\ \bibinfo {author} {\bibfnamefont {G.~H.}\ \bibnamefont
  {Booth}},\ }\bibfield  {title} {\enquote {\bibinfo {title} {Rigorous wave
  function embedding with dynamical fluctuations},}\ }\href {\doibase
  10.1103/PhysRevB.98.235132} {\bibfield  {journal} {\bibinfo  {journal} {Phys.
  Rev. B}\ }\textbf {\bibinfo {volume} {98}},\ \bibinfo {pages} {235132}
  (\bibinfo {year} {2018})}\BibitemShut {NoStop}%
\bibitem [{\citenamefont {Fertitta}\ and\ \citenamefont
  {Booth}(2019)}]{Fertitta2019}%
  \BibitemOpen
  \bibfield  {author} {\bibinfo {author} {\bibfnamefont {E.}~\bibnamefont
  {Fertitta}}\ and\ \bibinfo {author} {\bibfnamefont {G.~H.}\ \bibnamefont
  {Booth}},\ }\bibfield  {title} {\enquote {\bibinfo {title} {Energy-weighted
  density matrix embedding of open correlated chemical fragments},}\ }\href
  {\doibase 10.1063/1.5100290} {\bibfield  {journal} {\bibinfo  {journal} {The
  Journal of Chemical Physics}\ }\textbf {\bibinfo {volume} {151}},\ \bibinfo
  {pages} {014115} (\bibinfo {year} {2019})}\BibitemShut {NoStop}%
\bibitem [{\citenamefont {Li~Manni}\ \emph {et~al.}(2014)\citenamefont
  {Li~Manni}, \citenamefont {Carlson}, \citenamefont {Luo}, \citenamefont {Ma},
  \citenamefont {Olsen}, \citenamefont {Truhlar},\ and\ \citenamefont
  {Gagliardi}}]{limanni2014a}%
  \BibitemOpen
  \bibfield  {author} {\bibinfo {author} {\bibfnamefont {G.}~\bibnamefont
  {Li~Manni}}, \bibinfo {author} {\bibfnamefont {R.~K.}\ \bibnamefont
  {Carlson}}, \bibinfo {author} {\bibfnamefont {S.}~\bibnamefont {Luo}},
  \bibinfo {author} {\bibfnamefont {D.}~\bibnamefont {Ma}}, \bibinfo {author}
  {\bibfnamefont {J.}~\bibnamefont {Olsen}}, \bibinfo {author} {\bibfnamefont
  {D.~G.}\ \bibnamefont {Truhlar}}, \ and\ \bibinfo {author} {\bibfnamefont
  {L.}~\bibnamefont {Gagliardi}},\ }\bibfield  {title} {\enquote {\bibinfo
  {title} {Multiconfiguration pair-density functional theory},}\ }\href
  {\doibase 10.1021/ct500483t} {\bibfield  {journal} {\bibinfo  {journal}
  {Journal of Chemical Theory and Computation}\ }\textbf {\bibinfo {volume}
  {10}},\ \bibinfo {pages} {3669--3680} (\bibinfo {year} {2014})},\ \bibinfo
  {note} {pMID: 26588512}\BibitemShut {NoStop}%
\bibitem [{\citenamefont {Monkhorst}(1977)}]{Monkhorst1977}%
  \BibitemOpen
  \bibfield  {author} {\bibinfo {author} {\bibfnamefont {H.~J.}\ \bibnamefont
  {Monkhorst}},\ }\bibfield  {title} {\enquote {\bibinfo {title} {Calculation
  of properties with the coupled-cluster method},}\ }\href@noop {} {\bibfield
  {journal} {\bibinfo  {journal} {Int. J. Quantum Chem.}\ }\textbf {\bibinfo
  {volume} {S11}},\ \bibinfo {pages} {421--432} (\bibinfo {year}
  {1977})}\BibitemShut {NoStop}%
\bibitem [{\citenamefont {Dalgaard}\ and\ \citenamefont
  {Monkhorst}(1983)}]{Dalgaard1983}%
  \BibitemOpen
  \bibfield  {author} {\bibinfo {author} {\bibfnamefont {E.}~\bibnamefont
  {Dalgaard}}\ and\ \bibinfo {author} {\bibfnamefont {H.}~\bibnamefont
  {Monkhorst}},\ }\bibfield  {title} {\enquote {\bibinfo {title} {{Some aspects
  of the time-dependent coupled-cluster approach to dynamic response
  functions}},}\ }\href@noop {} {\bibfield  {journal} {\bibinfo  {journal}
  {Phys. Rev. A}\ }\textbf {\bibinfo {volume} {28}} (\bibinfo {year}
  {1983})}\BibitemShut {NoStop}%
\bibitem [{\citenamefont {Christiansen}, \citenamefont {J{\o}rgensen},\ and\
  \citenamefont {H{\"a}ttig}(1998)}]{Christiansen:1998p36}%
  \BibitemOpen
  \bibfield  {author} {\bibinfo {author} {\bibfnamefont {O.}~\bibnamefont
  {Christiansen}}, \bibinfo {author} {\bibfnamefont {P.}~\bibnamefont
  {J{\o}rgensen}}, \ and\ \bibinfo {author} {\bibfnamefont {C.}~\bibnamefont
  {H{\"a}ttig}},\ }\bibfield  {title} {\enquote {\bibinfo {title} {Response
  functions from fourier component variational perturbation theory applied to a
  time-averaged quasienergy},}\ }\href@noop {} {\bibfield  {journal} {\bibinfo
  {journal} {Int. J. Quantum Chem.}\ }\textbf {\bibinfo {volume} {68}},\
  \bibinfo {pages} {1--52} (\bibinfo {year} {1998})}\BibitemShut {NoStop}%
\bibitem [{\citenamefont {Helgaker}\ \emph {et~al.}(2012)\citenamefont
  {Helgaker}, \citenamefont {Coriani}, \citenamefont {J{\o}rgensen},
  \citenamefont {Kristensen}, \citenamefont {Olsen},\ and\ \citenamefont
  {Ruud}}]{Helgaker2012}%
  \BibitemOpen
  \bibfield  {author} {\bibinfo {author} {\bibfnamefont {T.}~\bibnamefont
  {Helgaker}}, \bibinfo {author} {\bibfnamefont {S.}~\bibnamefont {Coriani}},
  \bibinfo {author} {\bibfnamefont {P.}~\bibnamefont {J{\o}rgensen}}, \bibinfo
  {author} {\bibfnamefont {K.}~\bibnamefont {Kristensen}}, \bibinfo {author}
  {\bibfnamefont {J.}~\bibnamefont {Olsen}}, \ and\ \bibinfo {author}
  {\bibfnamefont {K.}~\bibnamefont {Ruud}},\ }\bibfield  {title} {\enquote
  {\bibinfo {title} {{Recent advances in wave function-based methods of
  molecular-property calculations.}}}\ }\href@noop {} {\bibfield  {journal}
  {\bibinfo  {journal} {Chem. Rev.}\ }\textbf {\bibinfo {volume} {112}},\
  \bibinfo {pages} {543--631} (\bibinfo {year} {2012})}\BibitemShut {NoStop}%
\bibitem [{\citenamefont {Silver}, \citenamefont {Sivia},\ and\ \citenamefont
  {Gubernatis}(1990)}]{silver1990}%
  \BibitemOpen
  \bibfield  {author} {\bibinfo {author} {\bibfnamefont {R.~N.}\ \bibnamefont
  {Silver}}, \bibinfo {author} {\bibfnamefont {D.~S.}\ \bibnamefont {Sivia}}, \
  and\ \bibinfo {author} {\bibfnamefont {J.~E.}\ \bibnamefont {Gubernatis}},\
  }\bibfield  {title} {\enquote {\bibinfo {title} {Maximum-entropy method for
  analytic continuation of quantum monte carlo data},}\ }\href {\doibase
  10.1103/PhysRevB.41.2380} {\bibfield  {journal} {\bibinfo  {journal} {Phys.
  Rev. B}\ }\textbf {\bibinfo {volume} {41}},\ \bibinfo {pages} {2380--2389}
  (\bibinfo {year} {1990})}\BibitemShut {NoStop}%
\bibitem [{\citenamefont {Jarrell}\ and\ \citenamefont
  {Gubernatis}(1996)}]{jarell1996}%
  \BibitemOpen
  \bibfield  {author} {\bibinfo {author} {\bibfnamefont {M.}~\bibnamefont
  {Jarrell}}\ and\ \bibinfo {author} {\bibfnamefont {J.}~\bibnamefont
  {Gubernatis}},\ }\href@noop {} {\bibfield  {journal} {\bibinfo  {journal}
  {Phys. Rep.}\ }\textbf {\bibinfo {volume} {269}},\ \bibinfo {pages} {133}
  (\bibinfo {year} {1996})}\BibitemShut {NoStop}%
\bibitem [{exc()}]{excited_state_output}%
  \BibitemOpen
  \href@noop {} {}\bibinfo {note} {See output files {\tt
  output\_file\_excited\_state\_be2\_b1g.txt} and {\tt
  stats\_file\_excited\_state\_be2\_b1g.txt} for the excited state calculation
  and the files {\tt output\_file\_real\_time\_be2\_b1g.txt}, and {\tt
  fft\_spectrum\_be2\_b1g.txt} for the real-time calculation in the
  supplemental material~\cite{supplement}}\BibitemShut {NoStop}%
\bibitem [{\citenamefont {Kramida}\ and\ \citenamefont
  {Martin}(1997)}]{kramida1997}%
  \BibitemOpen
  \bibfield  {author} {\bibinfo {author} {\bibfnamefont {A.}~\bibnamefont
  {Kramida}}\ and\ \bibinfo {author} {\bibfnamefont {W.~C.}\ \bibnamefont
  {Martin}},\ }\bibfield  {title} {\enquote {\bibinfo {title} {A compilation of
  energy levels and wavelengths for the spectrum of neutral beryllium (be
  i)},}\ }\href {\doibase 10.1063/1.555999} {\bibfield  {journal} {\bibinfo
  {journal} {Journal of Physical and Chemical Reference Data}\ }\textbf
  {\bibinfo {volume} {26}},\ \bibinfo {pages} {1185--1194} (\bibinfo {year}
  {1997})},\ \Eprint {http://arxiv.org/abs/https://doi.org/10.1063/1.555999}
  {https://doi.org/10.1063/1.555999} \BibitemShut {NoStop}%
\bibitem [{\citenamefont {Kato}(1957)}]{Kato57}%
  \BibitemOpen
  \bibfield  {author} {\bibinfo {author} {\bibfnamefont {T.}~\bibnamefont
  {Kato}},\ }\bibfield  {title} {\enquote {\bibinfo {title} {{On the
  Eigenfunctions of Many-Particle Systems in Quantum Mechanics}},}\ }\href@noop
  {} {\bibfield  {journal} {\bibinfo  {journal} {Commun. Pure Appl. Math.}\
  }\textbf {\bibinfo {volume} {{10}}},\ \bibinfo {pages} {{151--177}} (\bibinfo
  {year} {1957})}\BibitemShut {NoStop}%
\bibitem [{\citenamefont {Jastrow}(1955)}]{jastrow55}%
  \BibitemOpen
  \bibfield  {author} {\bibinfo {author} {\bibfnamefont {R.}~\bibnamefont
  {Jastrow}},\ }\bibfield  {title} {\enquote {\bibinfo {title} {{Many-body
  problems with strong forces}},}\ }\href@noop {} {\bibfield  {journal}
  {\bibinfo  {journal} {Phys. Rev.}\ }\textbf {\bibinfo {volume} {{98}}},\
  \bibinfo {pages} {1479--1484} (\bibinfo {year} {1955})}\BibitemShut {NoStop}%
\bibitem [{\citenamefont {Fournais}\ \emph {et~al.}(2009)\citenamefont
  {Fournais}, \citenamefont {Hoffmann-Ostenhof}, \citenamefont
  {Hoffmann-Ostenhof},\ and\ \citenamefont {{{\O}stergaard
  S{\o}rensen}}}]{FHHO09}%
  \BibitemOpen
  \bibfield  {author} {\bibinfo {author} {\bibfnamefont {S.}~\bibnamefont
  {Fournais}}, \bibinfo {author} {\bibfnamefont {M.}~\bibnamefont
  {Hoffmann-Ostenhof}}, \bibinfo {author} {\bibfnamefont {T.}~\bibnamefont
  {Hoffmann-Ostenhof}}, \ and\ \bibinfo {author} {\bibfnamefont
  {T.}~\bibnamefont {{{\O}stergaard S{\o}rensen}}},\ }\bibfield  {title}
  {\enquote {\bibinfo {title} {Analytic structure of many-body coulombic wave
  functions},}\ }\href@noop {} {\bibfield  {journal} {\bibinfo  {journal}
  {Commun. Math. Phys.}\ }\textbf {\bibinfo {volume} {289}},\ \bibinfo {pages}
  {291--310} (\bibinfo {year} {2009})}\BibitemShut {NoStop}%
\bibitem [{\citenamefont {Boys}\ and\ \citenamefont {Handy}(1969)}]{Boys69}%
  \BibitemOpen
  \bibfield  {author} {\bibinfo {author} {\bibfnamefont {S.~F.}\ \bibnamefont
  {Boys}}\ and\ \bibinfo {author} {\bibfnamefont {N.~C.}\ \bibnamefont
  {Handy}},\ }\bibfield  {title} {\enquote {\bibinfo {title} {The determination
  of energies and wavefunctions with full electronic correlation},}\
  }\href@noop {} {\bibfield  {journal} {\bibinfo  {journal} {Proc. Roy. Soc.}\
  }\textbf {\bibinfo {volume} {{A 310}}},\ \bibinfo {pages} {43--61} (\bibinfo
  {year} {1969})}\BibitemShut {NoStop}%
\bibitem [{\citenamefont {{K. E. Schmidt and J. W.
  Moskowitz}}(1990)}]{Schmidt90}%
  \BibitemOpen
  \bibfield  {author} {\bibinfo {author} {\bibnamefont {{K. E. Schmidt and J.
  W. Moskowitz}}},\ }\href@noop {} {\bibfield  {journal} {\bibinfo  {journal}
  {J. Chem. Phys.}\ }\textbf {\bibinfo {volume} {93}},\ \bibinfo {pages} {4172}
  (\bibinfo {year} {1990})}\BibitemShut {NoStop}%
\bibitem [{\citenamefont {Aquilante}\ \emph {et~al.}(2016)\citenamefont
  {Aquilante}, \citenamefont {Autschbach}, \citenamefont {Carlson},
  \citenamefont {Chibotaru}, \citenamefont {Delcey}, \citenamefont {De~Vico},
  \citenamefont {Fdez.~Galv{\'a}n}, \citenamefont {Ferr{\'e}}, \citenamefont
  {Frutos}, \citenamefont {Gagliardi}, \citenamefont {Garavelli}, \citenamefont
  {Giussani}, \citenamefont {Hoyer}, \citenamefont {Li~Manni}, \citenamefont
  {Lischka}, \citenamefont {Ma}, \citenamefont {Malmqvist}, \citenamefont
  {M{\"u}ller}, \citenamefont {Nenov}, \citenamefont {Olivucci}, \citenamefont
  {Pedersen}, \citenamefont {Peng}, \citenamefont {Plasser}, \citenamefont
  {Pritchard}, \citenamefont {Reiher}, \citenamefont {Rivalta}, \citenamefont
  {Schapiro}, \citenamefont {Segarra-Mart{\`i}}, \citenamefont {Stenrup},
  \citenamefont {Truhlar}, \citenamefont {Ungur}, \citenamefont {Valentini},
  \citenamefont {Vancoillie}, \citenamefont {Veryazov}, \citenamefont
  {Vysotskiy}, \citenamefont {Weingart}, \citenamefont {Zapata},\ and\
  \citenamefont {Lindh}}]{molcas-general}%
  \BibitemOpen
  \bibfield  {author} {\bibinfo {author} {\bibfnamefont {F.}~\bibnamefont
  {Aquilante}}, \bibinfo {author} {\bibfnamefont {J.}~\bibnamefont
  {Autschbach}}, \bibinfo {author} {\bibfnamefont {R.~K.}\ \bibnamefont
  {Carlson}}, \bibinfo {author} {\bibfnamefont {L.~F.}\ \bibnamefont
  {Chibotaru}}, \bibinfo {author} {\bibfnamefont {M.~G.}\ \bibnamefont
  {Delcey}}, \bibinfo {author} {\bibfnamefont {L.}~\bibnamefont {De~Vico}},
  \bibinfo {author} {\bibfnamefont {I.}~\bibnamefont {Fdez.~Galv{\'a}n}},
  \bibinfo {author} {\bibfnamefont {N.}~\bibnamefont {Ferr{\'e}}}, \bibinfo
  {author} {\bibfnamefont {L.~M.}\ \bibnamefont {Frutos}}, \bibinfo {author}
  {\bibfnamefont {L.}~\bibnamefont {Gagliardi}}, \bibinfo {author}
  {\bibfnamefont {M.}~\bibnamefont {Garavelli}}, \bibinfo {author}
  {\bibfnamefont {A.}~\bibnamefont {Giussani}}, \bibinfo {author}
  {\bibfnamefont {C.~E.}\ \bibnamefont {Hoyer}}, \bibinfo {author}
  {\bibfnamefont {G.}~\bibnamefont {Li~Manni}}, \bibinfo {author}
  {\bibfnamefont {H.}~\bibnamefont {Lischka}}, \bibinfo {author} {\bibfnamefont
  {D.}~\bibnamefont {Ma}}, \bibinfo {author} {\bibfnamefont {P.-{\AA}.}\
  \bibnamefont {Malmqvist}}, \bibinfo {author} {\bibfnamefont {T.}~\bibnamefont
  {M{\"u}ller}}, \bibinfo {author} {\bibfnamefont {A.}~\bibnamefont {Nenov}},
  \bibinfo {author} {\bibfnamefont {M.}~\bibnamefont {Olivucci}}, \bibinfo
  {author} {\bibfnamefont {T.~B.}\ \bibnamefont {Pedersen}}, \bibinfo {author}
  {\bibfnamefont {D.}~\bibnamefont {Peng}}, \bibinfo {author} {\bibfnamefont
  {F.}~\bibnamefont {Plasser}}, \bibinfo {author} {\bibfnamefont
  {B.}~\bibnamefont {Pritchard}}, \bibinfo {author} {\bibfnamefont
  {M.}~\bibnamefont {Reiher}}, \bibinfo {author} {\bibfnamefont
  {I.}~\bibnamefont {Rivalta}}, \bibinfo {author} {\bibfnamefont
  {I.}~\bibnamefont {Schapiro}}, \bibinfo {author} {\bibfnamefont
  {J.}~\bibnamefont {Segarra-Mart{\`i}}}, \bibinfo {author} {\bibfnamefont
  {M.}~\bibnamefont {Stenrup}}, \bibinfo {author} {\bibfnamefont {D.~G.}\
  \bibnamefont {Truhlar}}, \bibinfo {author} {\bibfnamefont {L.}~\bibnamefont
  {Ungur}}, \bibinfo {author} {\bibfnamefont {A.}~\bibnamefont {Valentini}},
  \bibinfo {author} {\bibfnamefont {S.}~\bibnamefont {Vancoillie}}, \bibinfo
  {author} {\bibfnamefont {V.}~\bibnamefont {Veryazov}}, \bibinfo {author}
  {\bibfnamefont {V.~P.}\ \bibnamefont {Vysotskiy}}, \bibinfo {author}
  {\bibfnamefont {O.}~\bibnamefont {Weingart}}, \bibinfo {author}
  {\bibfnamefont {F.}~\bibnamefont {Zapata}}, \ and\ \bibinfo {author}
  {\bibfnamefont {R.}~\bibnamefont {Lindh}},\ }\bibfield  {title} {\enquote
  {\bibinfo {title} {{Molcas 8: New capabilities for multiconfigurational
  quantum chemical calculations across the periodic table}},}\ }\href {\doibase
  10.1002/jcc.24221} {\bibfield  {journal} {\bibinfo  {journal} {J. Comput.
  Chem.}\ }\textbf {\bibinfo {volume} {37}},\ \bibinfo {pages} {506} (\bibinfo
  {year} {2016})},\ \Eprint
  {http://arxiv.org/abs/https://onlinelibrary.wiley.com/doi/pdf/10.1002/jcc.24221}
  {https://onlinelibrary.wiley.com/doi/pdf/10.1002/jcc.24221} \BibitemShut
  {NoStop}%
\bibitem [{\citenamefont {Werner}\ \emph
  {et~al.}(2012{\natexlab{a}})\citenamefont {Werner}, \citenamefont {Knowles},
  \citenamefont {Knizia}, \citenamefont {Manby},\ and\ \citenamefont
  {Sch{\"u}tz}}]{molpro-general-1}%
  \BibitemOpen
  \bibfield  {author} {\bibinfo {author} {\bibfnamefont {H.-J.}\ \bibnamefont
  {Werner}}, \bibinfo {author} {\bibfnamefont {P.~J.}\ \bibnamefont {Knowles}},
  \bibinfo {author} {\bibfnamefont {G.}~\bibnamefont {Knizia}}, \bibinfo
  {author} {\bibfnamefont {F.~R.}\ \bibnamefont {Manby}}, \ and\ \bibinfo
  {author} {\bibfnamefont {M.}~\bibnamefont {Sch{\"u}tz}},\ }\bibfield  {title}
  {\enquote {\bibinfo {title} {\texttt{Molpro}: a general-purpose quantum
  chemistry program package},}\ }\href@noop {} {\bibfield  {journal} {\bibinfo
  {journal} {Wiley Interdiscip. Rev. Comput. Mol. Sci.}\ }\textbf {\bibinfo
  {volume} {2}},\ \bibinfo {pages} {242} (\bibinfo {year}
  {2012}{\natexlab{a}})}\BibitemShut {NoStop}%
\bibitem [{\citenamefont {Werner}\ \emph {et~al.}(2015)\citenamefont {Werner},
  \citenamefont {Knowles}, \citenamefont {Knizia}, \citenamefont {Manby},
  \citenamefont {{Sch\"{u}tz}} \emph {et~al.}}]{molpro-general-2}%
  \BibitemOpen
  \bibfield  {author} {\bibinfo {author} {\bibfnamefont {H.-J.}\ \bibnamefont
  {Werner}}, \bibinfo {author} {\bibfnamefont {P.~J.}\ \bibnamefont {Knowles}},
  \bibinfo {author} {\bibfnamefont {G.}~\bibnamefont {Knizia}}, \bibinfo
  {author} {\bibfnamefont {F.~R.}\ \bibnamefont {Manby}}, \bibinfo {author}
  {\bibfnamefont {M.}~\bibnamefont {{Sch\"{u}tz}}},  \emph {et~al.},\
  }\href@noop {} {\enquote {\bibinfo {title} {\texttt{MOLPRO}, version 2015.1,
  a package of \emph{ab initio} programs},}\ } (\bibinfo {year} {2015}),\
  \bibinfo {note} {see \detokenize{http://www.molpro.net}}\BibitemShut
  {NoStop}%
\bibitem [{\citenamefont {Smeyers}\ and\ \citenamefont
  {Doreste-Suarez}(1973)}]{hphf}%
  \BibitemOpen
  \bibfield  {author} {\bibinfo {author} {\bibfnamefont {Y.~G.}\ \bibnamefont
  {Smeyers}}\ and\ \bibinfo {author} {\bibfnamefont {L.}~\bibnamefont
  {Doreste-Suarez}},\ }\bibfield  {title} {\enquote {\bibinfo {title}
  {{Half-Projected and Projected Hartree-Fock Calculations for Singlet Ground
  States. I. four-Electron Atomic Systems}},}\ }\href {\doibase
  10.1002/qua.560070406} {\bibfield  {journal} {\bibinfo  {journal} {Int. J.
  Quantum Chem.}\ }\textbf {\bibinfo {volume} {7}},\ \bibinfo {pages} {687}
  (\bibinfo {year} {1973})},\ \Eprint
  {http://arxiv.org/abs/https://onlinelibrary.wiley.com/doi/pdf/10.1002/qua.560070406}
  {https://onlinelibrary.wiley.com/doi/pdf/10.1002/qua.560070406} \BibitemShut
  {NoStop}%
\bibitem [{\citenamefont {Helgaker}, \citenamefont {J{\o}rgensen},\ and\
  \citenamefont {Olsen}(2000)}]{helgaker}%
  \BibitemOpen
  \bibfield  {author} {\bibinfo {author} {\bibfnamefont {T.}~\bibnamefont
  {Helgaker}}, \bibinfo {author} {\bibfnamefont {P.}~\bibnamefont
  {J{\o}rgensen}}, \ and\ \bibinfo {author} {\bibfnamefont {J.}~\bibnamefont
  {Olsen}},\ }\href@noop {} {\emph {\bibinfo {title} {{Molecular
  Electronic-Structure Theory}}}}\ (\bibinfo  {publisher} {John Wiley \&
  Sons},\ \bibinfo {address} {Chichester},\ \bibinfo {year} {2000})\BibitemShut
  {NoStop}%
\bibitem [{\citenamefont {Booth}\ \emph {et~al.}(2011)\citenamefont {Booth},
  \citenamefont {Cleland}, \citenamefont {Thom},\ and\ \citenamefont
  {Alavi}}]{hphf-fciqmc}%
  \BibitemOpen
  \bibfield  {author} {\bibinfo {author} {\bibfnamefont {G.~H.}\ \bibnamefont
  {Booth}}, \bibinfo {author} {\bibfnamefont {D.}~\bibnamefont {Cleland}},
  \bibinfo {author} {\bibfnamefont {A.~J.~W.}\ \bibnamefont {Thom}}, \ and\
  \bibinfo {author} {\bibfnamefont {A.}~\bibnamefont {Alavi}},\ }\bibfield
  {title} {\enquote {\bibinfo {title} {{Breaking the carbon dimer: The
  challenges of multiple bond dissociation with full configuration interaction
  quantum Monte Carlo methods}},}\ }\href {\doibase 10.1063/1.3624383}
  {\bibfield  {journal} {\bibinfo  {journal} {J. Chem. Phys.}\ }\textbf
  {\bibinfo {volume} {135}},\ \bibinfo {pages} {084104} (\bibinfo {year}
  {2011})},\ \Eprint {http://arxiv.org/abs/https://doi.org/10.1063/1.3624383}
  {https://doi.org/10.1063/1.3624383} \BibitemShut {NoStop}%
\bibitem [{\citenamefont {Booth}, \citenamefont {Smart},\ and\ \citenamefont
  {Alavi}(2014)}]{linear-scaling-fciqmc}%
  \BibitemOpen
  \bibfield  {author} {\bibinfo {author} {\bibfnamefont {G.~H.}\ \bibnamefont
  {Booth}}, \bibinfo {author} {\bibfnamefont {S.~D.}\ \bibnamefont {Smart}}, \
  and\ \bibinfo {author} {\bibfnamefont {A.}~\bibnamefont {Alavi}},\ }\bibfield
   {title} {\enquote {\bibinfo {title} {{Linear-scaling and parallelisable
  algorithms for stochastic quantum chemistry}},}\ }\href {\doibase
  10.1080/00268976.2013.877165} {\bibfield  {journal} {\bibinfo  {journal}
  {Mol. Phys.}\ }\textbf {\bibinfo {volume} {112}},\ \bibinfo {pages} {1855}
  (\bibinfo {year} {2014})},\ \Eprint
  {http://arxiv.org/abs/https://doi.org/10.1080/00268976.2013.877165}
  {https://doi.org/10.1080/00268976.2013.877165} \BibitemShut {NoStop}%
\bibitem [{\citenamefont {Gel'fand}\ and\ \citenamefont
  {Cetlin}(1950{\natexlab{a}})}]{gelfand-1}%
  \BibitemOpen
  \bibfield  {author} {\bibinfo {author} {\bibfnamefont {I.~M.}\ \bibnamefont
  {Gel'fand}}\ and\ \bibinfo {author} {\bibfnamefont {M.~L.}\ \bibnamefont
  {Cetlin}},\ }\bibfield  {title} {\enquote {\bibinfo {title}
  {Finite-dimensional representations of the group of unimodular matrices},}\
  }\href@noop {} {\bibfield  {journal} {\bibinfo  {journal} {Dokl. Akad. Nauk}\
  }\textbf {\bibinfo {volume} {71}},\ \bibinfo {pages} {825} (\bibinfo {year}
  {1950}{\natexlab{a}})}\BibitemShut {NoStop}%
\bibitem [{\citenamefont {Gel'fand}\ and\ \citenamefont
  {Cetlin}(1950{\natexlab{b}})}]{gelfand-2}%
  \BibitemOpen
  \bibfield  {author} {\bibinfo {author} {\bibfnamefont {I.~M.}\ \bibnamefont
  {Gel'fand}}\ and\ \bibinfo {author} {\bibfnamefont {M.~L.}\ \bibnamefont
  {Cetlin}},\ }\bibfield  {title} {\enquote {\bibinfo {title}
  {Finite-dimensional representations of the group of orthogonal matrices},}\
  }\href@noop {} {\bibfield  {journal} {\bibinfo  {journal} {Dokl. Akad. Nauk}\
  }\textbf {\bibinfo {volume} {71}},\ \bibinfo {pages} {1017} (\bibinfo {year}
  {1950}{\natexlab{b}})},\ \bibinfo {note} {amer. Math. Soc. Transl. 64, 116
  (1967)}\BibitemShut {NoStop}%
\bibitem [{\citenamefont {Gel'fand}(1950)}]{gelfand-3}%
  \BibitemOpen
  \bibfield  {author} {\bibinfo {author} {\bibfnamefont {I.~M.}\ \bibnamefont
  {Gel'fand}},\ }\bibfield  {title} {\enquote {\bibinfo {title} {The center of
  an infinitesimal group ring},}\ }\href@noop {} {\bibfield  {journal}
  {\bibinfo  {journal} {Mat. Sb.}\ }\textbf {\bibinfo {volume} {26(68)}},\
  \bibinfo {pages} {103} (\bibinfo {year} {1950})}\BibitemShut {NoStop}%
\bibitem [{\citenamefont {Paldus}(1974)}]{Paldus1974}%
  \BibitemOpen
  \bibfield  {author} {\bibinfo {author} {\bibfnamefont {J.}~\bibnamefont
  {Paldus}},\ }\bibfield  {title} {\enquote {\bibinfo {title} {Group
  theoretical approach to the configuration interaction and perturbation theory
  calculations for atomic and molecular systems},}\ }\href {\doibase
  10.1063/1.1681883} {\bibfield  {journal} {\bibinfo  {journal} {J. Chem.
  Phys.}\ }\textbf {\bibinfo {volume} {61}},\ \bibinfo {pages} {5321} (\bibinfo
  {year} {1974})}\BibitemShut {NoStop}%
\bibitem [{\citenamefont {Paldus}(1975)}]{Paldus1975}%
  \BibitemOpen
  \bibfield  {author} {\bibinfo {author} {\bibfnamefont {J.}~\bibnamefont
  {Paldus}},\ }\bibfield  {title} {\enquote {\bibinfo {title} {A pattern
  calculus for the unitary group approach to the electronic correlation
  problem},}\ }\href {\doibase 10.1002/qua.560090823} {\bibfield  {journal}
  {\bibinfo  {journal} {Int. J. Quantum Chem.}\ }\textbf {\bibinfo {volume}
  {9}},\ \bibinfo {pages} {165} (\bibinfo {year} {1975})},\ \Eprint
  {http://arxiv.org/abs/https://onlinelibrary.wiley.com/doi/pdf/10.1002/qua.560090823}
  {https://onlinelibrary.wiley.com/doi/pdf/10.1002/qua.560090823} \BibitemShut
  {NoStop}%
\bibitem [{\citenamefont {Paldus}(1976)}]{Paldus1976}%
  \BibitemOpen
  \bibfield  {author} {\bibinfo {author} {\bibfnamefont {J.}~\bibnamefont
  {Paldus}},\ }\bibfield  {title} {\enquote {\bibinfo {title} {{Unitary-group
  approach to the many-electron correlation problem: Relation of Gelfand and
  Weyl tableau formulations}},}\ }\href {\doibase 10.1103/PhysRevA.14.1620}
  {\bibfield  {journal} {\bibinfo  {journal} {Phys. Rev. A}\ }\textbf {\bibinfo
  {volume} {14}},\ \bibinfo {pages} {1620} (\bibinfo {year}
  {1976})}\BibitemShut {NoStop}%
\bibitem [{\citenamefont {Shavitt}(1977)}]{Shavitt1977}%
  \BibitemOpen
  \bibfield  {author} {\bibinfo {author} {\bibfnamefont {I.}~\bibnamefont
  {Shavitt}},\ }\bibfield  {title} {\enquote {\bibinfo {title} {Graph
  theoretical concepts for the unitary group approach to the many-electron
  correlation problem},}\ }\href {\doibase 10.1002/qua.560120819} {\bibfield
  {journal} {\bibinfo  {journal} {Int. J. Quantum Chem.}\ }\textbf {\bibinfo
  {volume} {12}},\ \bibinfo {pages} {131} (\bibinfo {year} {1977})},\ \Eprint
  {http://arxiv.org/abs/https://onlinelibrary.wiley.com/doi/pdf/10.1002/qua.560120819}
  {https://onlinelibrary.wiley.com/doi/pdf/10.1002/qua.560120819} \BibitemShut
  {NoStop}%
\bibitem [{\citenamefont {Shavitt}(1978)}]{Shavitt1978}%
  \BibitemOpen
  \bibfield  {author} {\bibinfo {author} {\bibfnamefont {I.}~\bibnamefont
  {Shavitt}},\ }\bibfield  {title} {\enquote {\bibinfo {title} {Matrix element
  evaluation in the unitary group approach to the electron correlation
  problem},}\ }\href {\doibase 10.1002/qua.560140803} {\bibfield  {journal}
  {\bibinfo  {journal} {Int. J. Quantum Chem.}\ }\textbf {\bibinfo {volume} {14
  S12}},\ \bibinfo {pages} {5} (\bibinfo {year} {1978})}\BibitemShut {NoStop}%
\bibitem [{\citenamefont {Paldus}(1981)}]{Paldus1981}%
  \BibitemOpen
  \bibfield  {author} {\bibinfo {author} {\bibfnamefont {J.}~\bibnamefont
  {Paldus}},\ }\bibfield  {title} {\enquote {\bibinfo {title} {{Unitary Group
  Approach to Many-Electron Correlation Problem}},}\ }in\ \href {\doibase
  10.1007/978-3-642-93163-5_1} {\emph {\bibinfo {booktitle} {{The Unitary Group
  for the Evaluation of Electronic Energy Matrix Elements}}}},\ \bibinfo
  {editor} {edited by\ \bibinfo {editor} {\bibfnamefont {J.}~\bibnamefont
  {Hinze}}}\ (\bibinfo  {publisher} {Springer Berlin Heidelberg},\ \bibinfo
  {address} {Berlin, Heidelberg},\ \bibinfo {year} {1981})\ p.~\bibinfo {pages}
  {1}\BibitemShut {NoStop}%
\bibitem [{\citenamefont {Shavitt}(1981)}]{Shavitt1981}%
  \BibitemOpen
  \bibfield  {author} {\bibinfo {author} {\bibfnamefont {I.}~\bibnamefont
  {Shavitt}},\ }\bibfield  {title} {\enquote {\bibinfo {title} {{The Graphical
  Unitary Group Approach and Its Application to Direct Configuration
  Interaction Calculations}},}\ }in\ \href {\doibase
  10.1007/978-3-642-93163-5_2} {\emph {\bibinfo {booktitle} {The Unitary Group
  for the Evaluation of Electronic Energy Matrix Elements}}},\ \bibinfo
  {editor} {edited by\ \bibinfo {editor} {\bibfnamefont {J.}~\bibnamefont
  {Hinze}}}\ (\bibinfo  {publisher} {Springer Berlin Heidelberg},\ \bibinfo
  {address} {Berlin, Heidelberg},\ \bibinfo {year} {1981})\ p.~\bibinfo {pages}
  {51}\BibitemShut {NoStop}%
\bibitem [{\citenamefont {Dobrautz}(2019)}]{dobrautz-phd}%
  \BibitemOpen
  \bibfield  {author} {\bibinfo {author} {\bibfnamefont {W.}~\bibnamefont
  {Dobrautz}},\ }\emph {\bibinfo {title} {Development of Full Configuration
  Interaction Quantum Monte Carlo Methods for Strongly Correlated Electron
  Systems}},\ \href {\doibase 10.18419/opus-10593} {Ph.D. thesis},\ \bibinfo
  {school} {University of Stuttgart} (\bibinfo {year} {2019})\BibitemShut
  {NoStop}%
\bibitem [{\citenamefont {Hachmann}, \citenamefont {Cardoen},\ and\
  \citenamefont {Chan}(2006)}]{hydrogen-chain-2}%
  \BibitemOpen
  \bibfield  {author} {\bibinfo {author} {\bibfnamefont {J.}~\bibnamefont
  {Hachmann}}, \bibinfo {author} {\bibfnamefont {W.}~\bibnamefont {Cardoen}}, \
  and\ \bibinfo {author} {\bibfnamefont {G.~K.-L.}\ \bibnamefont {Chan}},\
  }\bibfield  {title} {\enquote {\bibinfo {title} {Multireference correlation
  in long molecules with the quadratic scaling density matrix renormalization
  group},}\ }\href {\doibase 10.1063/1.2345196} {\bibfield  {journal} {\bibinfo
   {journal} {J. Chem. Phys.}\ }\textbf {\bibinfo {volume} {125}},\ \bibinfo
  {pages} {144101} (\bibinfo {year} {2006})},\ \Eprint
  {http://arxiv.org/abs/https://doi.org/10.1063/1.2345196}
  {https://doi.org/10.1063/1.2345196} \BibitemShut {NoStop}%
\bibitem [{\citenamefont {Motta}\ \emph {et~al.}(2017)\citenamefont {Motta},
  \citenamefont {Ceperley}, \citenamefont {Chan}, \citenamefont {Gomez},
  \citenamefont {Gull}, \citenamefont {Guo}, \citenamefont {Jim\'enez-Hoyos},
  \citenamefont {Lan}, \citenamefont {Li}, \citenamefont {Ma}, \citenamefont
  {Millis}, \citenamefont {Prokof'ev}, \citenamefont {Ray}, \citenamefont
  {Scuseria}, \citenamefont {Sorella}, \citenamefont {Stoudenmire},
  \citenamefont {Sun}, \citenamefont {Tupitsyn}, \citenamefont {White},
  \citenamefont {Zgid},\ and\ \citenamefont {Zhang}}]{hydrogen-chain}%
  \BibitemOpen
  \bibfield  {author} {\bibinfo {author} {\bibfnamefont {M.}~\bibnamefont
  {Motta}}, \bibinfo {author} {\bibfnamefont {D.~M.}\ \bibnamefont {Ceperley}},
  \bibinfo {author} {\bibfnamefont {G.~K.-L.}\ \bibnamefont {Chan}}, \bibinfo
  {author} {\bibfnamefont {J.~A.}\ \bibnamefont {Gomez}}, \bibinfo {author}
  {\bibfnamefont {E.}~\bibnamefont {Gull}}, \bibinfo {author} {\bibfnamefont
  {S.}~\bibnamefont {Guo}}, \bibinfo {author} {\bibfnamefont {C.~A.}\
  \bibnamefont {Jim\'enez-Hoyos}}, \bibinfo {author} {\bibfnamefont {T.~N.}\
  \bibnamefont {Lan}}, \bibinfo {author} {\bibfnamefont {J.}~\bibnamefont
  {Li}}, \bibinfo {author} {\bibfnamefont {F.}~\bibnamefont {Ma}}, \bibinfo
  {author} {\bibfnamefont {A.~J.}\ \bibnamefont {Millis}}, \bibinfo {author}
  {\bibfnamefont {N.~V.}\ \bibnamefont {Prokof'ev}}, \bibinfo {author}
  {\bibfnamefont {U.}~\bibnamefont {Ray}}, \bibinfo {author} {\bibfnamefont
  {G.~E.}\ \bibnamefont {Scuseria}}, \bibinfo {author} {\bibfnamefont
  {S.}~\bibnamefont {Sorella}}, \bibinfo {author} {\bibfnamefont {E.~M.}\
  \bibnamefont {Stoudenmire}}, \bibinfo {author} {\bibfnamefont
  {Q.}~\bibnamefont {Sun}}, \bibinfo {author} {\bibfnamefont {I.~S.}\
  \bibnamefont {Tupitsyn}}, \bibinfo {author} {\bibfnamefont {S.~R.}\
  \bibnamefont {White}}, \bibinfo {author} {\bibfnamefont {D.}~\bibnamefont
  {Zgid}}, \ and\ \bibinfo {author} {\bibfnamefont {S.}~\bibnamefont {Zhang}}
  (\bibinfo {collaboration} {Simons Collaboration on the Many-Electron
  Problem}),\ }\bibfield  {title} {\enquote {\bibinfo {title} {{Towards the
  Solution of the Many-Electron Problem in Real Materials: Equation of State of
  the Hydrogen Chain with State-of-the-Art Many-Body Methods}},}\ }\href
  {\doibase 10.1103/PhysRevX.7.031059} {\bibfield  {journal} {\bibinfo
  {journal} {Phys. Rev. X}\ }\textbf {\bibinfo {volume} {7}},\ \bibinfo {pages}
  {031059} (\bibinfo {year} {2017})}\BibitemShut {NoStop}%
\bibitem [{\citenamefont {Pariser}\ and\ \citenamefont
  {Parr}(1953{\natexlab{a}})}]{ppp-1}%
  \BibitemOpen
  \bibfield  {author} {\bibinfo {author} {\bibfnamefont {R.}~\bibnamefont
  {Pariser}}\ and\ \bibinfo {author} {\bibfnamefont {R.~G.}\ \bibnamefont
  {Parr}},\ }\bibfield  {title} {\enquote {\bibinfo {title} {{A Semi-Empirical
  Theory of the Electronic Spectra and Electronic Structure of Complex
  Unsaturated Molecules. I.}}}\ }\href {\doibase 10.1063/1.1698929} {\bibfield
  {journal} {\bibinfo  {journal} {J. Chem. Phys.}\ }\textbf {\bibinfo {volume}
  {21}},\ \bibinfo {pages} {466} (\bibinfo {year} {1953}{\natexlab{a}})},\
  \Eprint {http://arxiv.org/abs/https://doi.org/10.1063/1.1698929}
  {https://doi.org/10.1063/1.1698929} \BibitemShut {NoStop}%
\bibitem [{\citenamefont {Pariser}\ and\ \citenamefont
  {Parr}(1953{\natexlab{b}})}]{ppp-2}%
  \BibitemOpen
  \bibfield  {author} {\bibinfo {author} {\bibfnamefont {R.}~\bibnamefont
  {Pariser}}\ and\ \bibinfo {author} {\bibfnamefont {R.~G.}\ \bibnamefont
  {Parr}},\ }\bibfield  {title} {\enquote {\bibinfo {title} {{A Semi-Empirical
  Theory of the Electronic Spectra and Electronic Structure of Complex
  Unsaturated Molecules. II}},}\ }\href {\doibase 10.1063/1.1699030} {\bibfield
   {journal} {\bibinfo  {journal} {J. Chem. Phys.}\ }\textbf {\bibinfo {volume}
  {21}},\ \bibinfo {pages} {767} (\bibinfo {year} {1953}{\natexlab{b}})},\
  \Eprint {http://arxiv.org/abs/https://doi.org/10.1063/1.1699030}
  {https://doi.org/10.1063/1.1699030} \BibitemShut {NoStop}%
\bibitem [{\citenamefont {Gutzwiller}(1963)}]{Gutzwiller1963}%
  \BibitemOpen
  \bibfield  {author} {\bibinfo {author} {\bibfnamefont {M.~C.}\ \bibnamefont
  {Gutzwiller}},\ }\bibfield  {title} {\enquote {\bibinfo {title} {{Effect of
  Correlation on the Ferromagnetism of Transition Metals}},}\ }\href {\doibase
  10.1103/PhysRevLett.10.159} {\bibfield  {journal} {\bibinfo  {journal} {Phys.
  Rev. Lett.}\ }\textbf {\bibinfo {volume} {10}},\ \bibinfo {pages} {159}
  (\bibinfo {year} {1963})}\BibitemShut {NoStop}%
\bibitem [{\citenamefont {Chan}\ and\ \citenamefont
  {Head-Gordon}(2002)}]{block-dmrg-1}%
  \BibitemOpen
  \bibfield  {author} {\bibinfo {author} {\bibfnamefont {G.~K.-L.}\
  \bibnamefont {Chan}}\ and\ \bibinfo {author} {\bibfnamefont {M.}~\bibnamefont
  {Head-Gordon}},\ }\bibfield  {title} {\enquote {\bibinfo {title} {{Highly
  correlated calculations with a polynomial cost algorithm: A study of the
  density matrix renormalization group}},}\ }\href {\doibase 10.1063/1.1449459}
  {\bibfield  {journal} {\bibinfo  {journal} {J. Chem. Phys.}\ }\textbf
  {\bibinfo {volume} {116}},\ \bibinfo {pages} {4462} (\bibinfo {year}
  {2002})},\ \Eprint {http://arxiv.org/abs/https://doi.org/10.1063/1.1449459}
  {https://doi.org/10.1063/1.1449459} \BibitemShut {NoStop}%
\bibitem [{\citenamefont {Sharma}\ and\ \citenamefont
  {Chan}(2012)}]{block-dmrg-2}%
  \BibitemOpen
  \bibfield  {author} {\bibinfo {author} {\bibfnamefont {S.}~\bibnamefont
  {Sharma}}\ and\ \bibinfo {author} {\bibfnamefont {G.~K.-L.}\ \bibnamefont
  {Chan}},\ }\bibfield  {title} {\enquote {\bibinfo {title} {Spin-adapted
  density matrix renormalization group algorithms for quantum chemistry},}\
  }\href {\doibase 10.1063/1.3695642} {\bibfield  {journal} {\bibinfo
  {journal} {J. Chem. Phys.}\ }\textbf {\bibinfo {volume} {136}},\ \bibinfo
  {pages} {124121} (\bibinfo {year} {2012})},\ \Eprint
  {http://arxiv.org/abs/https://doi.org/10.1063/1.3695642}
  {https://doi.org/10.1063/1.3695642} \BibitemShut {NoStop}%
\bibitem [{\citenamefont {Chan}\ and\ \citenamefont
  {Sharma}(2011)}]{chansharma2011}%
  \BibitemOpen
  \bibfield  {author} {\bibinfo {author} {\bibfnamefont {G.~K.-L.}\
  \bibnamefont {Chan}}\ and\ \bibinfo {author} {\bibfnamefont {S.}~\bibnamefont
  {Sharma}},\ }\bibfield  {title} {\enquote {\bibinfo {title} {{The Density
  Matrix Renormalization Group in Quantum Chemistry}},}\ }\href {\doibase
  10.1146/annurev-physchem-032210-103338} {\bibfield  {journal} {\bibinfo
  {journal} {Annu. Rev. Phys. Chem.}\ }\textbf {\bibinfo {volume} {62}},\
  \bibinfo {pages} {465} (\bibinfo {year} {2011})}\BibitemShut {NoStop}%
\bibitem [{\citenamefont {White}(1992)}]{white1992}%
  \BibitemOpen
  \bibfield  {author} {\bibinfo {author} {\bibfnamefont {S.~R.}\ \bibnamefont
  {White}},\ }\bibfield  {title} {\enquote {\bibinfo {title} {Density matrix
  formulation for quantum renormalization groups},}\ }\href {\doibase
  10.1103/PhysRevLett.69.2863} {\bibfield  {journal} {\bibinfo  {journal}
  {Phys. Rev. Lett.}\ }\textbf {\bibinfo {volume} {69}},\ \bibinfo {pages}
  {2863} (\bibinfo {year} {1992})}\BibitemShut {NoStop}%
\bibitem [{\citenamefont {Li~Manni}, \citenamefont {Dobrautz},\ and\
  \citenamefont {Alavi}(2020)}]{LiManni2019}%
  \BibitemOpen
  \bibfield  {author} {\bibinfo {author} {\bibfnamefont {G.}~\bibnamefont
  {Li~Manni}}, \bibinfo {author} {\bibfnamefont {W.}~\bibnamefont {Dobrautz}},
  \ and\ \bibinfo {author} {\bibfnamefont {A.}~\bibnamefont {Alavi}},\
  }\bibfield  {title} {\enquote {\bibinfo {title} {Compression of spin-adapted
  multiconfigurational wave functions in exchange-coupled polynuclear spin
  systems},}\ }\href {\doibase 10.1021/acs.jctc.9b01013} {\bibfield  {journal}
  {\bibinfo  {journal} {Journal of Chemical Theory and Computation}\ }\textbf
  {\bibinfo {volume} {16}},\ \bibinfo {pages} {2202--2215} (\bibinfo {year}
  {2020})},\ \bibinfo {note} {pMID: 32053374}\BibitemShut {NoStop}%
\bibitem [{\citenamefont {Reiher}\ \emph {et~al.}(2017)\citenamefont {Reiher},
  \citenamefont {Wiebe}, \citenamefont {Svore}, \citenamefont {Wecker},\ and\
  \citenamefont {Troyer}}]{Reiher2017a}%
  \BibitemOpen
  \bibfield  {author} {\bibinfo {author} {\bibfnamefont {M.}~\bibnamefont
  {Reiher}}, \bibinfo {author} {\bibfnamefont {N.}~\bibnamefont {Wiebe}},
  \bibinfo {author} {\bibfnamefont {K.~M.}\ \bibnamefont {Svore}}, \bibinfo
  {author} {\bibfnamefont {D.}~\bibnamefont {Wecker}}, \ and\ \bibinfo {author}
  {\bibfnamefont {M.}~\bibnamefont {Troyer}},\ }\bibfield  {title} {\enquote
  {\bibinfo {title} {Elucidating reaction mechanisms on quantum computers},}\
  }\href {\doibase 10.1073/pnas.1619152114} {\bibfield  {journal} {\bibinfo
  {journal} {Proc. Natl. Acad. Sci.}\ }\textbf {\bibinfo {volume} {114}},\
  \bibinfo {pages} {7555--7560} (\bibinfo {year} {2017})}\BibitemShut {NoStop}%
\bibitem [{sca()}]{scaling_output}%
  \BibitemOpen
  \href@noop {} {}\bibinfo {note} {See output files {\tt
  output\_file\_scaling\_with\_*\_cores.txt} and {\tt
  output\_file\_energy\_with\_8b\_walkers.txt} in the supplemental
  material~\cite{supplement}}\BibitemShut {NoStop}%
\bibitem [{\citenamefont {Li}\ \emph {et~al.}(2019)\citenamefont {Li},
  \citenamefont {Li}, \citenamefont {Dattani}, \citenamefont {Umrigar},\ and\
  \citenamefont {Chan}}]{Li2019}%
  \BibitemOpen
  \bibfield  {author} {\bibinfo {author} {\bibfnamefont {Z.}~\bibnamefont
  {Li}}, \bibinfo {author} {\bibfnamefont {J.}~\bibnamefont {Li}}, \bibinfo
  {author} {\bibfnamefont {N.~S.}\ \bibnamefont {Dattani}}, \bibinfo {author}
  {\bibfnamefont {C.~J.}\ \bibnamefont {Umrigar}}, \ and\ \bibinfo {author}
  {\bibfnamefont {G.~K.-L.}\ \bibnamefont {Chan}},\ }\bibfield  {title}
  {\enquote {\bibinfo {title} {The electronic complexity of the ground-state of
  the femo cofactor of nitrogenase as relevant to quantum simulations},}\
  }\href {\doibase 10.1063/1.5063376} {\bibfield  {journal} {\bibinfo
  {journal} {The Journal of Chemical Physics}\ }\textbf {\bibinfo {volume}
  {150}},\ \bibinfo {pages} {024302} (\bibinfo {year} {2019})},\ \Eprint
  {http://arxiv.org/abs/https://doi.org/10.1063/1.5063376}
  {https://doi.org/10.1063/1.5063376} \BibitemShut {NoStop}%
\bibitem [{loa()}]{load_imbalance_output}%
  \BibitemOpen
  \href@noop {} {}\bibinfo {note} {See output files {\tt
  output\_file\_load\_imbalance\_n*.txt} in the supplemental
  material\cite{supplement}}\BibitemShut {NoStop}%
\bibitem [{fem()}]{femoco_lb}%
  \BibitemOpen
  \href@noop {} {}\bibinfo {note} {The FeMoco calculations were performed
  before the introduction of the PCHB excitaiton generator and thus using the
  Cauchy-Schwartz excitation generator, which is expected to yield higher load
  imbalance. Therefore, the FeMoco calculations have higher load imbalance at
  all considered scales compared to the Cr$_2$ example.}\BibitemShut {Stop}%
\bibitem [{\citenamefont {{Hunter}}(2007)}]{matplotlib}%
  \BibitemOpen
  \bibfield  {author} {\bibinfo {author} {\bibfnamefont {J.~D.}\ \bibnamefont
  {{Hunter}}},\ }\bibfield  {title} {\enquote {\bibinfo {title} {Matplotlib: A
  2d graphics environment},}\ }\href {\doibase 10.1109/MCSE.2007.55} {\bibfield
   {journal} {\bibinfo  {journal} {Computing in Science Engineering}\ }\textbf
  {\bibinfo {volume} {9}},\ \bibinfo {pages} {90--95} (\bibinfo {year}
  {2007})}\BibitemShut {NoStop}%
\bibitem [{\citenamefont {Knowles}\ and\ \citenamefont
  {Handy}(1989)}]{Knowles1989}%
  \BibitemOpen
  \bibfield  {author} {\bibinfo {author} {\bibfnamefont {P.~J.}\ \bibnamefont
  {Knowles}}\ and\ \bibinfo {author} {\bibfnamefont {N.~C.}\ \bibnamefont
  {Handy}},\ }\bibfield  {title} {\enquote {\bibinfo {title} {A determinant
  based full configuration interaction program},}\ }\href {\doibase
  https://doi.org/10.1016/0010-4655(89)90033-7} {\bibfield  {journal} {\bibinfo
   {journal} {Computer Physics Communications}\ }\textbf {\bibinfo {volume}
  {54}},\ \bibinfo {pages} {75 -- 83} (\bibinfo {year} {1989})}\BibitemShut
  {NoStop}%
\bibitem [{\citenamefont {Werner}\ \emph
  {et~al.}(2012{\natexlab{b}})\citenamefont {Werner}, \citenamefont {Knowles},
  \citenamefont {Knizia}, \citenamefont {Manby},\ and\ \citenamefont
  {Sch{\"u}tz}}]{MOLPRO-WIREs}%
  \BibitemOpen
  \bibfield  {author} {\bibinfo {author} {\bibfnamefont {H.-J.}\ \bibnamefont
  {Werner}}, \bibinfo {author} {\bibfnamefont {P.~J.}\ \bibnamefont {Knowles}},
  \bibinfo {author} {\bibfnamefont {G.}~\bibnamefont {Knizia}}, \bibinfo
  {author} {\bibfnamefont {F.~R.}\ \bibnamefont {Manby}}, \ and\ \bibinfo
  {author} {\bibfnamefont {M.}~\bibnamefont {Sch{\"u}tz}},\ }\bibfield  {title}
  {\enquote {\bibinfo {title} {{Molpro: a general-purpose quantum chemistry
  program package}},}\ }\href@noop {} {\bibfield  {journal} {\bibinfo
  {journal} {WIREs Comput Mol Sci}\ }\textbf {\bibinfo {volume} {2}},\ \bibinfo
  {pages} {242--253} (\bibinfo {year} {2012}{\natexlab{b}})}\BibitemShut
  {NoStop}%
\bibitem [{\citenamefont {Werner}\ \emph {et~al.}(2019)\citenamefont {Werner},
  \citenamefont {Knowles}, \citenamefont {Knizia}, \citenamefont {Manby},
  \citenamefont {{Sch\"{u}tz}} \emph {et~al.}}]{MOLPRO_brief}%
  \BibitemOpen
  \bibfield  {author} {\bibinfo {author} {\bibfnamefont {H.-J.}\ \bibnamefont
  {Werner}}, \bibinfo {author} {\bibfnamefont {P.~J.}\ \bibnamefont {Knowles}},
  \bibinfo {author} {\bibfnamefont {G.}~\bibnamefont {Knizia}}, \bibinfo
  {author} {\bibfnamefont {F.~R.}\ \bibnamefont {Manby}}, \bibinfo {author}
  {\bibfnamefont {M.}~\bibnamefont {{Sch\"{u}tz}}},  \emph {et~al.},\ }\href
  {https://www.molpro.net} {\enquote {\bibinfo {title} {Molpro, version 2019.2,
  a package of ab initio programs},}\ } (\bibinfo {year} {2019})\BibitemShut
  {NoStop}%
\bibitem [{\citenamefont {Fdez.~Galv{\'a}n}\ \emph {et~al.}(2019)\citenamefont
  {Fdez.~Galv{\'a}n}, \citenamefont {Vacher}, \citenamefont {Alavi},
  \citenamefont {Angeli}, \citenamefont {Aquilante}, \citenamefont
  {Autschbach}, \citenamefont {Bao}, \citenamefont {Bokarev}, \citenamefont
  {Bogdanov}, \citenamefont {Carlson}, \citenamefont {Chibotaru}, \citenamefont
  {Creutzberg}, \citenamefont {Dattani}, \citenamefont {Delcey}, \citenamefont
  {Dong}, \citenamefont {Dreuw}, \citenamefont {Freitag}, \citenamefont
  {Frutos}, \citenamefont {Gagliardi}, \citenamefont {Gendron}, \citenamefont
  {Giussani}, \citenamefont {Gonz{\`a}lez}, \citenamefont {Grell},
  \citenamefont {Guo}, \citenamefont {Hoyer}, \citenamefont {Johansson},
  \citenamefont {Keller}, \citenamefont {Knecht}, \citenamefont {Kova{\u
  c}evi{\'c}}, \citenamefont {K{\"a}llman}, \citenamefont {Li~Manni},
  \citenamefont {Lundberg}, \citenamefont {Ma}, \citenamefont {Mai},
  \citenamefont {Malhado}, \citenamefont {Malmqvist}, \citenamefont
  {Marquetand}, \citenamefont {Mewes}, \citenamefont {Norell}, \citenamefont
  {Olivucci}, \citenamefont {Oppel}, \citenamefont {Phung}, \citenamefont
  {Pierloot}, \citenamefont {Plasser}, \citenamefont {Reiher}, \citenamefont
  {Sand}, \citenamefont {Schapiro}, \citenamefont {Sharma}, \citenamefont
  {Stein}, \citenamefont {S{\o}rensen}, \citenamefont {Truhlar}, \citenamefont
  {Ugandi}, \citenamefont {Ungur}, \citenamefont {Valentini}, \citenamefont
  {Vancoillie}, \citenamefont {Veryazov}, \citenamefont {Weser}, \citenamefont
  {Weso{\l}owski}, \citenamefont {Widmark}, \citenamefont {Wouters},
  \citenamefont {Zech}, \citenamefont {Zobel},\ and\ \citenamefont
  {Lindh}}]{OpenMolcas}%
  \BibitemOpen
  \bibfield  {author} {\bibinfo {author} {\bibfnamefont {I.}~\bibnamefont
  {Fdez.~Galv{\'a}n}}, \bibinfo {author} {\bibfnamefont {M.}~\bibnamefont
  {Vacher}}, \bibinfo {author} {\bibfnamefont {A.}~\bibnamefont {Alavi}},
  \bibinfo {author} {\bibfnamefont {C.}~\bibnamefont {Angeli}}, \bibinfo
  {author} {\bibfnamefont {F.}~\bibnamefont {Aquilante}}, \bibinfo {author}
  {\bibfnamefont {J.}~\bibnamefont {Autschbach}}, \bibinfo {author}
  {\bibfnamefont {J.~J.}\ \bibnamefont {Bao}}, \bibinfo {author} {\bibfnamefont
  {S.~I.}\ \bibnamefont {Bokarev}}, \bibinfo {author} {\bibfnamefont {N.~A.}\
  \bibnamefont {Bogdanov}}, \bibinfo {author} {\bibfnamefont {R.~K.}\
  \bibnamefont {Carlson}}, \bibinfo {author} {\bibfnamefont {L.~F.}\
  \bibnamefont {Chibotaru}}, \bibinfo {author} {\bibfnamefont {J.}~\bibnamefont
  {Creutzberg}}, \bibinfo {author} {\bibfnamefont {N.}~\bibnamefont {Dattani}},
  \bibinfo {author} {\bibfnamefont {M.~G.}\ \bibnamefont {Delcey}}, \bibinfo
  {author} {\bibfnamefont {S.~S.}\ \bibnamefont {Dong}}, \bibinfo {author}
  {\bibfnamefont {A.}~\bibnamefont {Dreuw}}, \bibinfo {author} {\bibfnamefont
  {L.}~\bibnamefont {Freitag}}, \bibinfo {author} {\bibfnamefont {L.~M.}\
  \bibnamefont {Frutos}}, \bibinfo {author} {\bibfnamefont {L.}~\bibnamefont
  {Gagliardi}}, \bibinfo {author} {\bibfnamefont {F.}~\bibnamefont {Gendron}},
  \bibinfo {author} {\bibfnamefont {A.}~\bibnamefont {Giussani}}, \bibinfo
  {author} {\bibfnamefont {L.}~\bibnamefont {Gonz{\`a}lez}}, \bibinfo {author}
  {\bibfnamefont {G.}~\bibnamefont {Grell}}, \bibinfo {author} {\bibfnamefont
  {M.}~\bibnamefont {Guo}}, \bibinfo {author} {\bibfnamefont {C.~E.}\
  \bibnamefont {Hoyer}}, \bibinfo {author} {\bibfnamefont {M.}~\bibnamefont
  {Johansson}}, \bibinfo {author} {\bibfnamefont {S.}~\bibnamefont {Keller}},
  \bibinfo {author} {\bibfnamefont {S.}~\bibnamefont {Knecht}}, \bibinfo
  {author} {\bibfnamefont {G.}~\bibnamefont {Kova{\u c}evi{\'c}}}, \bibinfo
  {author} {\bibfnamefont {E.}~\bibnamefont {K{\"a}llman}}, \bibinfo {author}
  {\bibfnamefont {G.}~\bibnamefont {Li~Manni}}, \bibinfo {author}
  {\bibfnamefont {M.}~\bibnamefont {Lundberg}}, \bibinfo {author}
  {\bibfnamefont {Y.}~\bibnamefont {Ma}}, \bibinfo {author} {\bibfnamefont
  {S.}~\bibnamefont {Mai}}, \bibinfo {author} {\bibfnamefont {J.~P.}\
  \bibnamefont {Malhado}}, \bibinfo {author} {\bibfnamefont {P.~{\r A}.}\
  \bibnamefont {Malmqvist}}, \bibinfo {author} {\bibfnamefont {P.}~\bibnamefont
  {Marquetand}}, \bibinfo {author} {\bibfnamefont {S.~A.}\ \bibnamefont
  {Mewes}}, \bibinfo {author} {\bibfnamefont {J.}~\bibnamefont {Norell}},
  \bibinfo {author} {\bibfnamefont {M.}~\bibnamefont {Olivucci}}, \bibinfo
  {author} {\bibfnamefont {M.}~\bibnamefont {Oppel}}, \bibinfo {author}
  {\bibfnamefont {Q.~M.}\ \bibnamefont {Phung}}, \bibinfo {author}
  {\bibfnamefont {K.}~\bibnamefont {Pierloot}}, \bibinfo {author}
  {\bibfnamefont {F.}~\bibnamefont {Plasser}}, \bibinfo {author} {\bibfnamefont
  {M.}~\bibnamefont {Reiher}}, \bibinfo {author} {\bibfnamefont {A.~M.}\
  \bibnamefont {Sand}}, \bibinfo {author} {\bibfnamefont {I.}~\bibnamefont
  {Schapiro}}, \bibinfo {author} {\bibfnamefont {P.}~\bibnamefont {Sharma}},
  \bibinfo {author} {\bibfnamefont {C.~J.}\ \bibnamefont {Stein}}, \bibinfo
  {author} {\bibfnamefont {L.~K.}\ \bibnamefont {S{\o}rensen}}, \bibinfo
  {author} {\bibfnamefont {D.~G.}\ \bibnamefont {Truhlar}}, \bibinfo {author}
  {\bibfnamefont {M.}~\bibnamefont {Ugandi}}, \bibinfo {author} {\bibfnamefont
  {L.}~\bibnamefont {Ungur}}, \bibinfo {author} {\bibfnamefont
  {A.}~\bibnamefont {Valentini}}, \bibinfo {author} {\bibfnamefont
  {S.}~\bibnamefont {Vancoillie}}, \bibinfo {author} {\bibfnamefont
  {V.}~\bibnamefont {Veryazov}}, \bibinfo {author} {\bibfnamefont
  {O.}~\bibnamefont {Weser}}, \bibinfo {author} {\bibfnamefont {T.~A.}\
  \bibnamefont {Weso{\l}owski}}, \bibinfo {author} {\bibfnamefont {P.-O.}\
  \bibnamefont {Widmark}}, \bibinfo {author} {\bibfnamefont {S.}~\bibnamefont
  {Wouters}}, \bibinfo {author} {\bibfnamefont {A.}~\bibnamefont {Zech}},
  \bibinfo {author} {\bibfnamefont {J.~P.}\ \bibnamefont {Zobel}}, \ and\
  \bibinfo {author} {\bibfnamefont {R.}~\bibnamefont {Lindh}},\ }\bibfield
  {title} {\enquote {\bibinfo {title} {Openmolcas: From source code to
  insight},}\ }\href {\doibase 10.1021/acs.jctc.9b00532} {\bibfield  {journal}
  {\bibinfo  {journal} {Journal of Chemical Theory and Computation}\ }\textbf
  {\bibinfo {volume} {15}},\ \bibinfo {pages} {5925--5964} (\bibinfo {year}
  {2019})},\ \bibinfo {note} {pMID: 31509407},\ \Eprint
  {http://arxiv.org/abs/https://doi.org/10.1021/acs.jctc.9b00532}
  {https://doi.org/10.1021/acs.jctc.9b00532} \BibitemShut {NoStop}%
\bibitem [{\citenamefont {Sun}\ \emph {et~al.}(2017)\citenamefont {Sun},
  \citenamefont {Berkelbach}, \citenamefont {Blunt}, \citenamefont {Booth},
  \citenamefont {Guo}, \citenamefont {Li}, \citenamefont {Liu}, \citenamefont
  {McClain}, \citenamefont {Sharma}, \citenamefont {Wouters},\ and\
  \citenamefont {Chan}}]{pyscf}%
  \BibitemOpen
  \bibfield  {author} {\bibinfo {author} {\bibfnamefont {Q.}~\bibnamefont
  {Sun}}, \bibinfo {author} {\bibfnamefont {T.~C.}\ \bibnamefont {Berkelbach}},
  \bibinfo {author} {\bibfnamefont {N.~S.}\ \bibnamefont {Blunt}}, \bibinfo
  {author} {\bibfnamefont {G.~H.}\ \bibnamefont {Booth}}, \bibinfo {author}
  {\bibfnamefont {S.}~\bibnamefont {Guo}}, \bibinfo {author} {\bibfnamefont
  {Z.}~\bibnamefont {Li}}, \bibinfo {author} {\bibfnamefont {J.}~\bibnamefont
  {Liu}}, \bibinfo {author} {\bibfnamefont {J.}~\bibnamefont {McClain}},
  \bibinfo {author} {\bibfnamefont {S.}~\bibnamefont {Sharma}}, \bibinfo
  {author} {\bibfnamefont {S.}~\bibnamefont {Wouters}}, \ and\ \bibinfo
  {author} {\bibfnamefont {G.~K.-L.}\ \bibnamefont {Chan}},\ }\bibfield
  {title} {\enquote {\bibinfo {title} {Pyscf: The python-based simulations of
  chemistry framework},}\ }\href@noop {} {\bibfield  {journal} {\bibinfo
  {journal} {WIREs Comput Mol Sci 2018}\ }\textbf {\bibinfo {volume} {8}},\
  \bibinfo {pages} {e1340} (\bibinfo {year} {2017})}\BibitemShut {NoStop}%
\bibitem [{\citenamefont {Kresse}\ and\ \citenamefont
  {Furthm\"uller}(1996)}]{Kresse1996}%
  \BibitemOpen
  \bibfield  {author} {\bibinfo {author} {\bibfnamefont {G.}~\bibnamefont
  {Kresse}}\ and\ \bibinfo {author} {\bibfnamefont {J.}~\bibnamefont
  {Furthm\"uller}},\ }\bibfield  {title} {\enquote {\bibinfo {title} {Efficient
  iterative schemes for ab initio total-energy calculations using a plane-wave
  basis set},}\ }\href {\doibase 10.1103/PhysRevB.54.11169} {\bibfield
  {journal} {\bibinfo  {journal} {Phys. Rev. B}\ }\textbf {\bibinfo {volume}
  {54}},\ \bibinfo {pages} {11169--11186} (\bibinfo {year} {1996})}\BibitemShut
  {NoStop}%
\bibitem [{\citenamefont {Sun}, \citenamefont {Yang},\ and\ \citenamefont
  {Chan}(2017)}]{Sun2017}%
  \BibitemOpen
  \bibfield  {author} {\bibinfo {author} {\bibfnamefont {Q.}~\bibnamefont
  {Sun}}, \bibinfo {author} {\bibfnamefont {J.}~\bibnamefont {Yang}}, \ and\
  \bibinfo {author} {\bibfnamefont {G.~K.-L.}\ \bibnamefont {Chan}},\
  }\bibfield  {title} {\enquote {\bibinfo {title} {A general second order
  complete active space self-consistent-field solver for large-scale
  systems},}\ }\href {\doibase https://doi.org/10.1016/j.cplett.2017.03.004}
  {\bibfield  {journal} {\bibinfo  {journal} {Chemical Physics Letters}\
  }\textbf {\bibinfo {volume} {683}},\ \bibinfo {pages} {291 -- 299} (\bibinfo
  {year} {2017})},\ \bibinfo {note} {ahmed Zewail (1946-2016) Commemoration
  Issue of Chemical Physics Letters}\BibitemShut {NoStop}%
\bibitem [{\citenamefont {Yanai}\ \emph {et~al.}(2009)\citenamefont {Yanai},
  \citenamefont {Kurashige}, \citenamefont {Ghosh},\ and\ \citenamefont
  {Chan}}]{Yanai2009}%
  \BibitemOpen
  \bibfield  {author} {\bibinfo {author} {\bibfnamefont {T.}~\bibnamefont
  {Yanai}}, \bibinfo {author} {\bibfnamefont {Y.}~\bibnamefont {Kurashige}},
  \bibinfo {author} {\bibfnamefont {D.}~\bibnamefont {Ghosh}}, \ and\ \bibinfo
  {author} {\bibfnamefont {G.~K.-L.}\ \bibnamefont {Chan}},\ }\bibfield
  {title} {\enquote {\bibinfo {title} {Accelerating convergence in iterative
  solution for large-scale complete active space self-consistent-field
  calculations},}\ }\href {\doibase 10.1002/qua.22099} {\bibfield  {journal}
  {\bibinfo  {journal} {International Journal of Quantum Chemistry}\ }\textbf
  {\bibinfo {volume} {109}},\ \bibinfo {pages} {2178--2190} (\bibinfo {year}
  {2009})}\BibitemShut {NoStop}%
\bibitem [{\citenamefont {Bogdanov}\ \emph {et~al.}(2018)\citenamefont
  {Bogdanov}, \citenamefont {Li~Manni}, \citenamefont {Sharma}, \citenamefont
  {Gunnarsson},\ and\ \citenamefont {Alavi}}]{Bogdanov2018}%
  \BibitemOpen
  \bibfield  {author} {\bibinfo {author} {\bibfnamefont {N.~A.}\ \bibnamefont
  {Bogdanov}}, \bibinfo {author} {\bibfnamefont {G.}~\bibnamefont {Li~Manni}},
  \bibinfo {author} {\bibfnamefont {S.}~\bibnamefont {Sharma}}, \bibinfo
  {author} {\bibfnamefont {O.}~\bibnamefont {Gunnarsson}}, \ and\ \bibinfo
  {author} {\bibfnamefont {A.}~\bibnamefont {Alavi}},\ }\bibfield  {title}
  {\enquote {\bibinfo {title} {New superexchange paths due to
  breathing-enhanced hopping in corner-sharing cuprates},}\ }\href {\doibase
  arXiv:1803.07026} {\bibfield  {journal} {\bibinfo  {journal} {Arxiv}\ }
  (\bibinfo {year} {2018}),\ arXiv:1803.07026}\BibitemShut {NoStop}%
\bibitem [{sup()}]{supplement}%
  \BibitemOpen
  \href@noop {} {}\bibinfo {note} {Supplemental material, available at
  supplement-url}\BibitemShut {NoStop}%
\bibitem [{\citenamefont {Walker}(1977)}]{Walker1977}%
  \BibitemOpen
  \bibfield  {author} {\bibinfo {author} {\bibfnamefont {A.~J.}\ \bibnamefont
  {Walker}},\ }\bibfield  {title} {\enquote {\bibinfo {title} {An efficient
  method for generating discrete random variables with general
  distributions},}\ }\href {\doibase 10.1145/355744.355749} {\bibfield
  {journal} {\bibinfo  {journal} {ACM Trans. Math. Softw.}\ }\textbf {\bibinfo
  {volume} {3}},\ \bibinfo {pages} {253--256} (\bibinfo {year}
  {1977})}\BibitemShut {NoStop}%
\end{thebibliography}
\end{document}